\pdfoutput=1   % to process with pdflatex

\documentclass[twocolappendix]{emulateapj}
\usepackage{apjfonts}
\usepackage[bookmarks=false, colorlinks=true, citecolor=blue, backref=true]{hyperref}
\newcommand{\sfr}{{\rm SFR}}
\newcommand{\hst}{{\it HST}}
\newcommand{\galex}{{\it GALEX}}
\newcommand{\ha}{H$\alpha$}
\newcommand{\HI}{{\ion{H}{1}}}

\shorttitle{STAR FORMATION IN RED SEQUENCE GALAXIES}
\shortauthors{SALIM ET AL.}

\begin{document}

\title{Galaxy-scale star formation on the red sequence: the
  continued growth of S0\lowercase{s} and the quiescence of ellipticals} 

%\title{Galaxy-scale star formation in red sequence early-type galaxies: the
%  continued growth of S0\lowercase{s} and the quiescence of ellipticals} 

\author{Samir Salim\altaffilmark{1},
Jerome J. Fang\altaffilmark{2},
R.\ Michael Rich\altaffilmark{3},
S.\ M.\ Faber\altaffilmark{2},
David A.\ Thilker\altaffilmark{4}}

\altaffiltext{1}{Department of Astronomy, Indiana University,
  Bloomington, IN 47404, salims@indiana.edu}
\altaffiltext{2}{University of California Observatories/Lick Observatory, 
University of California, Santa Cruz, CA 95064}
\altaffiltext{3}{Department of Physics and Astronomy, University of
  California, Los Angeles, CA 90095}
\altaffiltext{4}{Center for Astrophysical Sciences, The Johns Hopkins
  University, Baltimore, MD 21218}

\begin{abstract}
  This paper examines star formation (SF) in relatively massive,
  primarily early-type galaxies (ETGs) at $z\sim 0.1$. A sample is
  drawn from bulge-dominated \galex/SDSS galaxies on the optical red
  sequence with strong UV excess and yet quiescent SDSS
  spectra. High-resolution far-UV imaging of 27 such ETGs using \hst\
  ACS/SBC reveals structured UV morphology in 93\% of the sample,
  consistent with low-level ongoing SF ($\sim 0.5 M_{\odot} {\rm
    yr}^{-1}$). In 3/4 of the sample the SF is extended on galaxy
  scales (25--75 kpc), while the rest contains smaller (5--15 kpc) SF
  patches in the vicinity of an ETG--presumably gas-rich satellites
  being disrupted. Optical imaging reveals that {\it all} ETGs with
  galaxy-scale SF in our sample have old stellar disks (mostly S0
  type). None is classified as a true elliptical. In our sample,
  galaxy-scale SF takes the form of UV rings of varying sizes and
  morphologies. For the majority of such objects we conclude that the
  gas needed to fuel current SF has been accreted from the IGM,
  probably in a prolonged, quasi-static manner, leading in some cases
  to additional disk buildup. The remaining ETGs with galaxy-scale SF
  have UV and optical morphologies consistent with minor merger-driven
  SF or with the final stages of SF in fading spirals.  Our analysis
  excludes that {\it all} recent SF on the red sequence resulted from
  gas-rich mergers. We find further evidence that galaxy-scale SF is
  almost exclusively an S0 phenomenon ($\sim 20\%$ S0s have SF) by
  examining the {\it overall} optically red SDSS ETGs. Conclusion is
  that significant number of field S0s maintain or resume low-level SF
  because the preventive feedback is not in place or is
  intermittent. True ellipticals, on the other hand, stay entirely
  quiescent even in the field.
 \end{abstract}


%%%
\section{Introduction} \label{sec:intro}
%%%

%Bismillahir rahmanir rahim

The processes that control global star formation (SF) in galaxies are at
the core of many current studies that aim to understand galaxy formation and
evolution. Efforts are being made both to recognize a range of
relevant processes and, perhaps more challenging, to establish which
processes dominate in different types of galaxies, in different
environments, and at different cosmic epochs. Since at the present
time the galaxies are still growing and transforming, many of the
mechanisms of SF regulation can be studied at lower
redshifts where the data are of much higher quality. However, despite
major advances in this area, there are still many open questions.

One of the most enduring puzzles is presented by the evolution of
massive early-type galaxies (ETGs), which comprise elliptical and
lenticular (S0) galaxies. Their overall old stellar populations (e.g.,
\citealt{trager,delucia06}) and consequently the lack of current SF
are to first order at odds with the otherwise successful model of
hierarchical assembly of galaxies \citep{kauffmann96} in which the
massive galaxies formed most recently. While their morphological and
kinematical transformation from disk galaxies into spheroid-dominated
ones is very well explained by dissipative major mergers (especially
for massive ellipticals, \citealt{barnes}), the question of their
subsequent quiescence is a separate question, and has started to be
addressed only more recently with the introduction of various
non-stellar feedback processes, especially the feedback from active
galactic nuclei (AGN) (\citealt{springel}). It now appears that in
addition to shutting down of star formation more or less concurrently
with the morphological transformation, another feedback mechanism is
required to keep an ETG free from subsequent SF \citep{croton1}. This
requirement for a {\it maintenance} (or {\it preventive}) feedback is
especially strong for field ETGs, which could be expected to continue
accreting cold gas from the intergalactic environment (e.g.,
\citealt{gabor}). Such accretion is probably the primary source of gas
for actively star forming galaxies as well \citep{keres}, without
which they would not be able to sustain their observed star formation
rates (SFR) for longer than a few Gyr
\citep{larson80,kennicutt94,bauermeister}.

While ETGs are {\it mostly} quiescent almost by definition, the
question is are their {\it entirely} quiescent? What fraction is?  If
they exhibit SF, is it a new episode due to a fresh supply of gas,
i.e., is galaxy being ``rejuvenated'' or are we seeing remnants of the
original disk SF? Is, in the former case, the SF present because of
the failure or absence of preventive feedback mechanism? Does SF tend
to be wide-spread, as in spiral galaxies, or confined to circumnuclear
regions? The dominance of old populations in ETGs makes the detection
of relatively small amounts of SF intrinsically difficult. However, if
SF could be systematically detected in ETGs, or, more generally, on
the ``red sequence'', it would represent a potentially powerful way of
identifying processes that lead to or prevent star formation. In
contrast, these regulative processes are more difficult to study in
actively star-forming (``blue cloud'') galaxies because of the high
``background'' of normal SF.

The presence of SF, especially in the {\it central} regions, has been
firmly established in some nearby ETGs; e.g., in the SAURON survey of
48 S0s and ellipticals \citep{combes,temi,shapiro,crocker}. Star
formation on {\it galaxy scales}, which we refer to to as the {\it
  extended SF}, has received less attention, to some extent because of
the difficulties in detecting it using optical methods. Detection of
ionized emission can be challenging in ETGs where the strong continuum
from old stars lowers the equivalent widths. Also, ionized emission
can arise from a number of sources not associated with SF
\citep{sarzi10}. Mid-IR emission (e.g., 24 $\mu$m from {\it
  Spitzer}/MIPS) in ETGs can have an order of magnitude stronger
contribution from intermediate-age and older stellar populations than
from the young stars \citep{salim09,kelson}. The most promising method
is the UV emission from young massive stars, which is an order of
magnitude more sensitive to recent SF than the blue optical flux
\citep{kauff07}, and probes timescales that are closer to the current
SF than those probed by the blue light from less massive, longer lived
stars ($\sim 100$ Myr vs.\ $\sim 1$ Gyr)\footnote{We note that far-UV
  (FUV, $\lambda \sim 1500\AA$) is easier to interpret as SF than the
  longer wavelength near-UV (NUV, $\lambda \sim 2300\AA$, which in
  ETGs primarily comes from main sequence turn off stars and is
  strongly dependent on the stellar metallicity
  \citep{donas,dorman03,smith12}.}.

Large-scale detection and characterization of ETGs in the UV was
facilitated with the UV surveys of \galex \citep{martin05}. Initial
\galex\ studies (e.g., \citealt{yi,rich}) based their approach on
selecting large statistical samples of galaxies on the {\it red
  optical sequence}, which is where most ETGs are found. While most
{\it optical} red galaxies remained red in the UV-optical colors, a
significant fraction exhibited a UV excess, which \citet{yi}
interpreted as low levels of ongoing SF\footnote{This underlines the
  importance of specifying the type of color (optical vs.\ UV-optical)
  when referring to the red sequence, as the optical red sequence will
  contain both truly quiescent galaxies and those with small relative
  amounts of SF.}. Since these studies culled their samples from SDSS
spectroscopic survey in which galaxies are typically found at $z\sim
0.1$, little could be said about the morphology of the UV light
(\galex\ has a resolution of 5$"$) and consequently whether this UV
excess actually arose from young stars.

Subsequent efforts focused on the UV morphology to confirm the
presence of the {\it extended} SF in the optical red sequence and/or
among individual ETGs
\citep{donovan,thilker10,cortesehughes,sr2010,marino1,lemonias}. However,
many questions remained: what is the origin of the star-forming gas in
ETGs?; has this SF started recently or is it prolonged?; is SF related
to the processes of disk building?; how the SF relates to the two
types of ETGs: lenticulars (S0s) and ellipticals (Es). This paper and
the accompanying work (Fang et al.\ 2012, Paper II) address these
questions using a sample of 29 ETGs selected from SDSS and \galex\
surveys for which detailed far-UV images were obtained with the
\hst. The initial analysis of this sample was presented in
\citet{sr2010} (hereafter SR2010), where it was shown that these ETGs,
selected on the basis of the presence of a strong UV excess, exhibit
clear signatures of extended SF. The current work expands on the
morphological and size-related aspects of the analysis of the \hst\
sample and appends it with the analyses of the general population of
ETGs from SDSS and \galex. Paper II tackles star formation histories
of this sample using surface brightness photometry, and also discusses
the selection of more complete samples of ETGs with extended SF.

The paper is structured as follows. Selection of the \hst\ sample is
explained in \S\ \ref{sec:sample}, and the resulting UV observations
and optical imaging data from SDSS and WIYN are described in \S\
\ref{sec:data}. Results are given in the subsequent five sections. In
\S\ \ref{sec:uv}, \ref{sec:ha}, \ref{sec:opt} we provide morphological
analysis based on the UV, \ha, and optical imaging, respectively,
which shows that the majority of the sample has SF on large scales
(tens of kpc), and that the SF is found exclusively in S0s and not in
ellipticals. This is followed by two sections that place our sample in
context. In \S\ \ref{sec:uv_opt} we analyze the relation between the
UV and optical morphologies, showing that the star forming gas is
preferentially acquired subsequent to a galaxy getting onto a red
sequence, while in \S\ \ref{sec:s0_vs_e} we further explore
differences between S0 and elliptical galaxies in terms of SF and
determine the incidence rate of SF in the two types in the overall
population. Discussion of the results in view of evolutionary
scenarios is given in \S\ \ref{sec:disc}, and the findings are
summarized in \S\ \ref{sec:conclusions}. Cosmology parameters
$\Omega_m=0.3$, $\Omega_\Lambda=0.7$, $H_0= 70\, {\rm km\, s^{-1}\,
  Mpc^{-1}}$ are assumed throughout.

%%%
\section{Sample selection} \label{sec:sample}
%%%

The sample presented in this paper comprises 29 galaxies selected from
SDSS and \galex\ catalogs and observed with the Solar Blind Channel
(SBC), a UV detector of the Advanced Camera for Surveys (ACS) on board
the {\it Hubble Space Telescope}. These observations were taken in
program GO-11158 (PI Rich). No other observations with {\it HST}
instruments were carried out. While 30 galaxies were originally
selected to be observed with the {\it HST}, one target was not
successfully imaged. We refer to the resulting 29 as the \hst\
sample. Here we provide a more detailed description of the sample
selection presented in SR2010.

(1)\enspace SR2010 started with 67,883 galaxies from the main
spectroscopic survey of SDSS DR4 that lie within the 645 deg$^2$
overlap with \galex\ Medium-imaging survey (MIS) (internal data
release IR1.1), regardless of whether the galaxies are detected by
\galex\ or not. Additional details about this parent sample are given
in \citet{s07}.  

(2)\enspace To obtain galaxies with larger angular sizes and better UV
measurements, the selection was restricted to $z<0.12$, which left
30,860 galaxies.

(3)\enspace Both far-UV (FUV) and near-UV (NUV) detections were
required, plus \galex\ detections were required not to fall close to
the edge of the circular detector ($<0.55$ deg). Asking for NUV was
meant to eliminate any spurious FUV-only detections. These cuts brought
the sample to 17,574 galaxies.

(4)\enspace Next, SR2010 required that the optical light profile be
consistent with an ETG, by asking for an SDSS concentration (the ratio
of the radii enclosing 90\% and 50\% of the Petrosian flux in $r$
band\footnote{In SR2010 the concentration was erroneously reported to
  be based on $i$ band. The difference in the two measures is
  negligible.}) to be $C>2.5$ \citep{strateva,bernardi}. Such a cut
yields a very complete sample of ETGs, while allowing non-negligible
late-type contamination \citep{bernardi10,masters}, which was removed by
subsequent cuts. This step left 7096 galaxies.

(5) \enspace (a) In order to select only galaxies with no
spectroscopic signs of ongoing star formation or nuclear activity,
SR2010 used the emission-line classification of \citet{b04}, applied to
SDSS DR4 by MPA/JHU group, and accepted only what B04 call
`unclassifiable' galaxies, which is equivalent to a cut on H$\alpha$
flux S/N ratio of $\lesssim 3$. This resulted in 1660 galaxies. (b) While the
previous cut should have removed most Type 2 AGN, it was noticed that the
scaling of line flux errors suggested by the MPA/JHU group to
reproduce systematic errors leads to the overestimation of line {\it
  ratio} errors (because the systematic errors of lines that are close
to each other in wavelength are correlated), so additional galaxies
were excluded if the line ratios would place them in the LINER region
of the BPT diagram ($-0.2\leq$ log([\ion{N}{2}]/H$\alpha) <0.5$ and
$-0.3\leq$ log([\ion{O}{3}]/H$\beta) <0.8$), regardless of line SNR. This
cut left 986 galaxies. (c) For several dozen of the remaining
galaxies the formal H$\alpha$ SNR is low simply because the part of
the spectrum around \ha\ was corrupted. For them the H$\alpha$ cut is
replaced by asking that H$\beta$ flux SNR be $<3$. Removing these sources
left 954 galaxies. This represents the sample of apparently quiescent ETGs.

(6) \enspace From quiescent ETGs, SR2010 selected those with strong UV
excess, $\sim 2\times$ stronger than the maximum excess expected UV
upturn from old populations. They defined this cut as rest-frame color
FUV$-r<5.3$ (rest-frame colors derived using KCORRECT v4,
\citealt{blanton_roweis}; all magnitudes in AB system). This resulted
in 63 galaxies. This cut, like others, was not meant to produce a
complete sample, but rather one with minimal contamination from
late-type galaxies and a sample size that could be practically
observed with the HST.

(7)\enspace The final selection cut this number in half, to 30, by
applying visual inspection of SDSS postage stamp images ($gri$
composites) and SDSS spectra. Most of the candidates were discarded
because there was a nearby blue source that gave rise to the UV
excess. Other reasons for exclusion include non-early type interlopers
(spiral arms or the absence of bulge), and the presence of E+A
spectral features.

Most recent \galex\ data processing (GR6) as well as the FUV
measurements from the \hst\ UV images themselves (Paper II) have
revealed that two galaxies in the sample had greatly overestimated FUV
magnitudes in the original (IR1.1) \galex\ photometry (difference of
$\sim$1 and $\sim$2 mag). Their true magnitudes would not qualify
these objects as having a strong UV excess, so we discuss them apart
from the rest of the sample. This brings the total number of galaxies
that have a strong UV excess and for which \hst\ observations produced
detections to 27. We denote individual galaxies by a prefix SR,
followed by a sequential number (01 to 30), which indicates the rank
in terms of FUV$-r$ color in the original list, with 01 being the
bluest object. Note that \hst\ failed to observe one galaxy, but we
retain it in the numbering order (SR24).

\begin{figure}
\includegraphics[width=3in]{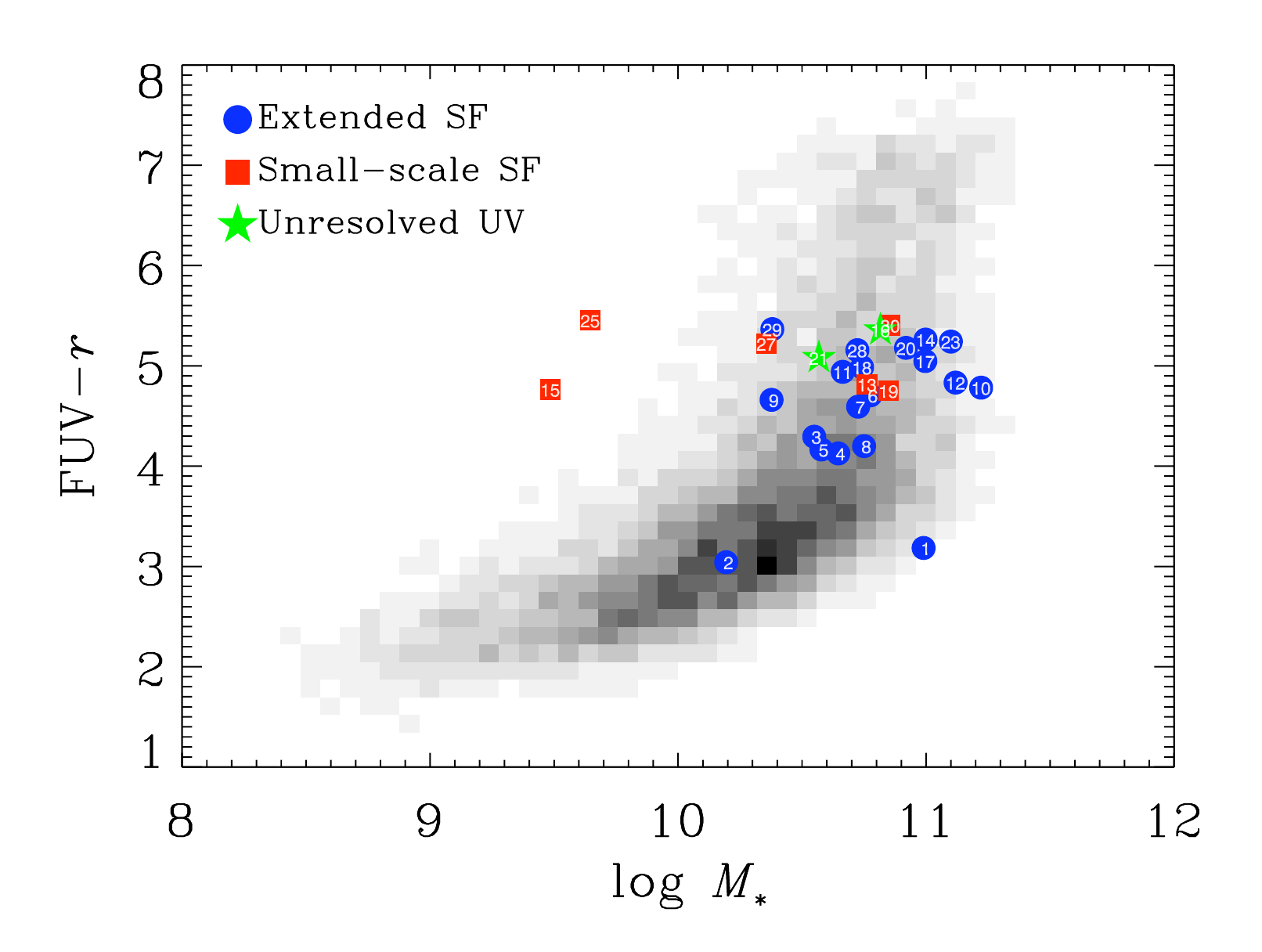}
\caption{UV-to-optical color vs.\ stellar mass diagram. The \hst\
  sample of ETGs with strong UV excess is shown as colored symbols,
  split into UV morphology classes: extended SF (blue dots),
  small-scale SF (red squares) and unresolved (green stars).  Numbers
  within symbols indicate object identifications (SR\#). Gray scale
  image represents the distribution of the underlying galaxies
  detected by SDSS and \galex. Truly quiescent ETGs are redder than
  the \hst\ sample, while the actively star-forming galaxies (``blue
  sequence'') are bluer than the \hst\ sample. The sample occupies the
  region of intermediate colors known as the green valley.}
\label{fig:mass_fuvr}
\end{figure}

The resulting sample has a general appearance of early-type galaxies
in SDSS $gri$ color composites and red optical colors ($g-r>0.7$
rest-frame, Figure \ref{fig:gr_fibssfr}) despite no explicit optical color
selection. Such colors place the sample fully within the {\it optical}
red sequence, yet the UV excess ensures that they occupy the
intermediate color region of the {\it UV-optical} color-magnitude
(mass) diagram (Figure \ref{fig:mass_fuvr}), known as the ``green
valley'' \citep{wyder}. Furthermore, none of the galaxies in the
resulting sample belongs to clusters from the GMBCG catalog of
\citet{hao} and its $z<0.1$ addendum.

In our selection, the general quiescence is ensured by removing
galaxies with emission lines in SDSS spectra. This criterion also
leads to a sample with heavily suppressed SF in the central ($\sim 5$
kpc) region covered by SDSS fiber. Is there a significant population
of ETGs with extended SF that is not covered by our selection, and is
such population likely to exhibit very different UV morphology from
that seen in our sample? We examine these issues in detail in the
Appendix. We conclude that our HST sample covers $\sim2/3$ of the
parameter space occupied by ETGs with extended SF. What the \hst\
sample does not cover are ETGs (still on the optical red sequence) in
which the extended emission is concurrent with the SF in the central
regions. Such galaxies may in some cases exhibit UV morphologies
different from those in our \hst\ sample; specifically there may be
more in-filled disks than what we observe. Any implications of these
selection effects will be discussed in appropriate sections.

%%%
\section{Data and observations} \label{sec:data}
%%%

%
\subsection{\hst\ UV imaging} \label{ssec:acs}

\hst\ ACS/SBC imaging of UV-strong ETGs represents the backbone of
this project. The SBC is a multi-anode micro-channel array with a
parallelogram-shaped field of view of $\sim35"\times 31"$ and $\sim
0\farcs03$/pix scale. The FWHM is however $\sim 0\farcs3$, which is
nevertheless an order of magnitude better that that of \galex,
allowing us to study the detailed spatial distribution of the UV
emission at typical redshifts of SDSS galaxies ($z \sim 0.1$). Imaging
was performed using the F125LP long-pass filter ($\lambda_{\rm eff}
=1459$ \AA), which has a similar effective wavelength as the \galex\
FUV filter.

A four-point dither strategy was employed to improve the sampling, and the
individual exposures were processed using MultiDrizzle tasks
\citep{koekemoer}. Each galaxy was observed for one orbit ($\sim$2.5
ks total, or $\sim$600 s per exposure). Below we describe in more
detail the reduction process, which was modified from the pipeline
reduction to address specific issues related to SBC data.

\subsubsection{SBC dark current subtraction}

\begin{figure}
\epsscale{1.0} \plotone{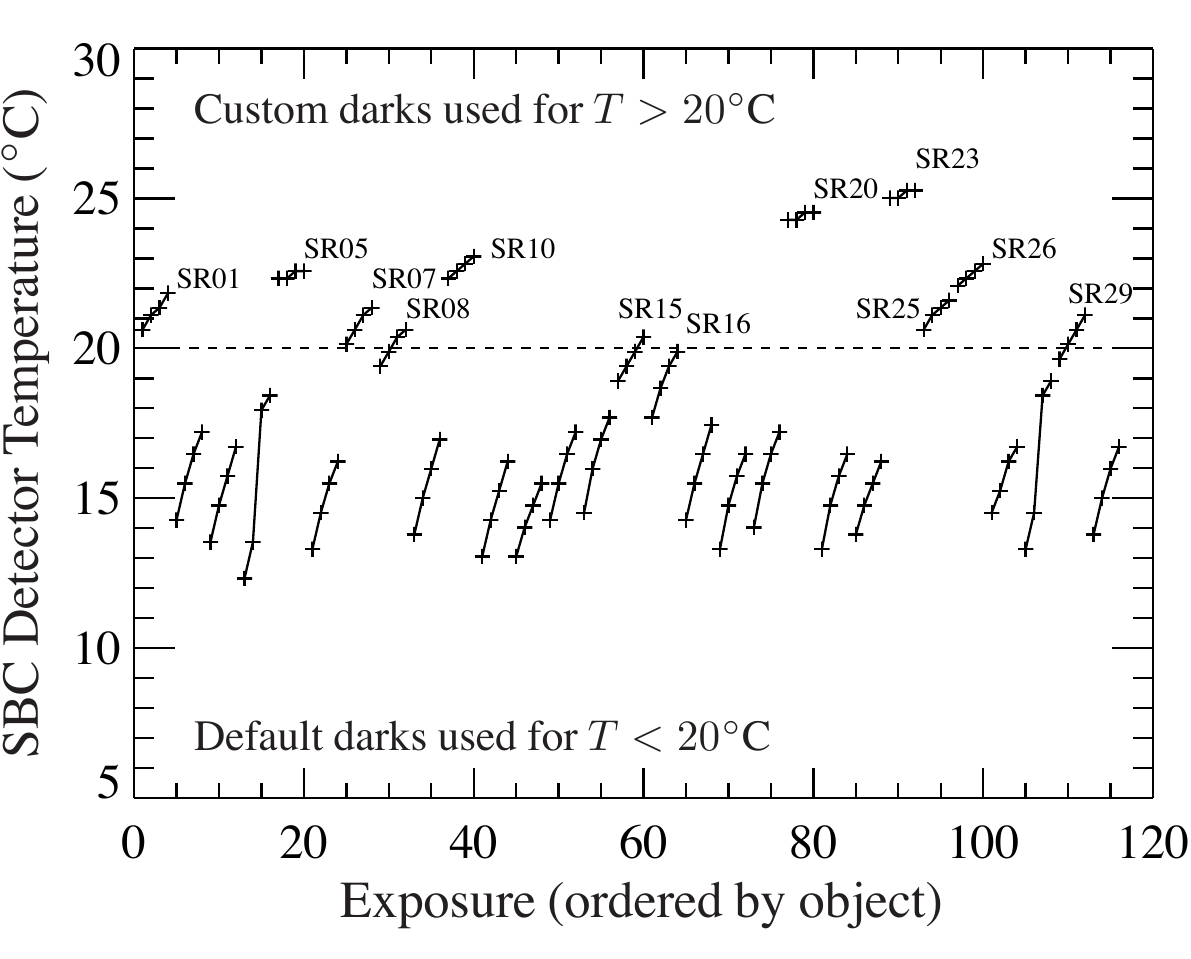}
\caption{SBC detector temperature for each individual exposure. The
  temperature increases over the course of each four-exposure set
  (connected points). The pipeline supplied dark files were used in
  reducing the raw exposures for which the detector temperature was
  cold ($T\lesssim 20\degr$C). Exposures for which $T\gtrsim 20\degr$C
  required the use of custom darks that accounted for the increased dark
  current. Galaxies for which custom darks were applied are labeled.}
\label{fig:sbc_temp}
\end{figure}

The SBC suffers from a temperature-dependent dark current that results
in a steadily increasing ``glow'' near the center of the detector as it
operates for extended periods (e.g., \citealt{teplitz}), i.e., over
the course of an orbit. Figure \ref{fig:sbc_temp} shows the detector
temperature (value extracted from the raw image FITS header keyword
{\sc MDECODT1}) for each exposure. While the majority of exposures
were taken at $T<20\degr$C, several were obtained at $T >
20\degr$C and suffer from severe dark current. The master dark files
provided by the data reduction pipeline only stack dark frames taken
when the detector is cold, $T\lesssim 20\degr$C, and are thus
inadequate for removing the increased dark current in exposures taken
at higher temperatures. To accurately subtract the dark current in
high-temperature frames we made custom dark images based on dark
frames taken at $T > 20\degr$C. Single dark exposures taken over a
range of temperatures are available from the \hst\ archive (see
\citealt{cox} for details). We constructed five custom dark images by
stacking several individual dark frames covering the temperature range
$20\degr{\rm C} < T < 25.5\degr$C and smoothing the resulting
stack. Each stack contains 5--8 dark frames spanning a narrow range in
detector temperature.

The individual raw science frames were then processed using the dark
image corresponding to the temperature of the detector at the time of
exposure. In a few raw frames the glow was not fully removed with the
dark image corresponding to a temperature from FITS header, so for
those a dark image corresponding to a higher temperature was used to
completely remove the dark current glow. In these cases the choice of
which dark to use was based on visually inspecting the resulting
processed science image and choosing the dark image that best removed
any residual dark current.

\subsubsection{Constructing ACS co-added images}

\begin{figure}
\epsscale{1.0} \plotone{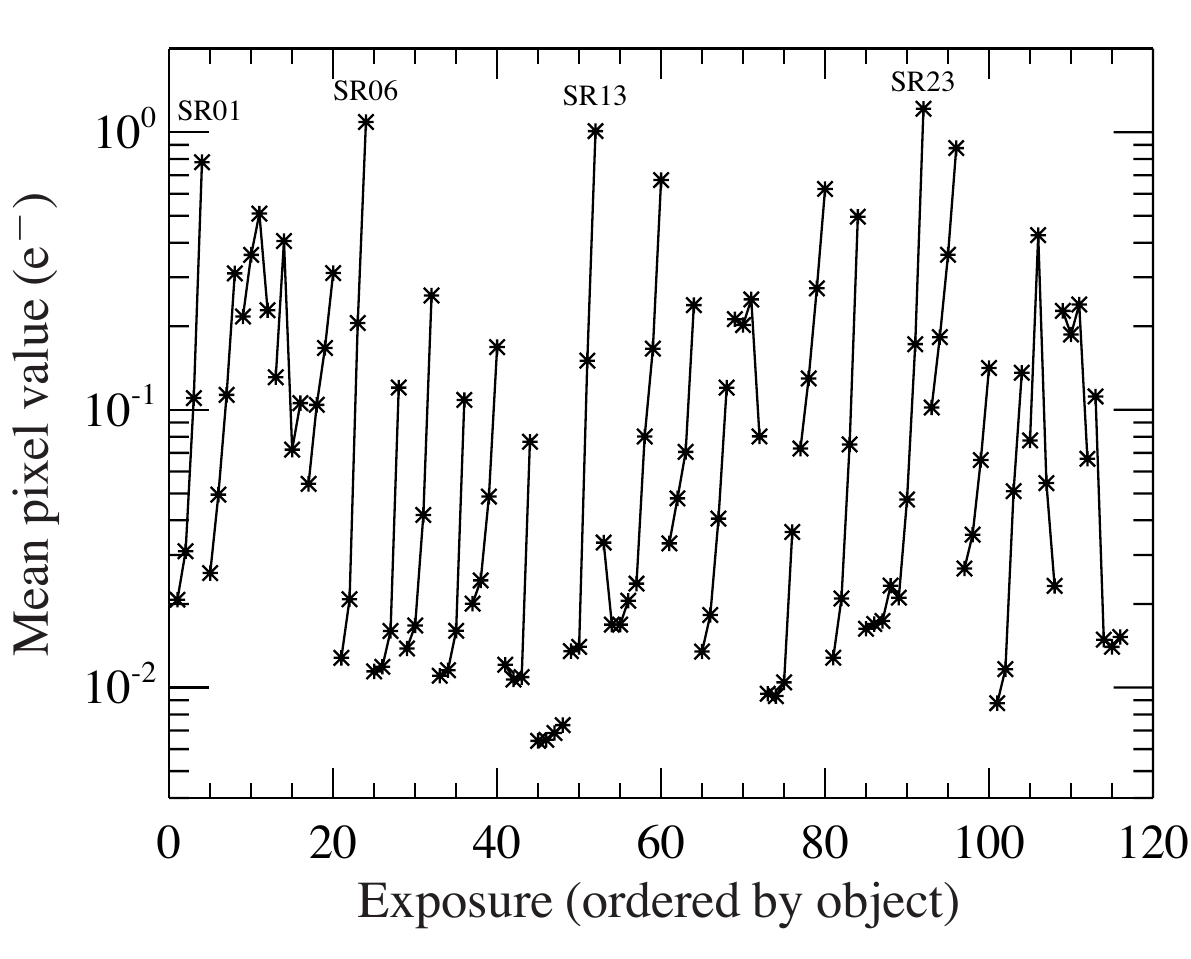}
\caption{Mean pixel value for individual raw exposures after the dark
  current subtraction. For clarity, each set of four exposures (one
  orbit) is connected by a solid line. Four targets have been labeled
  to guide the eye. Nearly all objects show an increase in the mean
  pixel value over the course of four exposures, reflecting the
  increase in the detected geocoronal background over the course of
  each orbit.}
\label{fig:sbc_bkg}
\end{figure}

The final science image is a stack of the four individual exposures
processed using MultiDrizzle, which corrects for geometrical
distortion and shifts and registers each exposure onto a common
coordinate system. In addition to dark current effects, SBC images
exhibit time-variable ``sky'' background, which in some exposures
dominates the signal. This background comes from the geocoronal
emission lines. While most of the geocoronal flux is in Ly$\alpha$ and
is therefore blocked by the choice of F125LP filter, some background
also arises from \ion{O}{1} at 1304\ \AA\, \citep{acs}. Geocoronal
emission is quite variable. This is demonstrated in Figure
\ref{fig:sbc_bkg}, where we plot the mean pixel value in each raw
exposure after masking out the object(s) in each. For each set of four
exposures, the mean value varies dramatically (up to two orders of
magnitude in some cases). Visual inspection of the exposures confirms
that the increase in the mean is due to the sky background, which
increasingly illuminates the exposures uniformly across the field of
view, rather than any residual dark current glow, which is
concentrated toward the center of each frame. Note that a large mean
value does not necessarily correlate with a high dark current in the
exposure (compare with Figure \ref{fig:sbc_temp}). Indeed, some
exposures taken at cold temperatures have very large mean pixel values
relative to the others. This reinforces the fact that the sky
background is dominated by scattered light rather than by the thermal
dark current of the detector. To account for the variable background,
we weighted each image by the inverse of the mean pixel value when
combining them in MultiDrizzle\footnote{Since the counts in SBC images
  have a Poisson distribution, variance is the same as the mean
  value.}. In this way co-added images have greater contributions from
exposures with lower sky background.

Since the data are dominated by Poisson statistics and most pixels
receive no counts we finally apply smoothing with a Gaussian kernel of
10 pix ($1\, \sigma$). This yields images with a PSF FWHM of $\sim
0\farcs6$, comparable to the seeing in our optical images. Smoothing
greatly improves our ability to see faint features while preserving
genuine details. \hst\ SBC imaging is presented in the first column of
imaging panels in Figures \ref{fig:panel1}--\ref{fig:panel10}.

\subsection{SDSS optical imaging} \label{ssec:sdss}

One of the goals of this work is to compare the UV and optical
morphologies of the \hst\ sample. For optical morphologies we draw
from two sources: SDSS $ugriz$ imaging and our WIYN 3.5-m $R$-band
imaging.  WIYN 3.5-m imaging is significantly deeper than that of SDSS
and has a better spatial resolution. However, it is not available for
the entire sample and thus needs to be supplemented by SDSS images.

SDSS imaging is performed in five filters, and each is exposed for
about 54 s. Typical seeing is $1\farcs4$.  To increase the effective
depth of SDSS images we co-add four bands ($griz$)
together\footnote{Co-adding of $griz$ images was performed without any
  weighting, as it produced very similar results (in terms of
  perceived depth) as the more elaborate co-addition of scaled
  variance-weighted images in 5 filters, with scaling based on the
  total flux in a given band.}. Since $u$ band is generally not
sensitive enough to detect extended galaxy structure (e.g., S0 disks)
at this redshift, we do not include it in the co-add. We show the
co-added SDSS images of each galaxy in the third column of imaging
panels in Figures \ref{fig:panel1}--\ref{fig:panel10}. SDSS images are
presented even when deeper WIYN images are available for two
reasons. First, to help evaluate the extent of features that may not
be detectable when only SDSS images are available, and second, to
present images that are similar to those that were originally used to
select the sample.

\subsection{WIYN 3.5-m optical imaging} \label{ssec:wiyn}

The optical imaging at the WIYN 3.5-m telescope focused on the 2/3 of
the sample having higher UV surface brightness as seen in the \hst\ UV
images. The imaging consisted of deep $R$-band observations
accompanied, for a subset of objects, by narrow-band \ha+[\ion{N}{2}]
observations. The goal of deep $R$ imaging was to facilitate more
accurate optical classification (including the presence of bars) and
to reveal any signatures of interaction (such as tidal features,
shells or dust lanes). Narrow-band imaging had the goal of searching
for \ha\ ionizing emission that would be indicative of ongoing SF
occurring on timescales shorter than those probed by UV bright
population. Analysis facilitated by these observations is presented in
\S\ \ref{sec:ha} and \ref{sec:opt}.

Observing took place over two runs (2010 Mar 22-24 UT and 2010 Aug
7-11 UT). Out of 8 total nights, weather permitted observing on
3.5. As a result, 18 galaxies were observed in $R$ and 13 in narrow
band. We used the MiniMo CCD camera with a pixel scale of
$0\farcs14$. Median total exposure times per galaxy were 30 minutes in
$R$ and 40 minutes in a narrow band (see Table
\ref{table}). Individual exposure times were 10 minutes in $R$ and 20
minutes in the narrow bands. Median seeing was $0\farcs8$ in $R$ and
$0\farcs7$ in narrow bands. We used the KPNO Combined Set of \ha\ narrow
band filters, choosing one among seven filters with central wavelength
closest to the wavelength of redshifted \ha. Filters have an approximate
FWHM of 80\ \AA. Scaled $R$ band images were used to subtract the
continuum light present in the narrow band.

Standard observing procedures were followed, however, no photometric
standards were acquired. Data reduction was performed using tasks in
IRAF and L.A.Cosmic \citep{lacosmic}. To facilitate continuum
subtraction we aligned all individual exposures of a single object to
the nearest integer pixel. For continuum subtraction of narrow band
images we also convolved all images of a given object to match the PSF
of the image with the poorest seeing (typically $0\farcs8$). Then the
images were co-added in each band with inverse variance as a weight. We
determined the scaling of the $R$ image to use for the continuum
following \citet{macchetto} by asking that the elliptical isophote at
$a=1\farcs4$, where we know emission lines contribute negligibly, have
the same intensity as in the narrow-band image.

To produce the $R$-band images with optimal depth and resolution,
which we show in the second column of imaging panels in Figures
\ref{fig:panel1}--\ref{fig:panel7}, we co-added images without any PSF
degradation, but weighted by inverse PSF FWHM. By comparing the image
of one galaxy that is in common to our sample and deep SDSS Stripe 82,
we conclude that the resulting depth is approximately 0.5 mag deeper
than that of Stripe 82, or $\mu_{R}\sim 26.5$ AB mag arcsec$^{-2}$.

Figures \ref{fig:panel1}--\ref{fig:panel7} show images of the same
object from three different sources: \hst, WIYN and SDSS. To align
\hst\ and SDSS images we used coordinate system already set in the
headers of FITS images. WIYN observations originally had coordinate
system zero points that were many arcminutes off from the true
value. To correct them we used the {\it imwcs} task from the WCSTools
package. This task matches the stars in the frame to known catalogs
and updates the FITS headers accordingly. The catalog to which the
matching was done was mostly GSC2. In several cases the task could not
obtain an astrometric solution and in those cases the astrometry
keywords were modified manually based on measured offsets and known
coordinates of the sample galaxy. The accuracy of registration is
typically within one arcsec.

%%%
\section{UV morphology of the \hst\ sample} 
\label{sec:uv}
%%%

We begin the presentation of the results with an overview of UV
morphologies of the \hst\ sample of strong-UV excess ETGs. This is a
more rigorous and more detailed description and classification than
the one provided in SR2010 and is accompanied with the display of all
\hst\ UV images.

As noted in SR2010, the great majority of the sample shows spatially
extended and/or patchy structures clearly indicative of star-forming
regions. Therefore, the conclusion was reached that for most of the
sample the UV excess is attributable to star formation. Inferred star
formation rates (SFRs) are typically $\sim 0.5 M_{\odot} {\rm
  yr}^{-1}$ (Table \ref{table}) \footnote{All estimates of stellar
  mass and SFR, as well as the specific SFR, come from \citet{s07} and
  are determined using Bayesian UV/optical SED fitting and
  extensive stellar population synthesis models that include dust
  attenuation. They are given in solar units and for Chabrier
  IMF.}. The extended/patchy UV emission is not consistent with the
emission from old hot populations (the classical UV upturn), which is
visible in the \hst\ UV images only as a relatively weak compact or
unresolved central source (see discussion in Paper II).

Based on the size of the UV emitting regions we
classify the sample into three main classes:

(1) ETGs with galaxy-scale (extended) star formation 

(2) ETGs with small-scale star formation

(3) ETGs with unresolved UV emission

We place the galaxy into class (1) if the UV emission is
present on scales similar to or larger than the apparent optical
extent of the galaxy in deep optical images, i.e., if the
ratio of the FUV diameter down to 27 AB mag arcsec$^{-2}$ to the
optical diameter corresponding to the 90\% of the Petrosian flux is
greater than one. For galaxies in our sample this typically translates
into UV diameters greater than 25 kpc. We refer to the part of the
sample that exhibits galaxy-scale SF as {\it extended star-forming
  early-type galaxies}, or ESF-ETGs. Out of 27 galaxies, 19 (70\%)
fall in this category, making it the dominant UV type (shown in
Figures \ref{fig:panel1}--\ref{fig:panel6}). 

Galaxies in class (2) show resolved UV-emitting regions that are
smaller than the optical extent of the galaxy, i.e., whose FUV to
optical size ratio is smaller than $\approx 1$. In our sample they
typically have UV diameters smaller than 15 kpc, and the UV emission
is often offset from the optical center. There are 6 such cases in our
sample (22\%), and they are shown in Figures \ref{fig:panel7} and
\ref{fig:panel8}. Finally, two galaxies (7\%) are classified as UV
unresolved, since none of the sources present within the optical
extent of the galaxy is resolved (Figure \ref{fig:panel9}). We proceed
by discussing the UV morphology of each class and introducing several
subclasses. Comparison with optical morphology will be presented
separately in \S\ \ref{sec:uv_opt}. 

\subsection{Extended star-forming ETGs (ESF-ETGs)} \label{ssec:ext-esf}

\begin{figure*}
\epsscale{1.0} \plotone{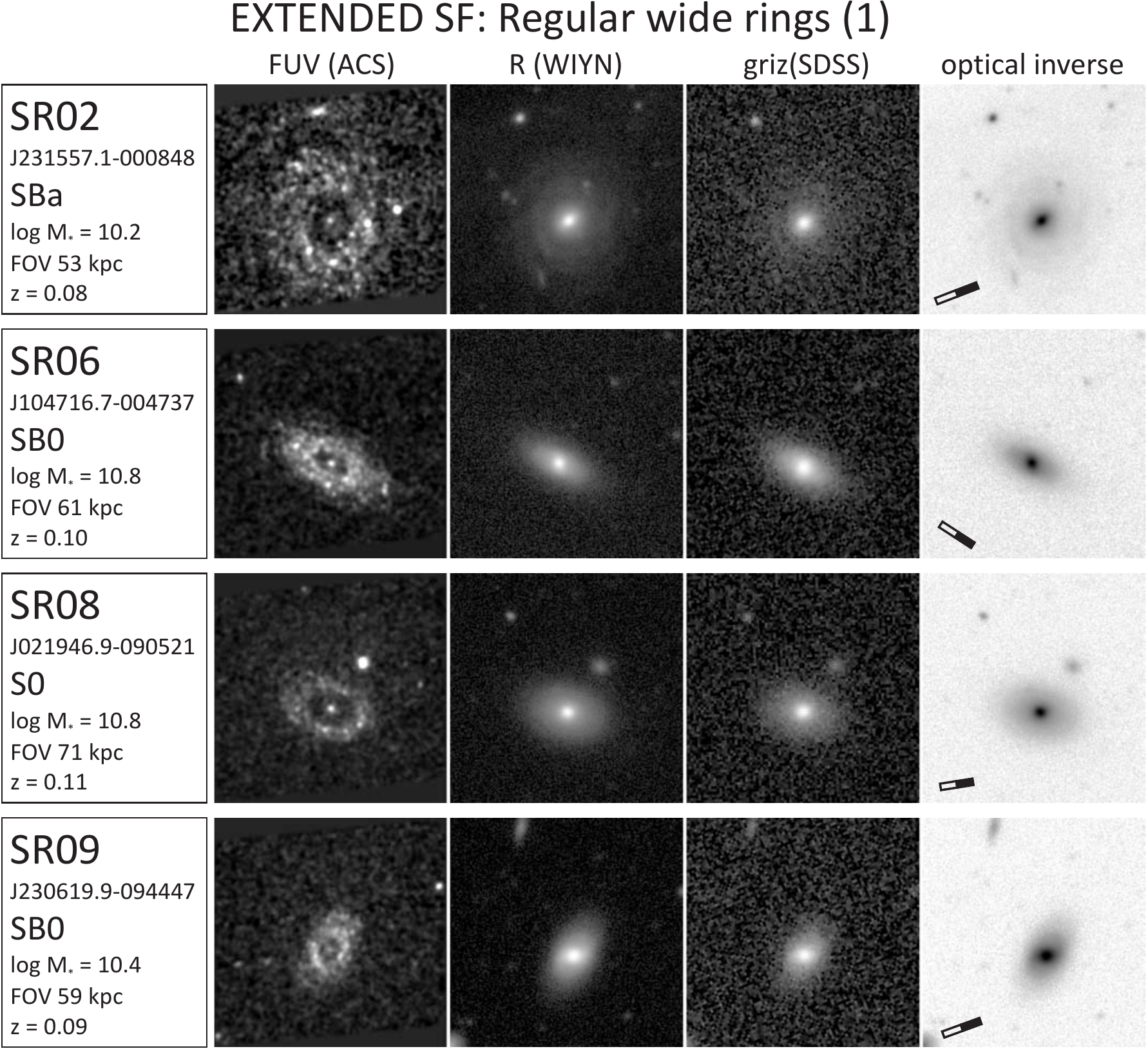}
\caption{Far-UV and optical images of early-type galaxies with
  extended SF and regular wide UV rings (first set of two). Data box
  on the left gives galaxy designation (SR stands for
  \citealt{sr2010}), SDSS ID based on J2000 coordinates, Hubble type,
  stellar mass, the extent of the horizontal field of view in kpc, and
  the redshift. All imaging panels span 35$"$. First imaging panel
  shows the \hst\ (ACS/SBC) far-UV image smoothed with a Gaussian
  kernel (FWHM = $0\farcs6$). Second panel shows the deep ($\sim 30$
  min) $R$-band image obtained with the 3.5-m WIYN telescope (when
  available). Third panel displays the SDSS image constructed by
  co-adding $griz$ exposures ($\sim 4$ min). The last panel shows the
  inverted optical image from WIYN, if available, or SDSS
  otherwise. The scale bar shows 10 kpc with the black end pointing
  towards North, with East being 90 degrees counter-clockwise. ACS/SBC
  field has a parallelogram shape, with the upper left and lower right
  slivers in rectangular panels not covered. Galaxies are not centered
  in ACS/SBC field, but are offset primarily downwards. Pixel
  intensity range was chosen in such way to avoid saturating the
  bright bulge. Linear scaling between the ADUs and pixel intensity
  was used for the FUV image, while optical images use asinh scaling
  \citep{lupton}, which allows the full dynamic range to be
  presented. This leads to the smaller, unsaturated appearance of the
  bulges compared to the views that would be obtained using the linear
  scaling. The ordering of galaxies follows the SR numbering, which in
  turn goes from the bluest (in FUV$-r$) to the reddest.}
\label{fig:panel1}
\end{figure*}

\begin{figure*}
\epsscale{1.0} \plotone{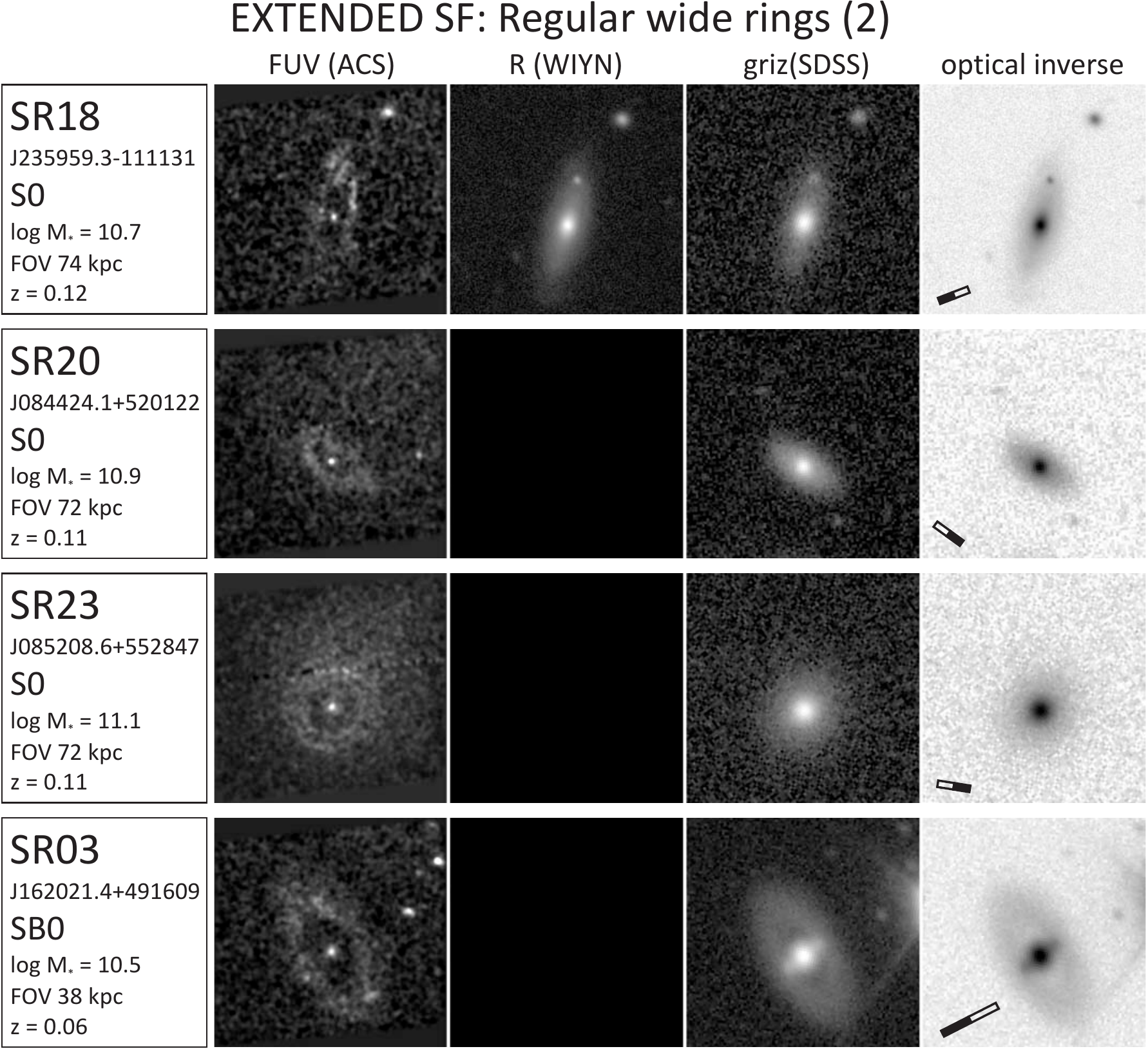}
\caption{Far-UV and optical images of early-type galaxies with extended SF
  and regular wide UV rings (second set). See Figure \ref{fig:panel1}
  for full description. SR03 does not follow the ordering scheme and
  is placed at the end because of its unique optical appearance. }
\label{fig:panel2}
\end{figure*}

\begin{figure*}
\epsscale{1.0} \plotone{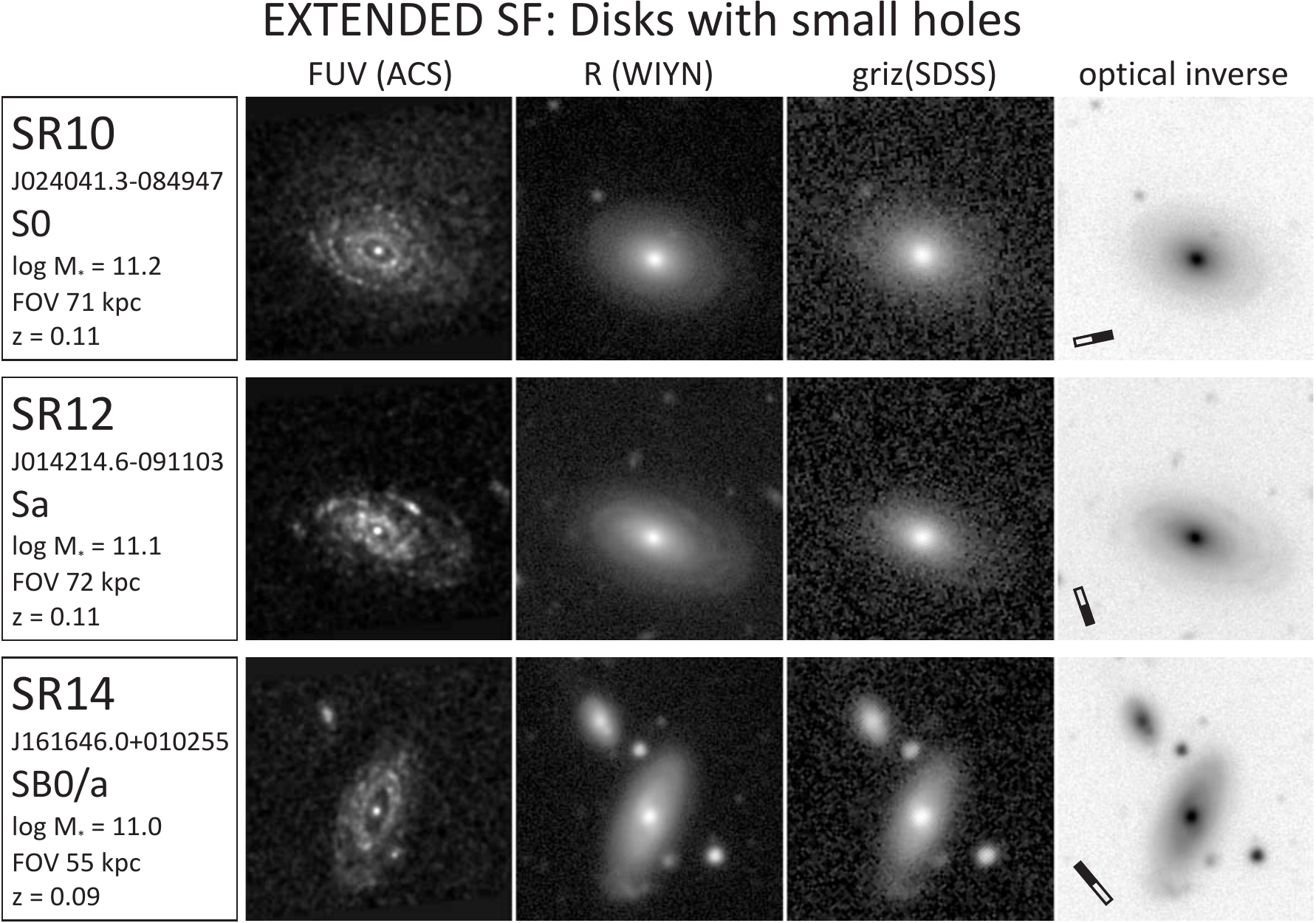}
\caption{Far-UV and optical images of early-type galaxies with extended SF
  and having UV disks with small central holes. See Figure \ref{fig:panel1} for
  full description.}
\label{fig:panel3}
\end{figure*}

\begin{figure*}
\epsscale{1.0} \plotone{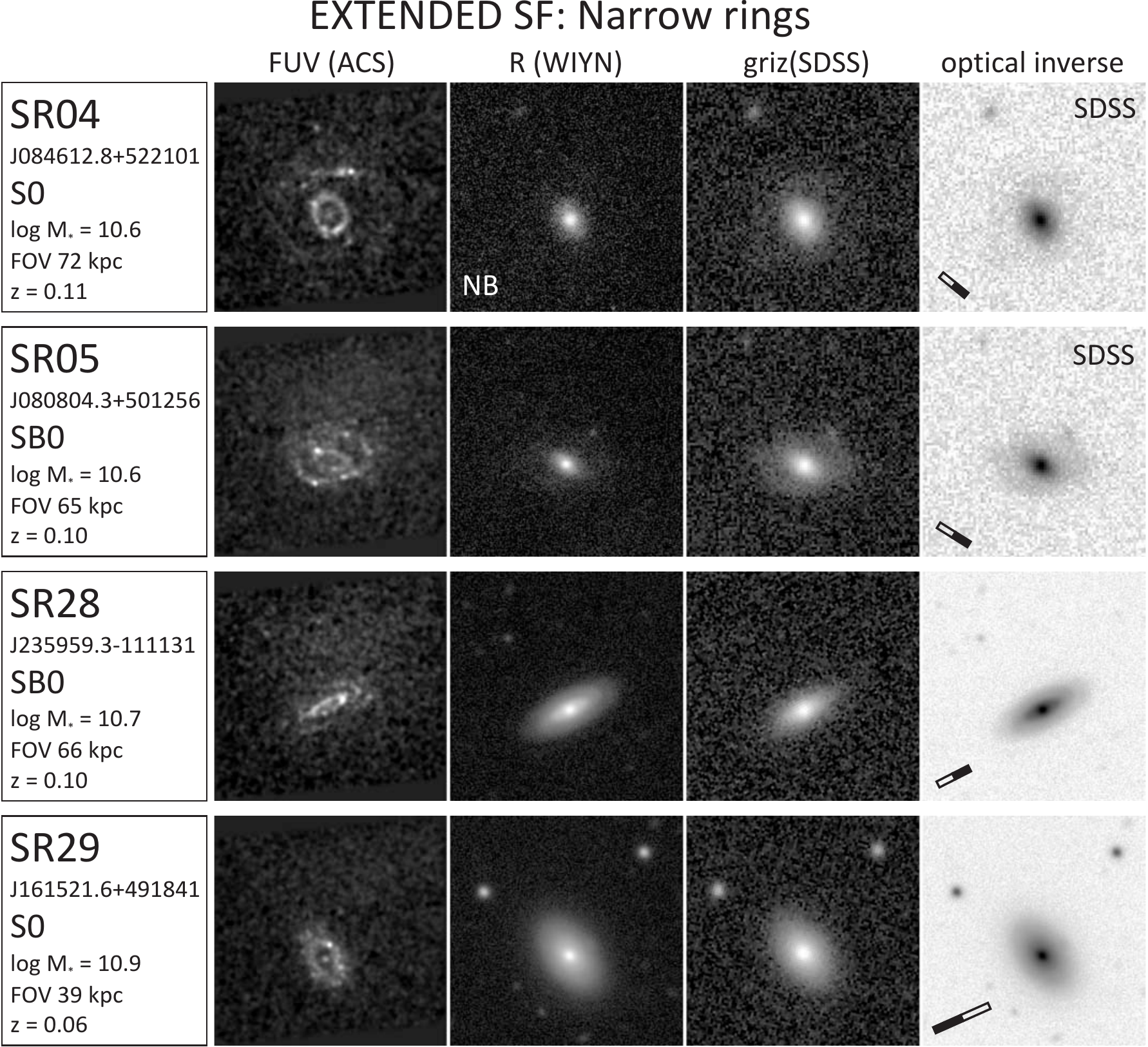}
\caption{Far-UV and optical images of early-type galaxies with
  extended SF in the form of narrow UV rings. See Figure
  \ref{fig:panel1} for full description. For SR04, the WIYN image does
  not show broad-band $R$ but narrow-band within $R$. For SR04 and
  SR05 the SDSS images are deeper than those from WIYN, so we show
  them instead of WIYN images in the inverted panels. See Figure \ref{fig:panel1}
  for full description.}
\label{fig:panel4}
\end{figure*}

\begin{figure*}
\epsscale{1.0} \plotone{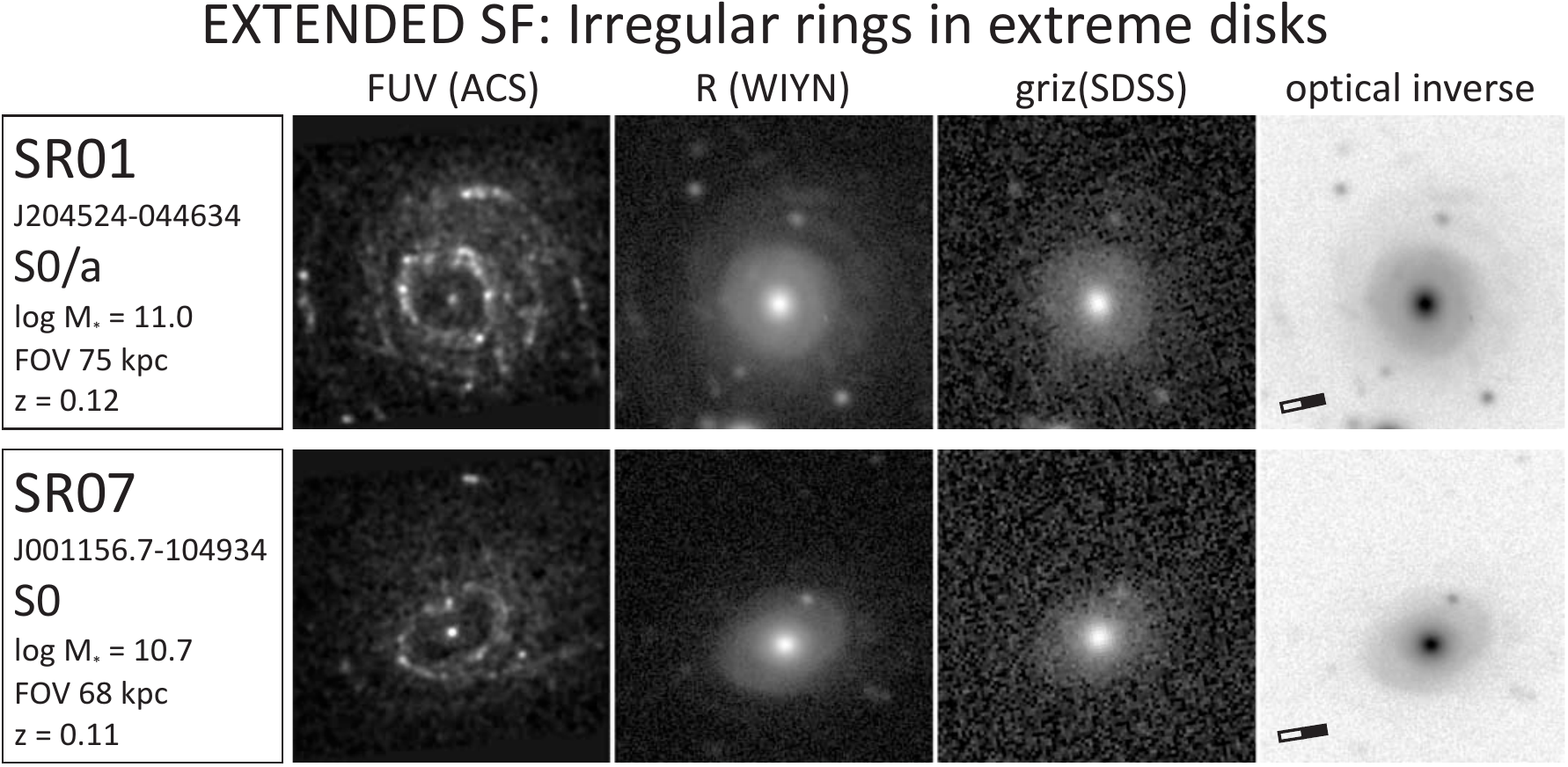}
\caption{Far-UV and optical images of early-type galaxies with
  extended SF and irregular UV rings within extreme disks. These
  galaxies share many properties with giant low surface brightness
  galaxies such as Malin 2 and UGC6614 (see \S\ \ref{sssec:glsb}). See
  Figure \ref{fig:panel1} for full description.}
\label{fig:panel5}
\end{figure*}

\begin{figure*}
\epsscale{1.0} \plotone{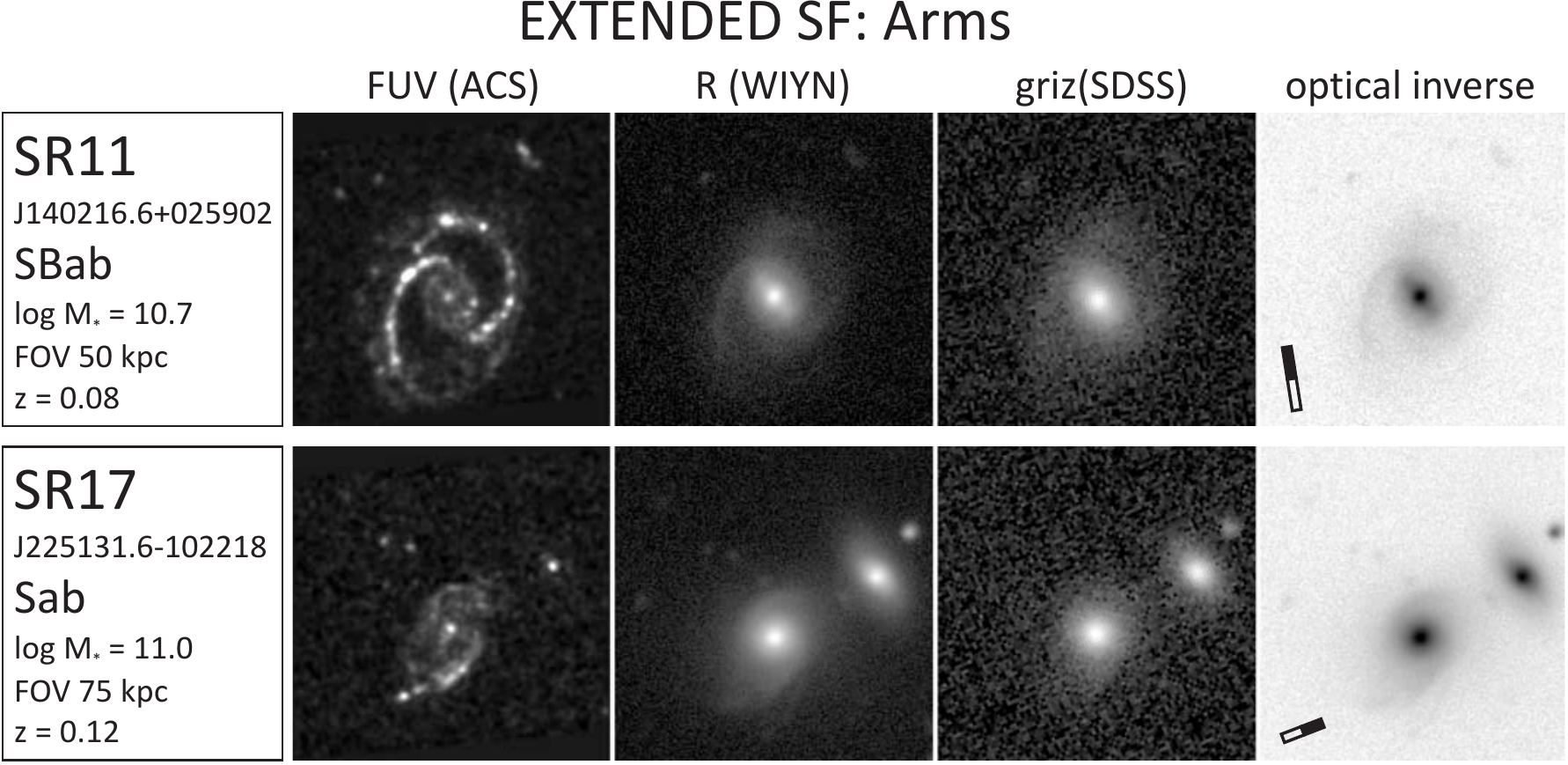}
\caption{Far-UV and optical images of early-type galaxies with extended SF
  and UV arms. In SR11 the arms form a pseudoring. Arms in SR17 appear
  to be tidal, formed by interaction with a companion. See Figure
  \ref{fig:panel1} for full description.}
\label{fig:panel6}
\end{figure*}

While star-forming rings are not a very common feature of normal
star forming galaxies (see for example the \galex\ Atlas of Nearby
Galaxies, \citealt{gildepaz07}), a quick glance at Figures
\ref{fig:panel1}--\ref{fig:panel6} reveals that they are the
ubiquitous among our ESF-ETGs (only one out of 19 galaxies in this
category does not possess a UV ring). Despite this common property
this class is far from uniform. To facilitate the discussion that will
attempt to relate different UV morphologies to their optical
properties and possible evolutionary scenarios, we proceed by dividing
the 19 ESF-ETGs into five subclasses. As any visual classification the
one presented here is to some extent subjective, and may represent
categorization of continuous properties, but we believe that it helps
systematize complex morphological information. Note that our
spectroscopic selection criterion would select against in-filled disks
because of their higher central SF. However, as we show in the
Appendix, such morphologies are the minority (1/3) compared
to those present in the \hst\ sample, so while they may not be
overwhelming, the UV rings will still dominate ($\gtrsim 2/3$) the
extended UV morphology in ETGs.

{\it Regular wide rings}. We start by focusing on the most populous
subclass within the extended star-forming class: ETGs with regular
wide rings. Eight galaxies belong in this group,
and their UV images are shown in the first column of imaging panels in Figures
\ref{fig:panel1} and \ref{fig:panel2}. All rings are characterized by
a relatively uniform, symmetric appearance. This regularity on large
scales does not imply a lack of features. For example, SR02 appears to
have flocculent structure in the ring, but unlike typical flocculent
spiral galaxies (e.g., NGC4414) this structure is restricted to the
ring, and is not present in the inner region. SR06 shows several
bright compact star forming regions within its ring. None of these
features, however, takes away from the overall regularity and axial
symmetry. A possible exception is the ring of SR18, which appears
incomplete, but is perhaps just much fainter on one side. SR08 has a
UV bright, compact companion whose FUV flux is blended with that of
the main galaxy in \galex\ photometry.  However, the UV flux of the
companion is actually two times smaller than that of the extended
emission of the main galaxy, so this ETGs remains a genuine strong
UV-excess source (with intrinsic FUV$-r=5.0$) even after the companion
flux is accounted for. The UV ring is lopsided, a possible result of
non-destructive interaction. Altogether, regular rings (with possible
exception of SR18) suggest that the SF is not a result of violent
processes such as galaxy interactions or mergers. Rather, the
distribution of the star-forming gas, whether it is internal or
external in origin, is related to more quiescent processes including
the secular effects of a bar that can lead to ring morphology
\citep{schwarz_pasa}. Rings in this subclass typically extend to
$32\pm4$ kpc\footnote{``Errors'' of average size indicate the standard
  deviation of sizes. Sizes represent diameters and were estimated
  visually. They are accurate to several kpc, depending on how well
  defined the edge of the UV light distribution is.}, and have surface
brightness in FUV between 26 and 28 AB mag arcsec$^{-2}$ (Paper II).

{\it Disks with small central holes}. We classify three galaxies as
belonging to the subclass of ESF-ETGs exhibiting disks with small
central holes or clearings (Figure \ref{fig:panel3}). This category is
similar to the one we just discussed. A distinction is being made
because the central clearing is much smaller than the extent of the
ring, which may point towards a different origin. Also, the overall
sizes are larger ($41\pm 6$ kpc). We also find that the mean star
formation rate (SFR) of this subclass is significantly higher than
that of regular wide rings (1.3 $M_{\odot}{\rm yr}^{-1}$ vs.\ 0.4
$M_{\odot}{\rm yr}^{-1}$). Certainly, there could be borderline cases
between these two categories, such as SR06 of the regular wide-ring
type.  Another distinct feature with respect to {\it regular wide
  rings} is that these disks are less uniform on smaller scales than
in the previous subclass. One can discern multiple concentric or
filamentary arm-like structures. Such features may indicate that the
UV structures, despite many similarities with wide rings, are closer
in origin to normal spiral disks. We will return to this point in \S\
\ref{sec:opt}.

{\it Narrow rings}. Four ESF-ETGs are classified as featuring narrow
rings (Figure \ref{fig:panel4}). The main characteristic of this
subclass is that the rings are both relatively narrow (1 to 2 kpc) and
small ($13\pm 2$ kpc). There is a greater diversity within this
subclass than in the previous two. Three of the four galaxies in this
group (SR04, 05, 28) show signs of an additional, outer ring. In SR04, the bright
streak above the middle of the frame may represent the bright part of
a very faint outer ring. Its narrow ring is relatively uniform. In
SR05 the outer ring is more complete but is somewhat offset from the
inner ring, a possible aftermath of an interaction. Both the inner and
the outer rings of SR05 have bright features in them. SR28 has an
incomplete outer ring, which may also be a part of a (former) spiral
arm. UV structure in SR28 is difficult to determine due to the high
inclination. The inner ring appears more complete but is not
uniform. Only SR29 shows a single, relatively regular ring (there is
only one bright region in it). What makes it belong to this category
rather than the {\it regular wide ring} class is its rather small size
(14 kpc) and narrowness. Non-uniformities seen in this group may
indicate that the UV structures were shaped by processes that involved
interactions.

{\it Irregular rings in extreme disks}. Only two galaxies are
classified as belonging to this remarkable subclass of ESF-ETGs: SR01
and SR07 (Figure \ref{fig:panel5}). Their UV structures are seen to
span an enormous extent of 75 and 68 kpc respectively. Both have
irregular inner rings of roughly the same size, surrounded by
filamentary structures in the very extended disk. Both the misshapen
inner ring and the extended UV structures point towards a different
origin of SF gas than in any of the previous categories.

{\it ESF-ETGs with arms}. The remaining two ESF-ETGs are placed in the
category exhibiting well-defined arms, although the two are quite
different in appearance and are each unique in our sample (Figure
\ref{fig:panel6}). SR11 presents one of the most stunning UV
morphologies in our sample, which is very different from what one
expects to see in a red sequence early-type galaxy. The most prominent
features are the two arms that emanate from an inner ring. The arms
almost appear to connect in an outer ring. This phenomenon has been
described in optical images of nearby galaxies as a ``pseudoring''
\citep{buta_crocker}, and we will discuss the mechanism for its
formation when describing the optical morphology of this galaxy (\S\
\ref{sec:opt}). The second armed galaxy, SR17, shows less regular
arms. Indeed, one arm is more prominently than the other, with
different extent and pitch angle. Optical evidence (\S\ \ref{sec:opt})
will show that this galaxy is interacting with its neighbor and that
these arms are probably tidally induced. SR17 is also the only ESF-ETG
for which there is no evidence of a UV ring. This absence and the
strong evidence for interaction may be related.

To summarize, the great majority of ESF-ETGs show relatively
undisturbed and regular UV morphology, but notable exceptions should
not be overlooked. Also, the presence of UV rings in all but one case
is intriguing. While our selection criteria would favor such
structures, they are likely intrinsically the most common type of
extended UV morphology in ETGs (see the Appendix).

\subsection{Small-scale star-forming ETGs} \label{ssec:sssf}

Six galaxies in our sample have resolved UV emission that is
significantly smaller than the optical extent of the galaxy. We
subdivide them into a group of four in which the UV emission is offset
from the galaxy center, and two where it is distributed centrally. The
latter two galaxies are also very different in terms of their much
lower stellar mass than the rest of the sample, which further
justifies this separation.

\begin{figure*}
\epsscale{1.0} \plotone{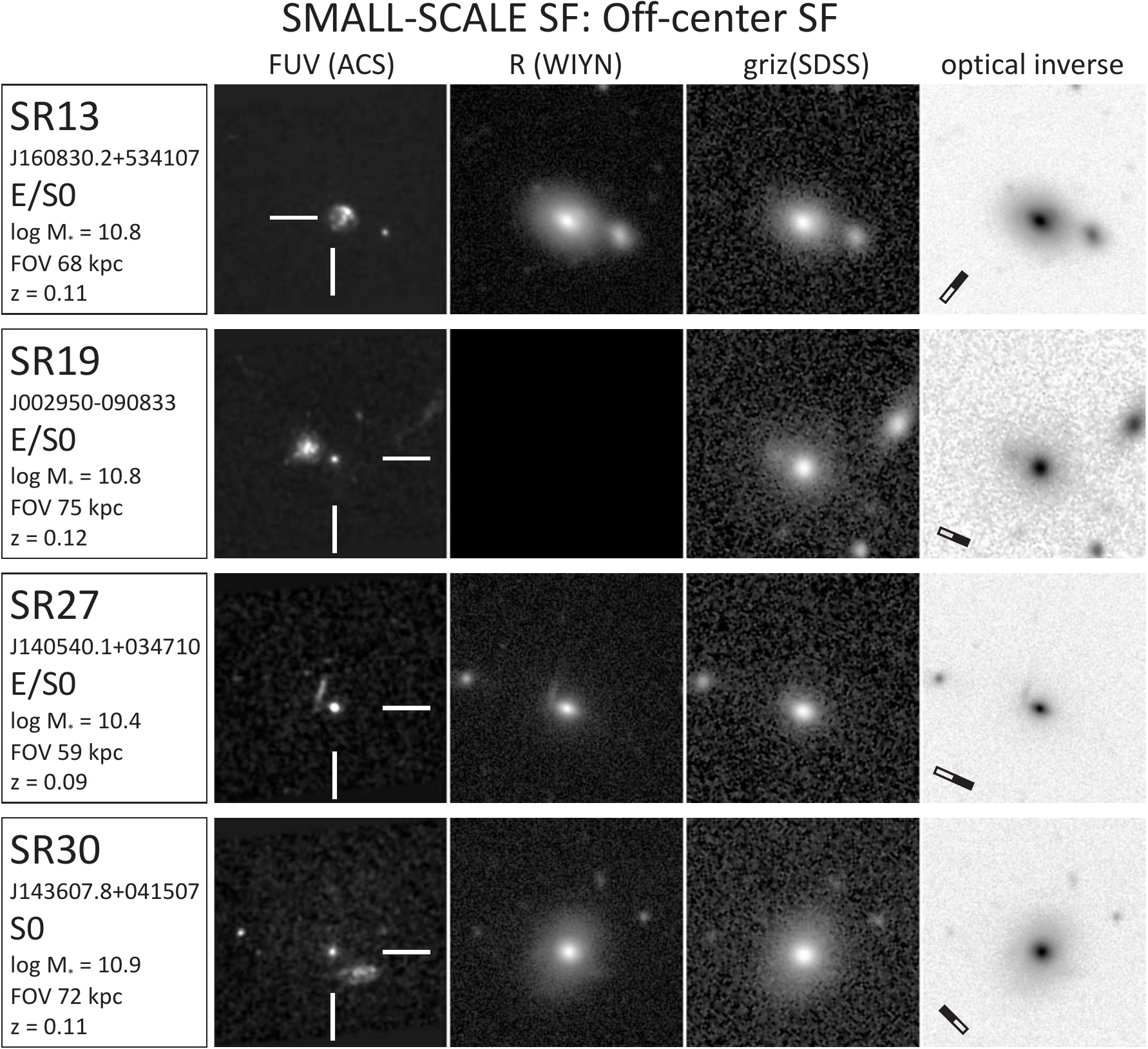}
\caption{Far-UV and optical images of early type galaxies with
  small-scale SF offset from the galaxy center. Crosshairs in the FUV
  images indicate the location of the optical center of the main,
  early-type galaxy. See Figure \ref{fig:panel1} for full
  description.}
\label{fig:panel7}
\end{figure*}

{\it Off-center small-scale SF}. We show the four galaxies classified
in this group in Figure \ref{fig:panel7}. While there are general
similarities in all four off-central cases (such as the size of the
star-forming patch of $\sim 10$ kpc), there are some interesting
differences. SR13 has a highly asymmetric star forming patch that
encompasses the optical center of the galaxy but is brightest 5 kpc in
projected separation from the center. Optical images shows an optical
enhancement coincident with the brightest UV region. There is also a
separate unresolved UV source corresponding to a likely companion
galaxy further out (some 15 kpc away). Next, in SR19, the star-forming
patch, located some 9 kpc from the center in projection, again shows
multiple bright UV clumps. An optical counterpart is faintly visible
too. The center of the main galaxy is an unresolved UV source that
appears relatively bright in the UV but nevertheless contains only
10\% of the total FUV flux. In SR27 the extended UV feature shows much
less structure and is quite narrow. One half of the UV emission comes
from this feature; the rest comes from an unresolved UV source at the
center. Finally, in SR30, the star-forming patch (at a projected
distance of 9 kpc) shows somewhat non-uniform, elongated structure and
accounts for 80\% of the UV flux. It is faintly visible in the optical
image as well. The star-forming patches in SR19, SR27 and SR30 could
be gas-rich dwarfs being disrupted and absorbed by the ETG, but not
yet having reached the center like SR13. After accounting for the UV
flux in off-center companions, the central unresolved emission in
SR19, SR27 and SR30 does not have a strong UV excess in FUV$-r$ and is
consistent with the old-star UV upturn.

Since the UV emission in this class is not centered on the optical
emission of the main optical galaxy, we need to ask if the off-center
structures represent physically associated components or companions,
or merely background or foreground ``contaminating'' galaxies that in
\galex\ low resolution images were blended with the main galaxy. In
SR27 and SR30 the UV morphology of the companions is relatively
undisturbed, so these are candidate interlopers. In order to estimate
how likely such a scenario is, we evaluate the incidence of blending
in \galex\ images using the following procedure. The FUV magnitude of
the main galaxies in this class (without the offset, potentially unrelated, UV flux)
is 24.1 mag at the faintest. Using the FUV galaxy counts from
\citet{xu} we ``generate'' galaxies down to FUV$=24.4$, the faintest
an ETG without SF would be at these redshifts, and distribute them
randomly over 10 sq.\ deg. The area of the mock catalog is arbitrary,
but it should be large enough to provide reliable statistics. Then we
go through this artificial catalog and identify pairs (or groups) that
are separated by $<5"$, which corresponds to the maximum separation of
the off-center UV ``companions''. Such pairs/groups would be blended
in \galex\ photometry which is derived from $5"$ FWHM images. We merge
these pairs/groups and recompute the FUV magnitude of the blend,
noting the number of components that comprise the blend. Out of 34,700
sources in the original catalog, 361 original pairs and one triple
become blended. Based on this simulation of blending we determine the
blending probability as the fraction of sources with the integrated
magnitude similar to that of the sample ($20.5<$FUV$<22.5$) that were
originally more than one source. We find this probability to be
3.1\%. Consequently, in our \hst\ sample of 30 galaxies we expect one
companion to be an interloper. Note, however, that this one interloper
will preferentially be found among the sources in which the UV flux of
the main galaxy itself is faint, such as those in the class of
small-scale star formers. Therefore, it is quite conceivable that
either SR27 or SR30 is a chance superposition of an ETG and an
unrelated UV-bright galaxy. Future observations could provide a
definitive answer, but the main conclusion drawn in this section, that
gas-rich satellites are consumed by ETGs, remains a likely explanation
for most of the objects in this class.

\begin{figure*}
\epsscale{1.0} \plotone{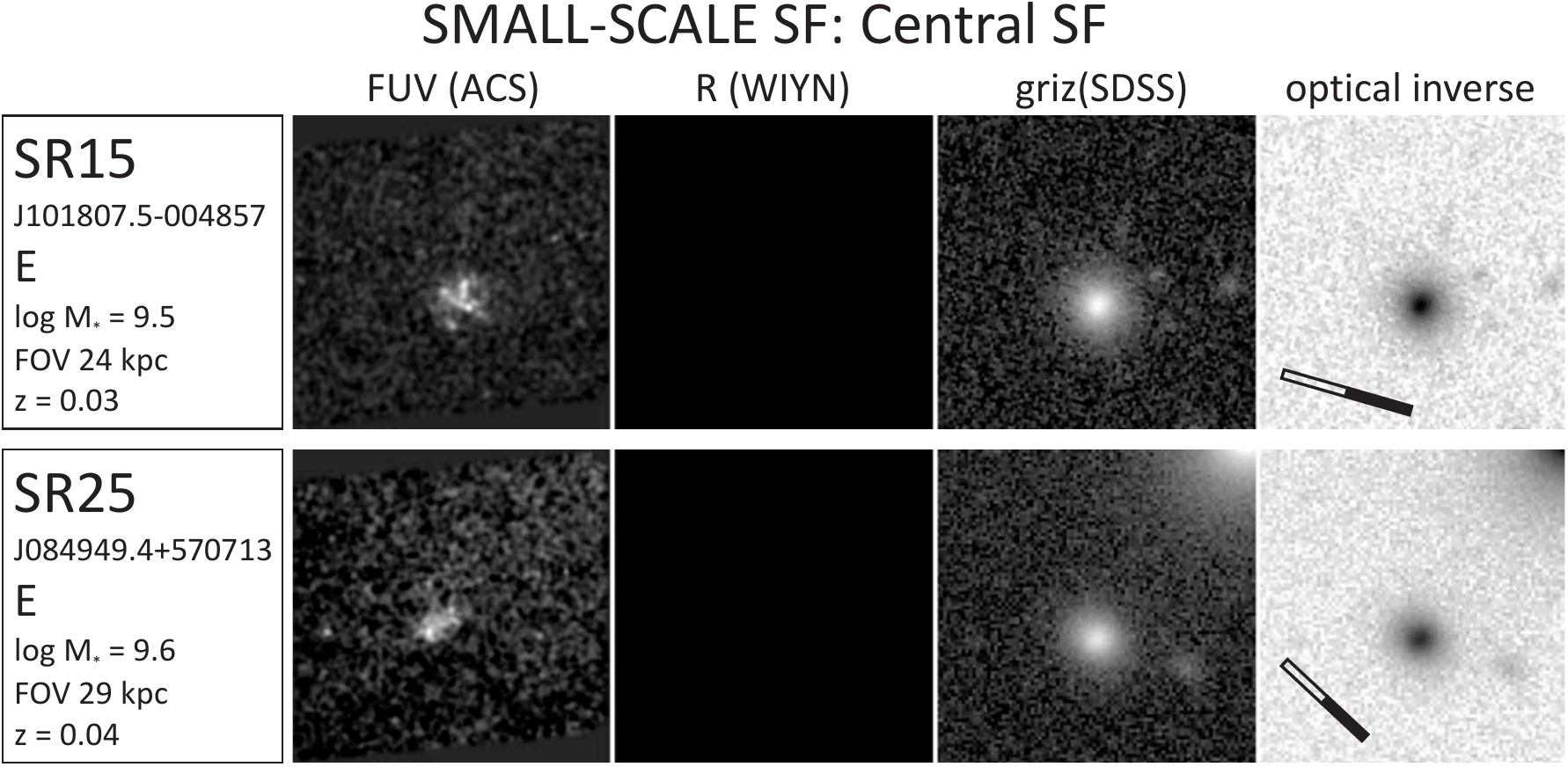}
\caption{Far-UV and optical images of early type galaxies with central
    small-scale SF. See Figure \ref{fig:panel1} for full description.}
\label{fig:panel8}
\end{figure*}

{\it Central small-scale SF}. Most of our sample galaxies are relatively
massive ETGs with stellar masses in the range $10.2 < \log M_* <
11.2$. Two exceptions are SR15 and SR25 with much lower masses of
$\log M_*=9.5$ and $\log M_*=9.6$ respectively (Figure
\ref{fig:mass_fuvr}). These are also among the faintest UV sources in
our sample. \hst\ UV imaging of them is shown in Figure
\ref{fig:panel8}, and they both feature central small-scale
star-forming regions 2.3 kpc across. Due to the high level of noise it
is difficult to say more about the structure of these regions; SR15
possibly shows several UV bright knots. These too may be minor merger
remnants, which can more easily boost the UV-optical color of a
low-mass main galaxy. The central UV emission appears quite diffuse,
which is probably why it led to no detectable emission in SDSS spectra.

To summarize, most small-scale structures in ETGs are non-uniform and
possibly related to minor mergers that do not lead to substantial
disruption of the main galaxy. Note that the relative dominance of
off-center structures with respect to central ones (four to two) does
not necessarily reflect the situation in the overall ETG population
because strong central SF is selected against in our sample.

\subsection{Unresolved UV sources and ETGs without strong UV excess} \label{ssec:compact}

\begin{figure*}
\epsscale{1.0} \plotone{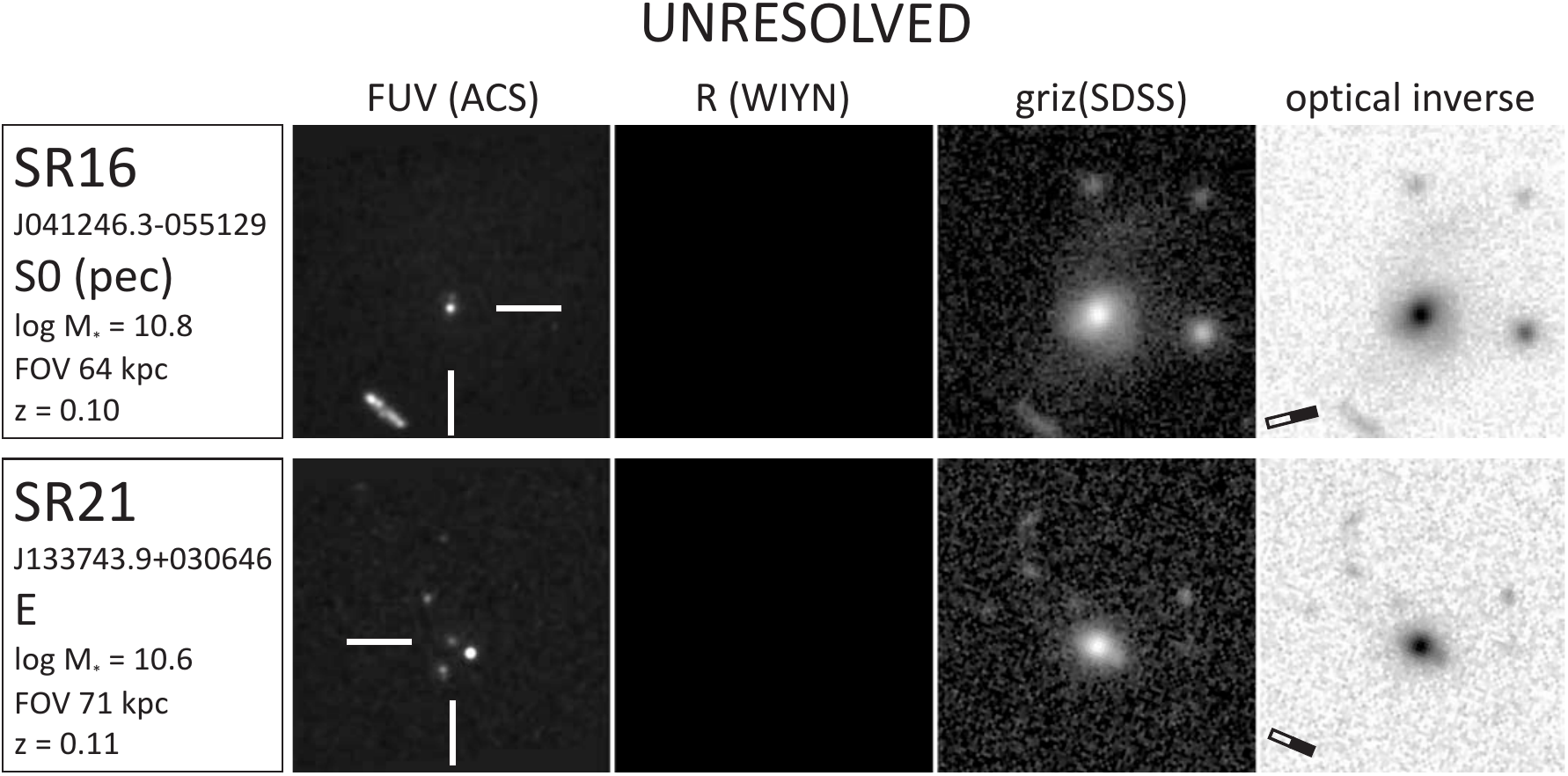}
\caption{Far-UV and optical images of early type galaxies with unresolved UV
  emission. See Figure \ref{fig:panel1} for full description.}
\label{fig:panel9}
\end{figure*}

{\it Unresolved UV sources}. There are two galaxies in the \hst\
sample in which we do not detect resolved structures indicative of
star formation. They are shown in Figure \ref{fig:panel9}. SR16 is
dominated by a bright unresolved central source. The color profile
measured in Paper II shows it to be bluer (FUV$-r\approx5.0$) than any
other center in the entire sample, and therefore least consistent with
an old population UV upturn. An AGN does not appear a likely
explanation given that the UV color is not very blue. Nevertheless, an
unusually strong classical UV upturn is probably the least problematic
explanation, since any unresolved nuclear star formation would be
detectable in the SDSS spectrum. The other galaxy in this class, SR21,
shows four unresolved sources within the optical extent of the galaxy,
of which the brightest is offset from the optical center of the main
galaxy and could be a compact burst of SF in a companion galaxy. As
measured in Paper II the fainter central source itself is too red to
be SF, so is probably from UV upturn as well.

Most galaxies classified as extended SF ETGs and small-scale SF ETGs
contain a compact UV source in addition to the more extended
emission. Despite their sometimes high UV surface brightness these
unresolved central sources contain only a small fraction of the total UV
flux. The nature of this emission is discussed in more detail in Paper
II. The FUV$-r$ colors of the central regions are consistent with
emission from old-population UV upturn. Apparently such stellar
emission appears rather compact even in $0\farcs3$ resolution FUV
imaging.

\begin{figure*}
\epsscale{1.0} \plotone{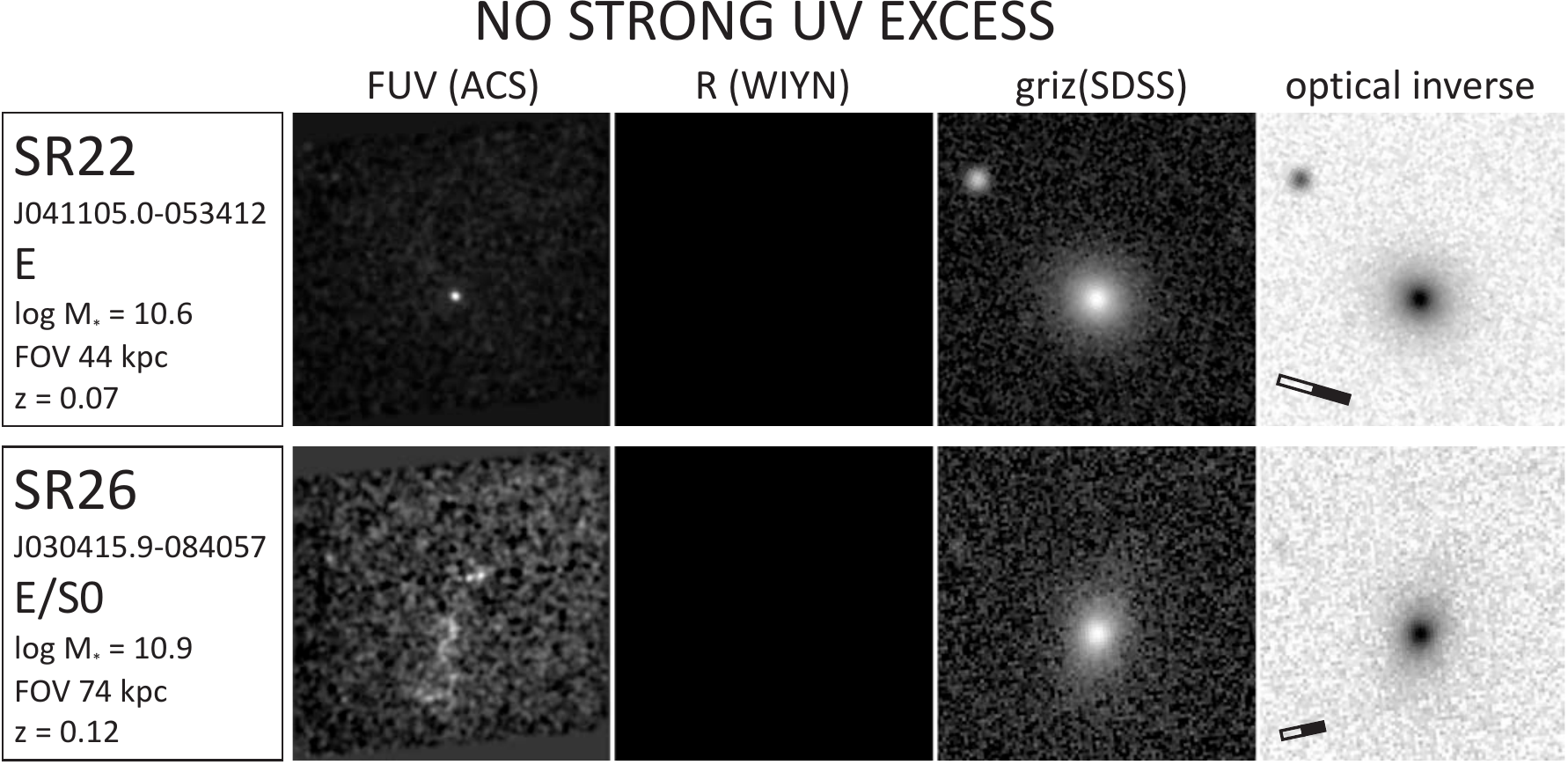}
\caption{Far-UV and optical images of early type galaxies with weaker UV
  excess than the rest of the sample, due to the original FUV
  magnitudes being underestimated (too bright). See Figure
  \ref{fig:panel1} for full description.}
\label{fig:panel10}
\end{figure*}

{\it ETGs with no strong UV excess}. Finally, we briefly discuss two
galaxies which entered the sample due to the erroneous original
\galex\ photometry. These are shown in Figure \ref{fig:panel10}. SR22
has unresolved UV morphology and, based on the latest \galex\ catalog
(GR6) and confirmed by measurements in Paper II, has FUV$-r\approx 6.7$,
consistent with a weak old-star UV upturn. SR26 also no longer
qualifies as having a strong UV excess -- its FUV flux is actually
very faint. It tentatively shows filamentary extended structure and is
thus possibly SF-related but appears quite unlike anything in our
``normal'' extended SF sample. The sources in this category will not
be analyzed in the rest of the paper.

%%%
\section{Ionized gas emission} \label{sec:ha}
%%%

The absence of detectable \ha\ emission in SDSS spectra of the sample
appears to contradict the star formation seen in the UV. SR2010
discussed two possibilities that would resolve this problem. One is
that no \ha\ emission is present in the central $3"$ region sampled by
SDSS fibers but that such emission exists further out. The fiber
diameter is one tenth of the size of panels in Figures
\ref{fig:panel1}--\ref{fig:panel10}. This explanation is certainly
consistent with the ring morphologies that dominate most of the sample
(see also the Appendix).  SR2010 provides an estimate that if one
assumes that the measured UV flux is due to SF, the expected SNR in
\ha\ would be some ten times higher than the observed limits. To this
argument we add one caveat. Such estimate did not take into account
the fact that the strong red continuum present in ETG bulges would act
as a high background and drive the SNR of the \ha\ flux
down. Therefore, if we now consider the \ha\ equivalent width (i.e.,
flux divided by the continuum) instead of the flux alone, we find that
the sample is still deficient in \ha\ with respect to what the UV
implies, but by a more moderate factor of $\sim 3$. Another proposed
explanation was that \ha\ is genuinely absent throughout the galaxy,
either because there is no ``instantaneous'' SF ($\sim 10^7$ yr),
while the UV would still trace the ``recent'' SF ($\sim 10^8$ yr), or
because the environment is not conducive to high-mass SF
\citep{krumholz}. The absence of \ha\ due to stochastic SF is not
expected because the SFRs are several orders of magnitude higher than
the expected onset of this phenomenon (e.g., \citealt{lee}). To test
these scenarios and establish if instantaneous SF is detectable at
large scales we conducted narrow-band \ha+[NII] imaging as described
in \S \ref{ssec:wiyn}.

\begin{figure}
\epsscale{1.0} \plotone{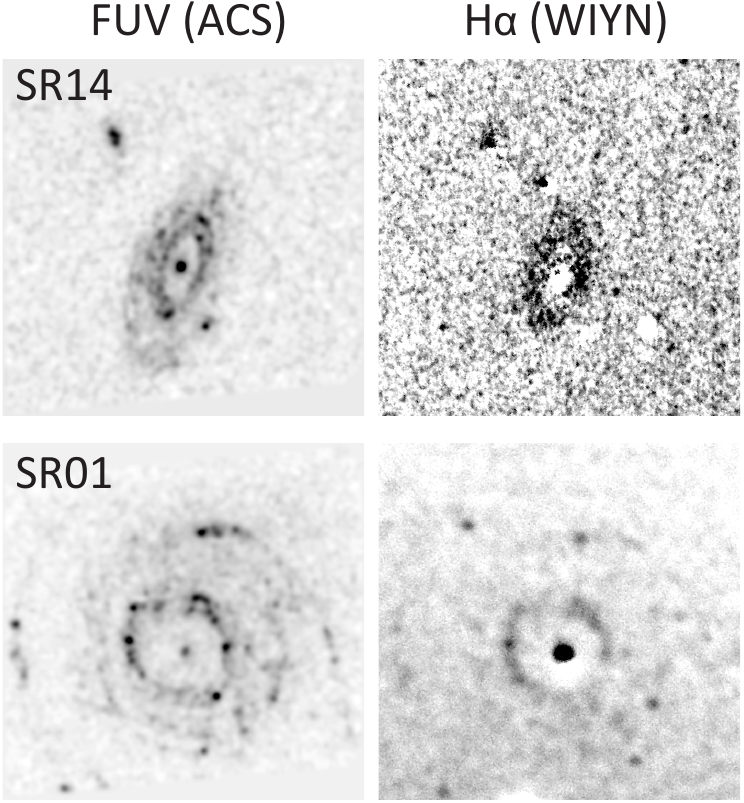}
\caption{Two early type galaxies with prominent \ha\ emission. On the
  left is the \hst\ ACS far-UV image, and on the right is the
  continuum subtracted \ha+[NII] narrow-band image. One other galaxy
  (not shown) has confident \ha\ detection out of 10 ESF-ETGs
  imaged. \ha\ is more likely to be detected in galaxies with brighter
  UV emission, indicating that the ionized gas emission is probably
  present in all UV emitters, but only detected in more luminous
  cases.}
\label{fig:ha}
\end{figure}

Narrow-band imaging was attempted for 12 galaxies in the sample: 10
belonging to the extended SF class (SR01, 02, 05, 06, 08, 11, 12, 14,
17 and 29) and two in the small-scale SF class (SR27 and 30). Out of
10 ESF-ETGs, we confidently detect \ha\ emission in three galaxies (SR
01, 12, 14), and tentatively in another three (SR02, 11, 17). Neither
of the small-scale star formers yielded a detection. Two of the
detections are shown in Figure \ref{fig:ha}. For SR14 (UV disk with a
central hole) \ha\ appears as a ring whose extent roughly follows the
brighter, inner portion of the UV disk. Another detection shown in
Figure \ref{fig:ha} is of SR01, classified as an irregular UV ring
within extreme disk. Again, the \ha\ and the Uv morphologies are
similar. Some compact \ha\ regions have no UV counterpart and may not
be associated with the galaxy.

On average the three galaxies with confident \ha\ detections are
brighter in the FUV than those with tentative detections, which in
turn are on average brighter than those with no detections. Therefore,
\ha\ detection appears primarily related to sensitivity and not the
intrinsic presence of ionized gas. The absence of \ha\ in the SDSS
spectra serves as a reminder that the information from fiber spectra
cannot always be extrapolated to entire galaxies.

%%%
\section{Optical morphology of the \hst\ sample} \label{sec:opt}
%%%

One of the primary reasons for obtaining new optical imaging in this
work was to be able to determine a more secure optical classification
than what is possible with SDSS images alone. At the typical redshift
of our sample ($z\approx 0.1$) distinguishing between Hubble types can
be challenging. Especially difficult is the distinction between S0s
(lenticulars) and pure ellipticals (Es) \citep{cheng_faber}. This
separation is often ambiguous even in nearby, well-studied galaxies
\citep{kormendy}.

Deeper and sharper WIYN optical images are available for 18 out of the
27 galaxies in the \hst\ sample\footnote{The two sources without
  strong UV excess are excluded from consideration.} and are shown in
the second column of panels in Figures
\ref{fig:panel1}--\ref{fig:panel7}. Considering that not all galaxies
have new imaging, we approached the classification in the following
way. We started by classifying based on the SDSS optical image ($griz$
co-add) alone (third column of panels in Figures
\ref{fig:panel1}--\ref{fig:panel7}), assigning preliminary Hubble
types and noting uncertainties. We used the criteria of
\citet{cheng_faber} to distinguish between Es and S0s.  The prime
criteria were elongation due to a moderately edge-on disk and, if the
galaxy was face on, the sharpness of the edge in the outer profile.  A
sharp fall-off was taken to represent the edge of an exponential disk,
and a gradual fall-off was taken to represent the slower decline in an
outer de Vaucouleurs or high-Sersic profile. Then, the classification
of 18 galaxies based on WIYN imaging was performed independently. This
classification was generally less ambiguous. Comparing the new
classification to the preliminary one we noticed that the new types
tend to be later; e.g., many galaxies that looked as if they could be
either S0s or Es from SDSS were now classified as S0s. In addition to
this, bars were revealed in several cases in which they were not
visible in SDSS images. Trying to correct for this ``low-resolution''
bias, we revisited the SDSS classification of galaxies having no WIYN
imaging and assigned final Hubble types. In several cases we still
could not decide between E and S0 type and left their type as ``E/S0''
classification. Hubble types are given in Figures
\ref{fig:panel1}--\ref{fig:panel7} as well as Table \ref{table}.

Overall, the sample is dominated by S0 galaxies (17, of which 15 are
S0 and 2 are S0/a). Four are later than S0 (2 Sa and 2 Sab), and six
are earlier (3 E/S0 and 3 E). Since the original SR2010 selection
visually screened the candidates and removed obvious late-type
interlopers, the relative lack of galaxies later than S0 is
expected. Even the four later-type galaxies are not typical Sa or Sb
spirals. Rather, they look like S0s with certain spiral-like
features. More surprising is the dearth of true ellipticals (spheroids
lacking disks) compared to S0s. We will discuss this absence in more
detail later (\S\ \ref{sec:s0_vs_e}).

An especially important optical morphological feature in the context of
this study is the presence of a stellar bar. Our sample is dominated by
star-forming rings and the classical mechanism for forming and
maintaining such rings is the secular dynamical effect of
a bar \citep{buta_combes}. We find evidence for bars in 8
galaxies. This represents 30\% of the full sample, or 38\% of galaxies
with secure disks (i.e., excluding E/S0s and pure Es). This rate is
roughly consistent with the 30\% observed in more nearby S0s
\citep{aguerri} but falls quite short of the fraction of the sample
exhibiting UV rings (86\% of galaxies with secure disks). Apparently,
bars are not the dominant mechanism for maintaining
rings in our sample.

Optical morphological disturbances are another feature that could
provide a clue to the origin of star-forming gas in our
sample. \citet{schawinski10} and \citet{kaviraj10} show that the
disturbed optical morphologies in ETGs, such as shells, tidal debris
or dust lanes, are correlated with a UV excess, thus connecting recent
SF to mergers. Is the opposite true as well, i.e., are ETGs with a UV
excess, such as our sample, necessarily morphologically disturbed? To
reach their conclusions, \citet{schawinski10} and \citet{kaviraj10}
relied on deep SDSS imaging of Stripe 82, which is 2 mag deeper than
the regular SDSS imaging. Our WIYN imaging offers similar, or slightly
higher, increase in depth over the nominal SDSS, so it should be
sensitive to features detected in these studies. However, we see no
evidence for shells or tidal debris in any of the 18 galaxies for
which there is WIYN $R$-band imaging. One galaxy (SR17) appears to
have tidally induced tails, and there is a nearby companion that shows
optical disturbance as well. Only one galaxy shows a dust lane (lower
left of SR08 disk, classified as regular wide ring, Figure
\ref{fig:panel1}), with two more (SR10 and SR14, both disks with small
central holes, Figure \ref{fig:panel3}) showing possible hints of dust
lanes. However, the latter look like spiral dust lanes, so they are not
merger signatures.

Overall, the optical morphologies in our sample are very relaxed. Even
in cases in which smaller companions are present there is no optical
evidence of strong disturbance.

%%%
\section{Relation between the optical and the UV morphologies} \label{sec:uv_opt}
%%%

In this section we examine the connection between the optical and UV
morphologies of the \hst\ sample. A strong correlation between UV and
optical morphologies would suggest that recent SF may be
preferentially found in certain types of ETGs. This would provide
useful insight into how SF is regulated in ETGs, possibly hinting at a
morphological dependence on feedback processes. Also, the relation
between optical features such as the bar and the UV rings would help
explain the latter. Finally, the analysis of the UV and optical sizes
of galaxies may reveal their star formation history.

\subsection{UV classes versus Hubble types}

\begin{figure}
\epsscale{1.0} \plotone{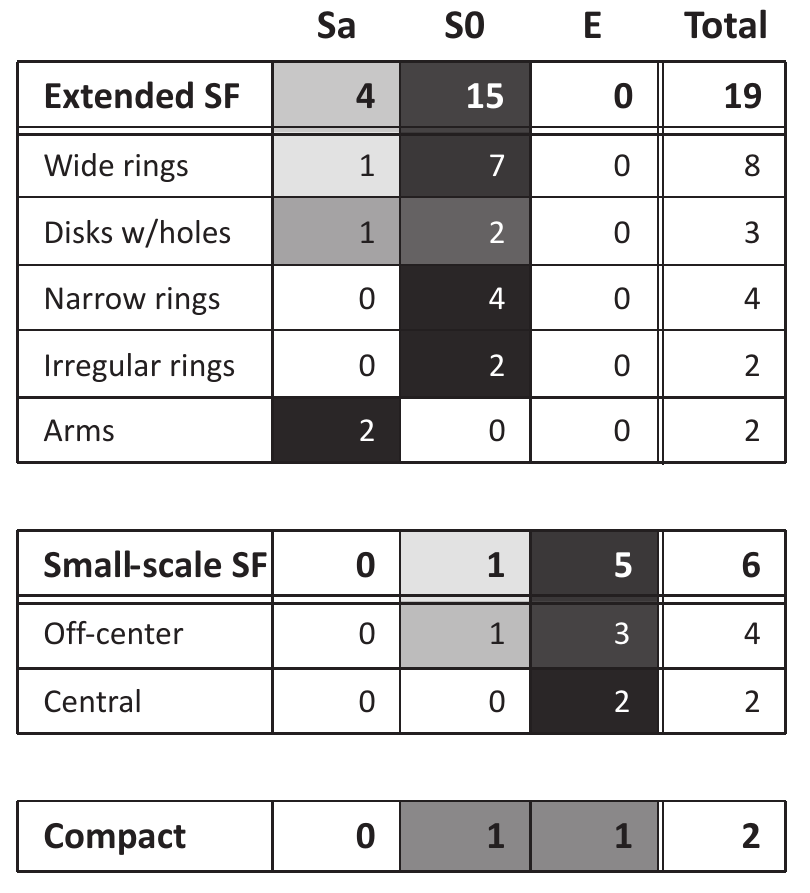}
\caption{Breakdown of Hubble types with respect to UV morphology class
  and subclass. Shades of gray represent relative frequency in each
  (sub)class. Barred and non-barred types are combined. Each type
  includes subtypes: Sa column includes two galaxies typed as Sab, S0
  includes two S0/a galaxies, and E includes four E/S0 (uncertain E or
  S0) cases. There is a clear trend such that galaxies with
  extended SF favor later Hubble types (S0 and later) while galaxies
  with small-scale SF tend to have earlier Hubble types.}
\label{fig:uv_opt_table}
\end{figure}

We represent the relation between UV and optical morphological types
as a schematic table in Figure \ref{fig:uv_opt_table}. Hubble types
are grouped into three categories: Sa (including Sab), S0 (including
S0/a), and E with E/S0 (i.e., uncertain E or S0). The number of
galaxies of each optical type and UV class/subclass is given in a
respective cell. The fraction of the given optical type with respect
to the total number of galaxies of a given UV (sub)class is shown by
the shading of the corresponding cell: light shading indicates lower
fraction. A striking result is that none of the 19 ETGs with extended
SF is classified as E/S0, and the majority (80\%) are classified as
S0/a. This provides a strong indication that the galaxy-scale SF may
not be happening in true ellipticals but only in galaxies already
containing disks, i.e., S0 and later.

We next discuss each UV morphological class and subclass in relation
to the optical types present in it. S0s dominate among regular wide
rings (88\%). The only wide ring galaxy that was classified as a later
type (SR02, SBa) does not have very prominent arms. Out of the three
disks with central holes, SR12 is classified as an Sa and the other
two as S0 and SB0/a. SR12 has well defined arms, while for SR14
(SB0/a) there is some ambiguity as to whether the arms are the
brighter portions of the ring faintly visible in the optical. Later
types appear somewhat more likely in this UV subclass. S0s are the
only optical type in which narrow or irregular rings occur, though the
numbers are small to asses the true incidence of later types. Finally,
both galaxies belonging to the UV {\it arms} subclass appear armed in
the optical as well and are therefore classified as later
types. However, these arms are much less prominent in the optical than
in the UV. In the case of SR17, as already discussed, the arms appear
to be tidal in origin, and both itself and the bright optical
companion show signs of disturbance. In the other galaxy with UV arms
(SR11) they almost close to form a pseudoring. In the optical this
closing is much less evident.

The situation is quite the opposite for galaxies classified as having
small-scale SF. Five out of six are classified as E/S0 or E; only one
is a clear S0. From this albeit small statistic we can speculate
that in the same way in which the extended SF in ETGs appears to favor
S0s, so could more compact and more central SF favor ellipticals. We
will return to this question in \S\ \ref{ssec:s0phen}.

For the unresolved UV sources, one (SR16) is a peculiar S0 and the
other (SR21) is an E.  Given that the source of UV in SR16 is not
clear (possibly old stars), we cannot place much weight on the fact
that this galaxy, unlike most others in the sample, does show some
evidence for disrupted morphology. For the other galaxy, the bright
unresolved UV source can be traced to the faint optical ``companion''.

\subsection{UV classes versus the presence of a bar}

\begin{figure}
\epsscale{0.7} \plotone{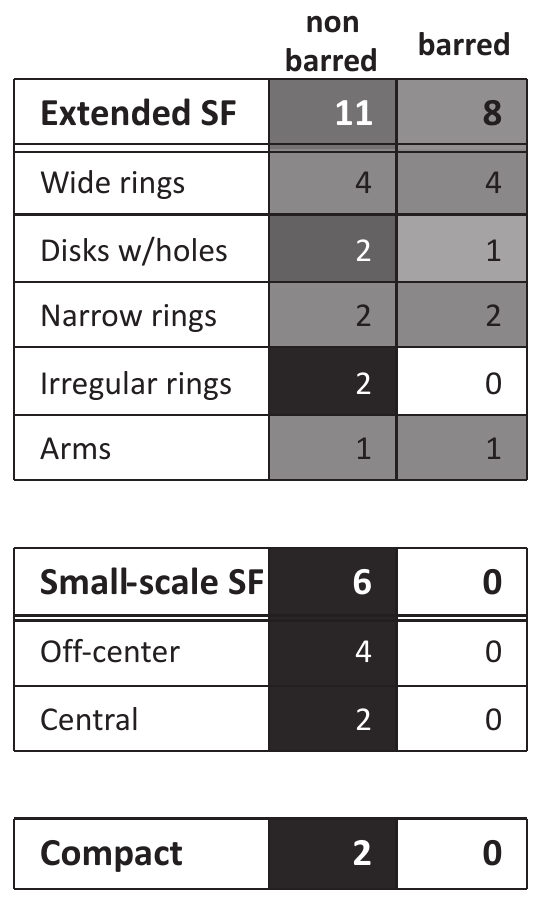}
\caption{Breakdown of barred vs.\ non-barred disks with respect to UV
  morphology class and subclass. Shades of gray represent relative
  frequency in each (sub)class.}
\label{fig:uv_bar_table}
\end{figure}

We continue our analysis of the relation between UV and optical
morphologies by focusing on the incidence of stellar bars in galaxies
of various UV classes. Bars are the primary mechanism believed to form
stellar rings, so their presence in our sample dominated by UV rings
would be of great importance. We represent this information in a
schematic table in Figure \ref{fig:uv_bar_table}.

We find no bar in any of the six small-scale star-forming ETGs or in
the two ETGs with unresolved UV sources. Given that most of these galaxies are
classified as E/S0s or Es, no bar is expected. The presence or absence of bars is
of much more significance for the 19 ESF-ETGs, of which 18 have rings,
which, if of non-collisional origin, are believed to require bars to
produce and perhaps also to maintain them.

Out of 8 galaxies exhibiting regular wide rings, whose smooth
morphology makes them prime candidates for resonance rings, one half
shows evidence for a bar and the other half does not. The presence or
absence of a bar, while notable in the optical, does not reflect
itself in any obvious difference in the UV. The UV rings have the same
extent regardless of the presence of a bar, and they appear to be
equally well defined. Of the galaxies with regular wide rings, one
stands out in terms of its optical morphology. SR03 is the only galaxy
in our sample that has a conspicuous {\it optical} ring that is
detached from the bulge of the galaxy. The ring appears to be held in
place by a very strong bar, one of only two such strong bars in our
sample. While there are several optical detached rings that have been
studied in local S0s (e.g., NGC1291 and NGC1543, both barred), neither
their optical nor UV morphologies fully match that of SR03. Namely, in
SR03 the bar extends all the way to the ring, making it presumably a
co-rotation ring, while in locally studied galaxies the co-rotation
ring is basically an inner ring, imbedded in a disk or surrounded by a
larger outer ring \citep{buta_combes,buta_morph}. An optical ring
means that, unlike other galaxies in our sample where the ring is
mostly a feature composed of young stars, here the older stars are
co-located in the ring, indicating that it has been present for most
of the lifetime of the galaxy, or that the dynamical processes were
able to migrate the stellar population alongside the star-forming gas.

Moving to other UV classes, only one of the three disks with small central
holes appears to have a bar (SR14). Galaxies with narrow rings are evenly
split between barred and non-barred, with no significant differences
in UV morphology. Neither of the galaxies classified as irregular
rings in extreme disks has a bar. Given the irregularity of the rings,
it is quite possible they are not of the resonant type, which means
that bars are not expected to play any part in their formation or
maintenance.

Of the two galaxies with UV arms, SR17 has no ring and no optical
bar. On the other hand, SR11, with arms forming a pseudoring, has a
strong bar. This bar has an associated inner ring, thus representing
a classical resonance co-rotation ring. A good local example of a
galaxy with very similar UV and optical morphologies is UGC12646
\citep{buta_combes}. Altogether SR11, with its bar, inner ring, and
outer pseudoring is very well explained by bar-induced dynamical
processes, convincingly reproduced in numerical simulations
\citep{schwarz_pasa,athanassoula}

To summarize, we find that among the ESF-ETGs the barred and
non-barred galaxies appear quite similar in the UV. The range of their
UV-based SFRs (determined using SED fitting, see \citealt{s07})
largely overlap, although barred galaxies on average have $\log {\rm
  SFR}=-0.34\pm0.06$, while non-barred have $\log {\rm
  SFR}=-0.16\pm0.07$, a somewhat significant difference ($2\,
\sigma$), but in the opposite direction from what might be
expected. Therefore, the presence of a bar probably does not lead to
higher SFRs in our sample.

%%%
\section{Implications of UV and optical sizes for star formation history \label{sec:sizes}}
%%%

In this section we will deduce some aspects of the star formation
history of the galaxies in our sample by analyzing their UV and
optical sizes. First we try to establish whether the current SF is
part of a long-term process that leads to disk buildup in ETGs, and
then we explore whether the relation between the optical and UV sizes
can point towards the origin of the gas responsible for SF.

\subsection{Current SF and the optical disk re-growth \label{ssec:optsize}}

What is the relation between the current SF and the optical disk? If
we assume that the optical disk requires prolonged SF to build up, the
question can be asked whether the current episode of SF visible in the
UV is related to a {\it renewed}, additional disk building {\it after}
the galaxy got onto the optical red sequence (i.e., are stars being
made beyond the ``original'' extent of the disk), or whether it is the
fading last phase of the original disk SF, confined within the ``expected'',
non-enhanced extent.

To try to answer this question we compare the {\it optical} sizes of
our \hst\ sample galaxies to an underlying population of {\it
  quiescent} disk-like ETGs. If the optical sizes are larger than
those of the comparison sample it would indicate that (a) additional
disk building took place after the galaxy arrived on the optical red
sequence (i.e., after the initial intense SF was finished). If, on the
other hand, the optical sizes of ESF-ETGs are comparable to those of
quiescent ETGs, it would indicate either (b) that the current episode
of SF has not lasted sufficiently long, or has been intense enough to
have produced much of the optical disk, or (c) that the SF that we see
is just the fading remnant of the SF that produced the original,
normal-sized disk. Case (a) would arise from processes such as ETG
rejuvenation through a steady, long-lasting IGM accretion; case (b)
would result from either very low level accretion or from a transient
minor-merger induced SF, while case (c) would be consistent with a
recent arrival onto the optical red sequence, i.e., the final phases
of SF quenching. For this test we use optical, $r$-band sizes to
minimize the ``artificial'' boosting of sizes of ESF-ETGs at shorter
wavelengths due to ongoing SF. In other words, the assumption is that
fading alone (case c) will not affect the optical size measured in the
$r$ band.

\begin{figure}
\epsscale{1.2} \plotone{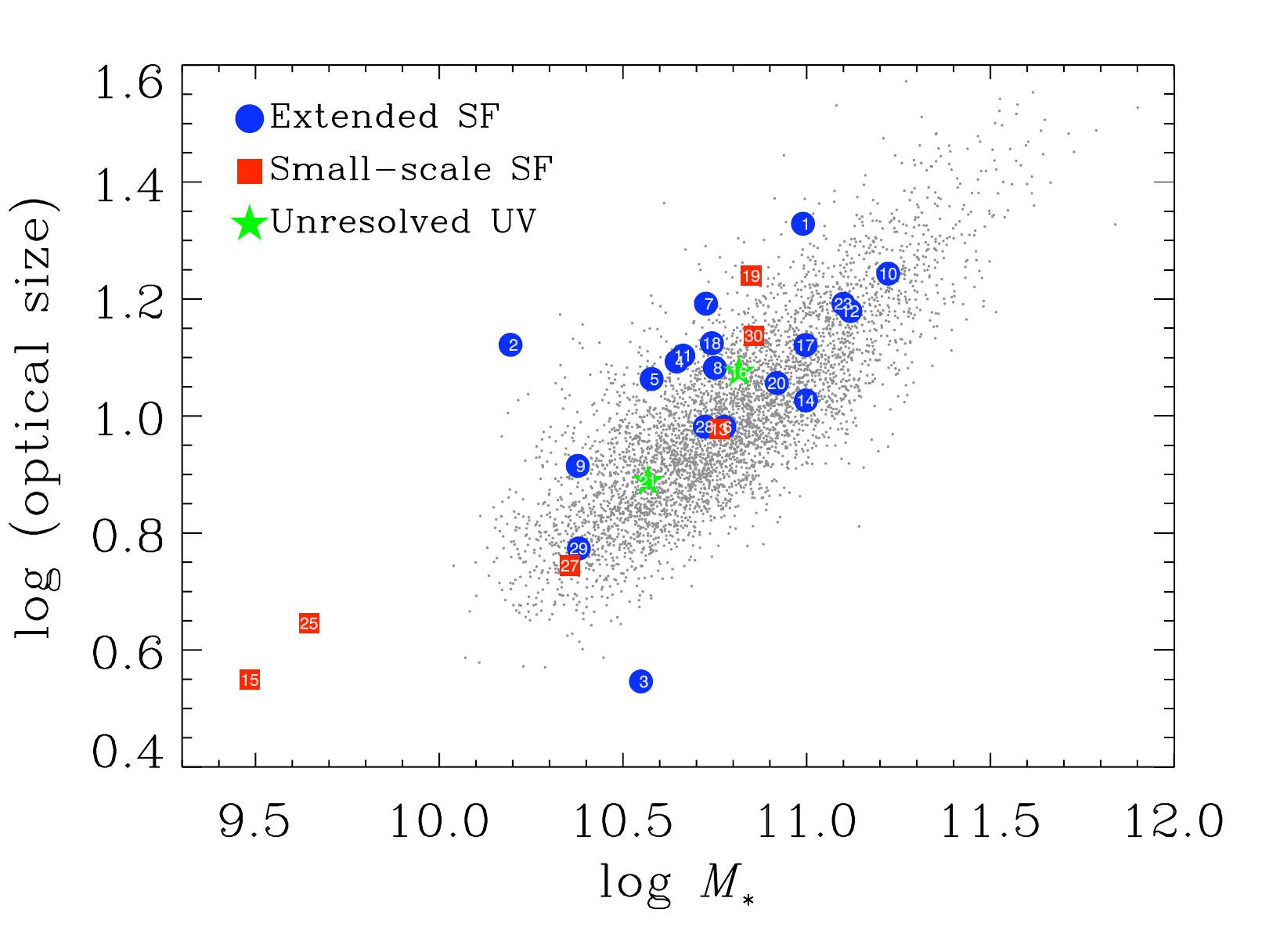}
\caption{Comparison of optical disk physical sizes of the \hst\ sample
  (various colored symbols) and the underlying population of quiescent
  disk-like ETGs, at a given stellar mass. Size is measured as the radius
  containing 90\% of the Petrosian flux in $r$ and is given
  in kpc. The \hst\ sample is split into UV morphology classes:
  extended SF (blue dots), small-scale SF (red squares) and unresolved
  (green stars).  Numbers inside symbols indicate object
  identifications (SR\#). The underlying sample is generally assumed
  not to have experienced additional disk growth after getting on the
  optical red sequence, allowing a comparison with the \hst\ star-forming
  sample to show any subsequent build-up. SR01, 02 and 07 are most
  significantly larger than the underlying population.}
\label{fig:optsize_mass}
\end{figure}

To characterize the optical disk sizes we use physical sizes
corresponding to radii encompassing 90\% of the Petrosian flux in
$r$-band (rest-frame $V$). We verify that this choice of radius is a
good measure of the full extent of disks by overlaying them on optical
images. To select the comparison sample of quiescent disk-like ETGs,
we choose those with the similar redshift range as the sample ESF-ETGs
($0.08<z<0.12$), with red UV-to-optical and optical colors to ensure
quiescence (FUV$-r>6$ [including UV non-detections] and $g-r>0.7$),
and with optical light concentrations like that of the sample
($2.5<C<3.2$). The upper cut is the maximum concentration of the \hst\
sample and is meant to reduce the contamination from pure
spheroids. Note that any remaining spheroid contamination will be
adding galaxies with sizes larger than disky ETGs, since at a given
mass we find that ETGs with $C>3.2$ are larger than those in the disky
range of $C$. The assumption is that most of the quiescent disk-like
ETGs did not experience additional disk building upon arrival on the
optical red sequence and can thus serve to show reference sizes that
result from the original disk building. The results are shown in
Figure \ref{fig:optsize_mass}, where we plot the physical sizes of the
\hst\ sample and the comparison population against stellar mass. Blue
circles represent ESF-ETGs, red squares are small-scale star formers,
and green stars are unresolved UV galaxies. Numbers within the symbols
provide galaxy IDs. Overall, ESF-ETGs tend to be larger than the
corresponding red-sequence ETGs, though they range from significantly
larger to similar in size, and even a few undersized\footnote{SR03,
  the galaxy with the detached optical ring, is the only galaxy in the
  \hst\ sample significantly below the trend. We verify that in this
  case the SDSS radius encompassed only the bulge/bar region and none
  of the disk/ring, therefore severely underestimating the optical
  extent of the galaxy. The actual size is some 3.6 times greater,
  making this one of the galaxies that is {\it larger} than what is
  typical for its mass.}. Formally, we find using a K-S test that the
distribution of the residuals of optical sizes of ESF-ETGs with
respect to a nominal size-mass relation defined by quiescent ETGs has
a probability of only $2.4\times10^{-3}$ to have been drawn from the
same size distribution as the quiescent ETGs. The galaxies that most
stand out are SR01, 02 and 07. These include both galaxies that are
classified as {\it irregular rings in extreme disks}, while the third
galaxy (SR02) is classified as a {\it regular wide ring}. While the
presence of both extreme disks among the ``oversized'' galaxies does
not appear to be a coincidence, there is no clear cut connection
between the UV class and the level of disk size enhancement. Of the
five ESF-ETGs that are visually identified as having some star
formation outside of where we can trace the optical extent (SR01, 02,
04, 06, 07), four have oversized optical disks, which suggests that
the current SF will lead to further growth of optical disks.

To summarize, we see evidence that a part of the sample ($\sim 1/2$)
experienced some level of disk building ($\sim 50\%$ increase in size)
with respect to other disk-like ETGs, and the SF that we see is very
likely related to that process (see also arguments in Paper II). This
new disk-building phase apparently proceeds in the same plane as the
existing disk, at its outskirts. The other half has disk sizes that
suggest that the currently observed SF has not played a major role in
disk building after most of the original SF ceased, either because
disk enhancement involving external gas has started more recently, or
because the SF is using up the original internal supply of gas.

\subsection{Comparison of UV and optical disk sizes}

While most of the UV extent in ESF-ETGs is found within the optical
extent of the galaxy, this does not mean that the UV light profile
follows that of optical light. As shown in the analysis of the surface
brightness profiles in Paper II, the optical profiles drop more
steeply outwards than the FUV profiles, which are mostly flat. In
other words, the FUV to optical colors become bluer away from the
center by some 2 to 4 magnitudes. We can address the connection
between the UV and optical, and therefore between younger and older
populations, using a morphological analysis as well. Specifically,
systematic differences in the {\it ratio} of UV to optical disk sizes
can help us distinguish between the fading of the original SF and the
possible subsequent episodes of SF, which would preferentially happen
at the outskirts of a galaxy. In the previous section, where we
compared the optical sizes of star-forming and quiescent disks, we
were only able to identify likely cases of significant buildup due to SF
subsequent to the arrival on the red sequence, while the small-level
build up (including accretion that started/resumed more recently) and
the disk fading would not have led to the increase in optical
sizes. Therefore, the analysis of the ratios of UV to optical disk
sizes can provide further information on the processes that lead to
SF.

\begin{figure}
\epsscale{2.2} 
\plottwo{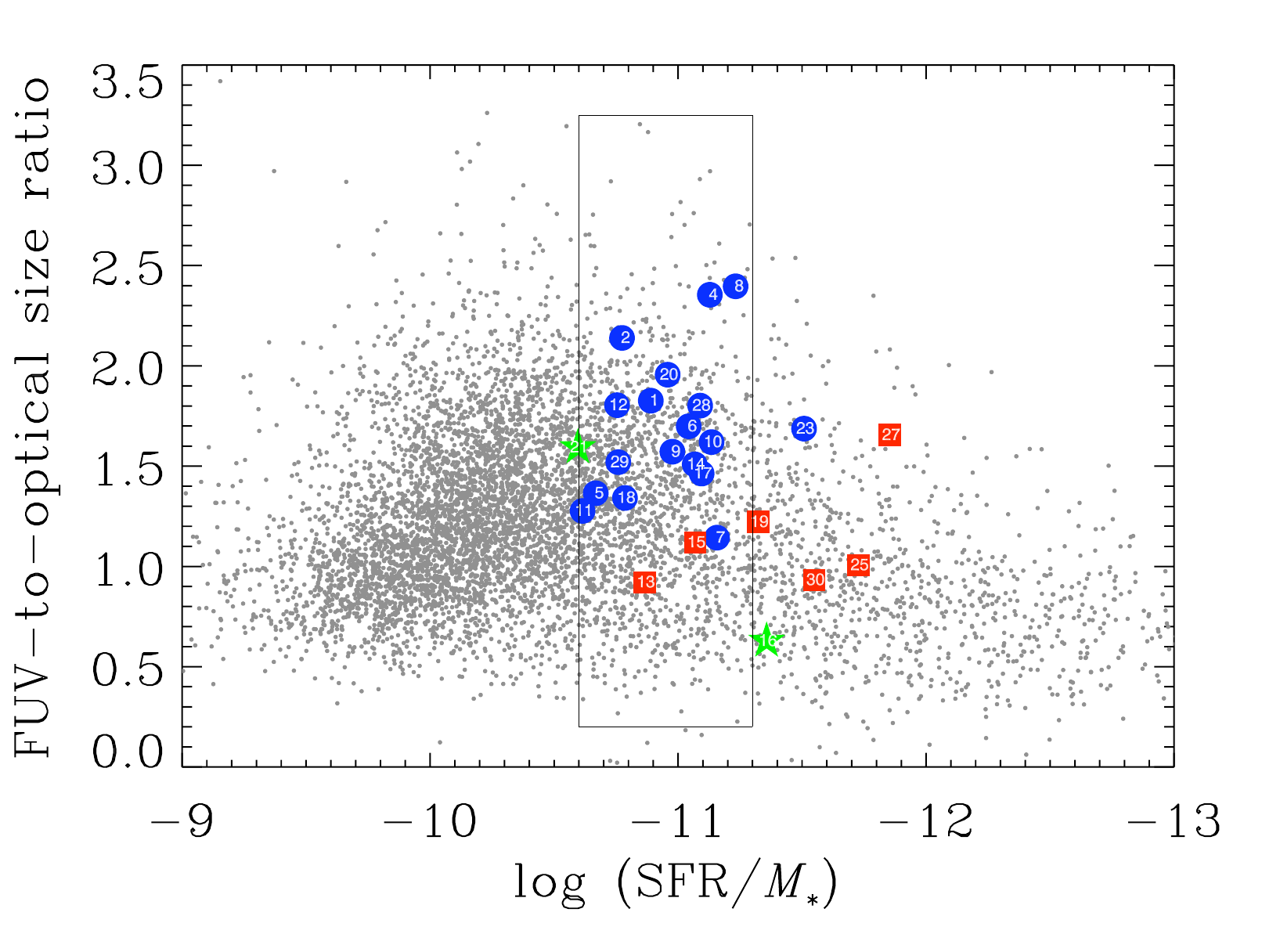}{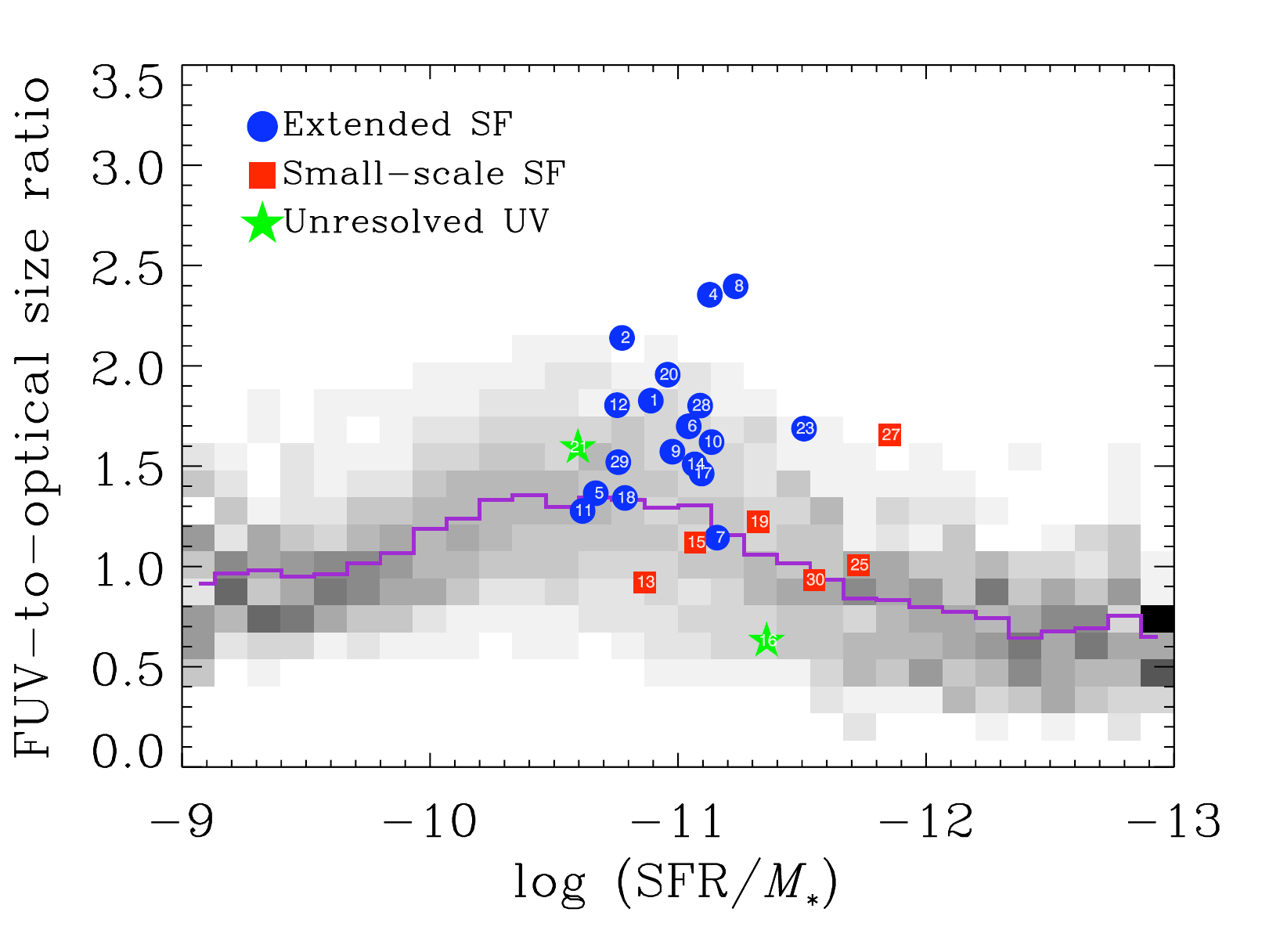}
\caption{Comparison of the far-UV to optical size ratios for the \hst\
  sample (colored symbols) and the underlying population of massive
  galaxies at the similar redshift ($10.2 < \log M_* < 11.2$,
  $0.08<z<0.12$), as a function of the specific dust-corrected
  SFR. FUV sizes are FWHMs measured by \galex, while optical sizes are
  90\% Petrosian flux diameters from SDSS. Top panel shows the
  underlying population as a scatter plot, while the bottom panel
  shows them as distributions normalized in each mass bin, with the
  purple line showing the median. Actively star-forming (blue cloud)
  galaxies are to the left of the sample, while the quiescent galaxies
  are to the right. ETGs with extended SF (blue symbols) tend to lie
  above the median size ratio. The part of the underlying sample used
  for calculating the mean comparison ratios, having the same specific
  SFR as the sample (see text), is indicated with a box in the upper
  panel. ESF-ETGs are systematically larger than other green valley
  (intermediate specific SFR) galaxies of similar mass.}
\label{fig:sizerat_ssfr}
\end{figure}

The next test consists of comparing the UV-to-optical size ratio of
the \hst\ sample of ESF-ETGs to the size ratio of other massive
galaxies with similar, intermediate specific SFRs (i.e., the
dust-corrected green valley population). The assumption now is that
most green valley galaxies are fading for the first time onto the red
sequence \citep{martin}. Thus, the size ratios similar to theirs would
support the fading star-forming scenario for the \hst\ sample (\S\
\ref{sec:disc}). Therefore, we select the comparison sample to have
the same mass range as the ESF-ETGs ($10.2 < \log M_* < 11.2$) and lie
in the similar redshift range ($0.08<z<0.12$). For the FUV size we
take the FWHM measurement from the \galex\ pipeline, and for the
optical size the 90\% Petrosian flux diameter in $r$\footnote{Note
  that because the FUV and optical size are not measured in the same
  way, a galaxy with a flat color profile need not necessarily have a
  size ratio of exactly 1.}. The results are shown in Figure
\ref{fig:sizerat_ssfr}a, where the size ratio is plotted against the
dust-corrected specific SFR. Blue cloud star-forming galaxies are to
the left of our sample, while the quiescent ones are to the right. Our
sample spans intermediate specific SFRs, as expected from the FUV$-r$
excess selection. Small-scale star forming ETGs (red squares) all tend
to have smaller UV to optical size ratios than ESF-ETGs, and also have
smaller specific SFRs. Unfortunately, at these redshifts \galex\ FUV
size measurements are quite uncertain for any individual galaxy, so
commenting on size ratios of individual galaxies may not be very
revealing. Instead, we check if {\it on average} the ESF-ETGs differ
in size ratio with respect to other galaxies of same specific SFR. To
perform the comparison, we focus on the specific SFR range of
$-11.3<\log ({\rm SFR}/M_*)<-10.6$ and size ratios between 0.1 and
3.3. These cuts leave two ESF-ETGs outside: SR03 with greatly
underestimated optical size (leading to unrealistic formal size ratio
of 4.9) and SR23, an outlier in terms of low specific SFR. Expanding
the cuts to include these objects would bias the size ratio of the
underlying sample towards even lower values. We calculate the
geometric mean of the ratios of ESF-ETGs to be $1.66\pm0.09$. The
corresponding value for the comparison sample is $1.25\pm0.01$. Monte
Carlo simulations of random drawings of 17 galaxies from the
underlying sample show that the mean ESF-ETGs size ratio is expected
to be larger than that of the underlying population by chance with a
probability of $3.8\times 10^{-4}$, corresponding to a 3.4$\sigma$
difference. Therefore, the UV sizes of ESF-ETGs are on average larger
than what is expected for transitioning galaxies with fading SF,
suggesting instead that the gas was accreted from an external
source. This result becomes easier to visualize when we show the
distribution of the underlying population in the conditional form
(normalized in each specific SFR bin) in Figure
\ref{fig:sizerat_ssfr}b. Statistically significant difference in size
ratios persist at the similar level even if we exclude 7 ESF-ETGs for
which we have already established significant disk enhancement from
the analysis in the previous section.

The analysis performed in this section is similar to that presented in
SR2010 using FUV physical sizes alone, but is more robust since it
avoids potential biases by comparing relative sizes and by selecting
the underlying population with the same mass as the \hst\ sample. The
results suggest that ESF-ETGs, while having the same dust-corrected
specific SFR as the more general population of green valley galaxies,
are different from them because their UV emission is distributed more
broadly than their optical emission, pointing to a different fueling
mechanism.

%%%
\section{Incidence of SF in S0\lowercase{s} vs.\ Ellipticals} \label{sec:s0_vs_e}
%%%

The selection criteria for our sample of ETGs were not explicitly
biased in favor of S0s or against true ellipticals, yet we see that
none of the ESF-ETGs is an elliptical. A question therefore arises
whether the galaxy-wide, extended SF is an exclusive S0 phenomenon. To
answer this we go beyond our sample to investigate the incidence of
the occurrence of SF among the general population of SDSS/\galex\
ETGs.

Since the classification of SDSS galaxies into Es vs.\ S0s is challenging
and not readily available in this redshift range, in this section we
resort to the fact that Es and S0s do not have the same distributions
of stellar mass, optical concentration, or axis ratio. We then
determine what fraction of ETGs with a given stellar mass, optical
light concentration, or axis ratio have extended SF, and whether
these trends indicate if the extended SF is primarily an S0 phenomenon
or not.

We define the underlying sample of early-type galaxies by requiring
high optical light concentration ($C>2.5$), red optical colors
($g-r>0.7$, rest-frame) and $z<0.12$, the same redshift cut as the
\hst\ sample. All our \hst\ ESF-ETGs are on the optical red sequence
($g-r>0.7$, Fig.\ 1 of SR2010), which basically means that any SF must
be at a low level (in the sense of low specific SFR, not necessarily
low {\it total} SFR). As in \S\
\ref{ssec:optsize} the underlying sample is not required to have a UV
detection, only to lie in the \galex\ footprint.

To select candidate ESF-ETGs among the ETGs defined above, we impose
two additional requirements: FUV$-r<6$, to select SF, and the
UV-to-optical size ratio $>1.15$ to select extended SF. The FUV$-r$
cut is not as blue as the one used to select the \hst\ sample, where
we required a {\it strong} UV excess of FUV$-r<5.3$. Studies of nearby
galaxies have shown that the {\it old star} UV upturn is not bluer than
FUV$-r=6$ \citep{donas}, so it is justified to use it as a cut to
select SF. The UV-to-optical size cut is simply chosen based on Figure
\ref{fig:sizerat_ssfr}, such that all ESF-ETGs from the \hst\ sample
lie above it. Note that no selection is based on the SDSS fiber
properties.

\begin{figure}
\epsscale{1.2} \plotone{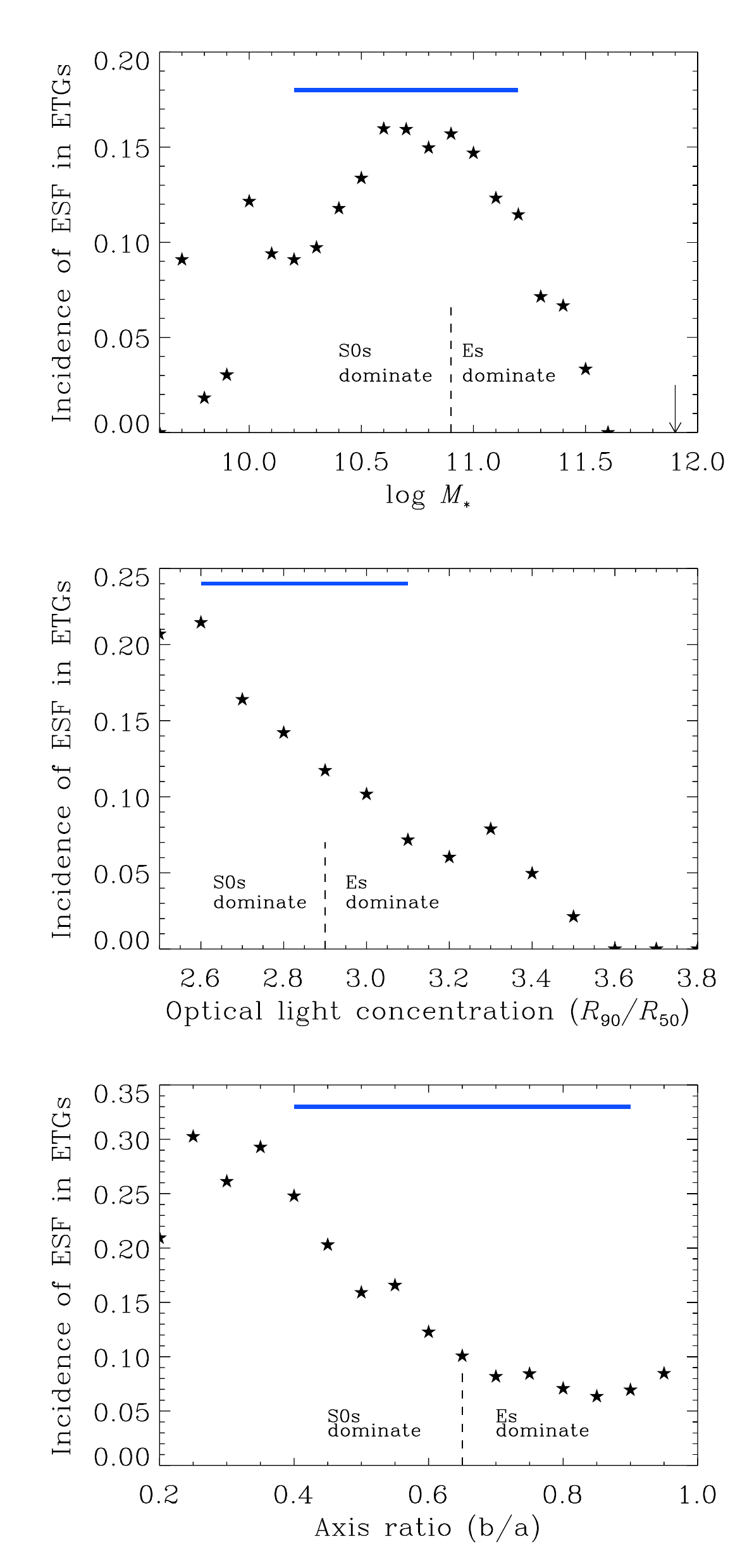}
\caption{Incidence of extended SF (i.e., SF which results in the UV extent
  exceeds the optical size) among ETGs as a function of stellar
  mass, optical light concentration, and optical axis ratio. ETGs are
  selected as concentrated galaxies on the optical red sequence. The
  blue bar represents the range of the \hst\ sample. Dashed lines in
  each panel show values above which ellipticals start to dominate
  over S0s (according to \citealt{cheng_faber,vulcani}). Trends are
  consistent with extended SF being more frequent in less massive,
  less concentrated galaxies that can appear more inclined, i.e., in
  galaxies of S0 type rather than true ellipticals.}
\label{fig:esf_etg}
\end{figure}

Figure \ref{fig:esf_etg} shows the incidence of ETGs (at the given
parameter value) that fulfill the extended SF criteria as a function
stellar mass, concentration and axis ratio. In the upper panel we see
that the incidence of extended SF among the ETGs peaks at 15\% for
masses between $10.6<\log M_*<11.1$. The range of masses of the \hst\
sample (blue bar) encompasses this peak. Above the mass corresponding
to the peak incidence the incidence relatively quickly declines and
reaches zero around $\log M_* \approx 11.6$, i.e., at masses which are
only 3 times higher than the masses at which SF was still at the
peak. While the sample contains ETGs with masses of up to $\log M_*
=11.9$, their numbers are too small to constrain the SF incidence,
i.e., it is consistent with zero. Interestingly, the decline of the SF
incidence starts at the same place at which ellipticals take over S0s
in terms of number densities (i.e., where the two stellar mass
functions cross, \citealt{cheng_faber,vulcani}). This is consistent
with the extended SF being primarily an S0 phenomenon. The drop at
lower masses, where S0s dominate, must be of different origin. One
possibility is that since towards the lower masses the fraction of
galaxies that are satellites increases \citep{mandelbaum}, and since
the satellites usually cannot maintain SF as they fall into a central
halo (e.g., \citealt{bekki}) the incidence of ETGs with extended SF
will decrease. This hypothesis will be tested in future work that will
consider the environments of ESF-ETGs.

Unlike the trend with mass, the incidence of extended SF among ETGs falls
monotonically with increasing concentration, again consistent with
a picture in which mainly S0s support extended SF. For the least
concentrated ETGs the rate is $\approx 20\%$. Finally, the SF
incidence vs.\ minor to major axis ratio also declines as the galaxies
get rounder in projection, but unlike the trend with concentration it
levels off as galaxies get round. This again is consistent with SF in
S0s, since some fraction of them will be seen face on. The fact that
the highest SF fraction ($\sim 25\%$) of any of the trends is reached
at lowest axis ratios, where true Es are absent ($b/a<0.45$,
\citealt{cheng_faber}) again argues for the connection with S0s.

Bolstered by these results, we attempt to derive an overall incidence
rate of extended SF among the ``pure'' S0s and Es. To select samples
with as little overlap as possible between S0s and Es, we take Es to
be ETGs that are more massive ($\log M_*>10.9$), more concentrated
($C>2.9$) and rounder ($b/a>0.65$) than the rest of ETGs, i.e.,
simultaneously using the cuts at which bulge-dominated ETGs start to
outnumber disk-like ETGs \citep{cheng_faber}. Applying these cuts
greatly decreases completeness, but this is not relevant since we are
only interested in the rate of incidence. Analogously, ``pure'' S0s
are selected as simultaneously not fulfilling these three
criteria. The results are very indicative. Massive ellipticals
selected in this way have extended SF in $3.8\%$ of cases, while the
fraction is $21\%$ for S0s. At face value, S0s are almost six times as
likely to exhibit extended SF. However, the more intrinsically rare
some objects are in a given group, the larger the relative
contamination. In this case, larger relative contamination will be for
Es. Since the above cuts produce only 55 ``pure'' Es with extended SF,
one can easily inspect SDSS color images of all of them and look for
non-elliptical interlopers. We find that a number of galaxies selected
as pure Es nevertheless show features incompatible with being an
actual elliptical, such as bars and obvious disks. Interestingly,
detached optical rings are seen in 5 objects (there was only one such
ring in our \hst\ sample). Altogether, it appears that at least 1/2 of
these galaxies are not ellipticals. Thus, the true rate of extended SF
in ellipticals is probably smaller than $2\%$, in stark contrast to
the S0 rate. The $21\%$ incidence rate for S0s is similar to that of
extended UV disks (XUVs) of late type spirals determined by
\citet{thilker07}, possibly suggesting a similar gas supply
mechanism(s).

The main focus of this work is on the phenomenon of the extended,
galaxy-scale SF. A few cases of more compact SF in the \hst\ sample
are probably not typical in the sense that we see them as being mostly
off-center, while in reality nuclear and small-scale SF may be as
common. To evaluate the incidence of such SF we use the above
methodology but ask that the UV to optical size ratio be smaller than
1.15. We find that the overall rates of small-scale SF are similar to
those of extended SF. Small-scale SF becomes more frequent than the
extended SF at the high-mass end ($\log M_*>11.2$), possibly
indicating that ellipticals are more efficient at preventing the
extended SF than the more concentrated central SF. Nevertheless, the
SF incidence reaches zero at the same high mass as in the case of the
extended SF. Other trends are similar to those of the extended SF,
with a possible increase of incidence at very high concentrations. We
find that the fraction of ``pure'' Es, with small-scale SF is $7.4\%$,
while it is $13\%$ for ``pure'' S0s. This is a smaller
disproportionality than in the case of extended SF. However, as in the
case of extended SF, the selection will be sensitive to contamination,
so the estimates provided here are more likely to be upper limits,
especially for ellipticals. Inspecting the SDSS images of ``pure'' Es
with small-scale SF we again see many that actually appear to have
extended SF (optical rings) and there is a large number of
non-elliptical interlopers.

%%%
\section{Discussion} \label{sec:disc}
%%%

The discussion will focus on exploring various mechanisms that can
lead to SF in the ETGs on the optical red sequence, and considering
which of them may be related with the UV morphologies present in our
sample. In general the galaxies on the red optical sequence are
considered ``red and dead'' because they have not have had high global
levels of SF in the recent past, yet, as we have seen, some of them
are active at a low relative level.

\subsection{Mechanisms driving SF in ETGs, and the source of molecular gas}

When considering the mechanisms driving SF in red sequence ETGs, it is
useful to first define two broad possibilities for the provenance of
the star-forming gas: {\it internal} and {\it external} (e.g.,
\citealt{wardleknapp}). Since galaxies in general grow by merging and
accretion, it is important to be precise by what is meant here. We
consider internal gas to be the material that was present in the disk
or the bulge of a galaxy at the time when it (first) arrived on the
{\it optical} red sequence, i.e., when the intense SF (having
high specific SFR) that used to keep it optically blue has ceased. The
internal gas would include whatever gas was left over from this {\it
  original} epoch of star formation, plus any gas subsequently released by the
existing stars (mostly as stellar winds in the evolved stages of
stellar evolution). In contrast, all of the gas that entered the
optical extent of an ETG after it got on the optical red sequence
would be considered external. This includes both the gas acquired from
merging with smaller, gas-rich galaxies and the gas accreted from the
IGM.

\begin{figure*}
\epsscale{1.2} \plotone{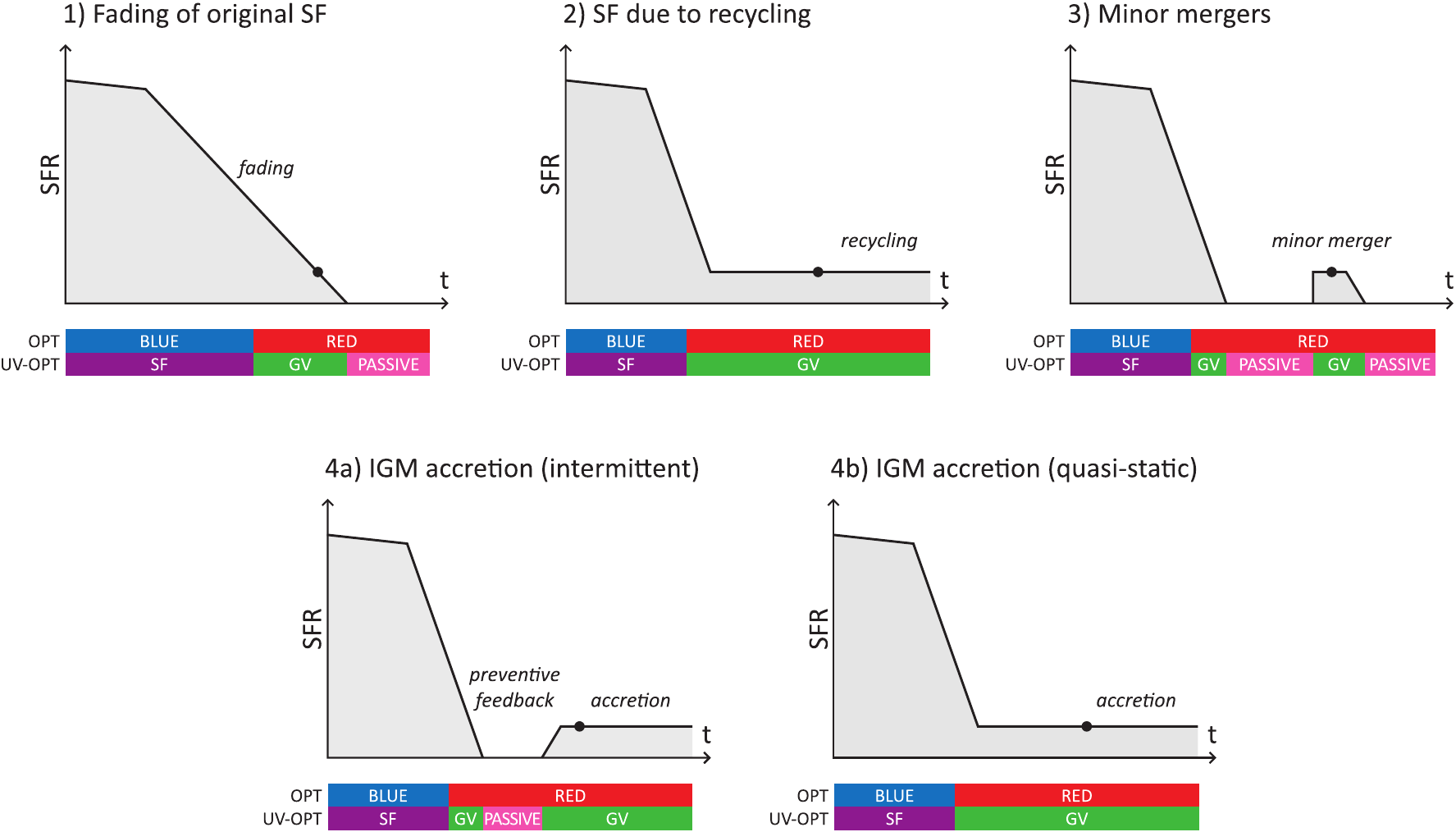}
\caption{Schematic representation of the SF histories associated with
  various mechanisms producing low-level star formation in optical red
  sequence field (and group) galaxies. For each model the dot
  represents the current moment in time. Bottom bars show the optical
  (upper bar) and the UV-optical color of a galaxy. In optical colors
  a galaxy is either in the blue cloud or the red sequence. Blue
  UV-optical colors represent active SF (labeled as SF) and their span
  coincides with optical blue span. However, unlike optical colors,
  UV-optical colors distinguish between low levels of SF (labeled GV
  for ``green valley'') and true quiescence ($\sfr=0$, labeled as
  PASSIVE). A galaxy leaves the blue cloud soon after the quenching
  starts. }
\label{fig:scheme}
\end{figure*}

In terms of the mechanisms driving the SF several scenarios have been
proposed in the literature. While they are not all mutually exclusive,
they are differentiated here to emphasize the dominant process. We
schematically illustrate these mechanisms in Figure
\ref{fig:scheme}. Each panel shows the SF history for the given
mechanism, as well as the optical classification (upper bar) into blue
cloud (BC) and red sequence (RS) and the UV-optical classification
into actively star-forming (SF), green valley (GV), and passive. Blue
cloud and actively star forming phases are coincident, while the
optical red sequence corresponds to either the green valley (some SF)
or the passive phase (no SF). A galaxy is no longer on blue sequence
soon after the commencement of quenching. In the discussion here we
are generally not interested in what causes the SF to start its
decline.

(1) {\it Fading of the original SF}. In this scenario any SF currently
present in red sequence galaxies is simply the final, low-level phase
of once-intense SF in a {\it disk} galaxy. In other words, intense SF
has been absent for some time and consequently the galaxy is now
optically red, but with enough UV from low-level SF to have ``green''
UV-optical color. In the UV-optical color-magnitude diagram (CMD) such
a galaxy would be considered a transitional galaxy on its way to
quiescence \citep{martin}. Note that in the optical CMD this galaxy is
already on the red sequence, but would have arrived there recently
because some SF is still present (see Figure \ref{fig:scheme}). In
this picture the remaining SF would be distributed across the galaxy
disk and possibly show fading spiral arms (similar to field ``red
spirals'' of \citealt{masters_redsp}). Since in this scenario the
spiral galaxy does not need to undergo a morphological transformation,
the reason it is fading can be assumed to be because of the lack of
new IGM accretion (e.g., massive-halo quenching of galaxy groups,
\citealt{dekel_birnboim}), possibly coupled with some other
non-destructive feedback mechanism that removes the gas. The origin of
gas in this model is internal because the galaxy is in the process of
exhausting the gas it had prior to arriving onto the optical red
sequence.

(2) {\it SF due to recycled gas}. In this scenario SF is quenched
primarily due to the lack of IGM accretion, but the quenching does not
proceed all the way to zero. New stars continue to form from the mass
loss of evolved stars. In this mechanism the new SF should mostly
follow the distribution of old stars, and is consequently more often
invoked to explain central SF in ETGs \citep{temi, shapiro}. Also,
unlike scenario (1), one would not necessarily expect to see fading
spiral arms because a morphological transformation into an S0 (large
bulge and featureless disk) could have preceded the quenching. The
main objection to this model is that all ETGs are expected to recycle
gas, yet most do not have either nuclear or widespread SF. This is
even more of an issue considering that the recycling mechanism should
be able to keep supplying the gas to be used in SF for very long
periods of time \citep{kennicutt94,temi,leitner}. The origin of gas in
this model is internal because SF does not require external input to
run its course.

(3) {\it Gas-rich minor mergers}. Minor mergers, being more frequent
than major mergers \citep{maller}, may represent an important way in
which galaxies grow and change in general \citep{bournaud07}. This
process is possibly as prevalent among the ETGs as among the general
galaxy population. Therefore, gas-rich minor mergers (typically
defined as mass ratio of 4:1 or higher) are difficult to dispute as a
mechanism that is taking place among ETGs, which may subsequently lead
to episodic SF in ETGs (Figure \ref{fig:scheme}). The key question is
how dominant minor merging is as the SF mechanism among ETGs.
\citet{kaviraj09} claim that principally the entire population of ETGs
having strong UV excess can be explained in this way. The success of
such explanations to a large extent depends on knowing the timescale
over which minor mergers lead to enhanced SF. A factor of a few
overestimate in timescale can turn this mechanism from dominant to
just one of the contributing mechanisms. One of the main goals of
Paper II is to explore episodic SF from the standpoint of stellar
populations. The origin of gas in this scenario is external.

The emphasis on mergers being {\it minor} stems from the fact that
major mergers would be producing more intense bursts, sufficient to
drive the galaxy all the way back into the blue cloud. Optically blue
ETGs have been identified in several studies based on SDSS data
\citep{kannappan,schawinski09}, and may be precursors to
post-starburst E+A galaxies. Because major mergers would probably
destroy stellar disks, and we see that disks are present in all of our
ESF-ETGs, they do not present a viable mechanism for the current SF,
though they may have happened in the past.

(4) {\it Intergalactic medium (IGM) accretion}. The IGM accretion
scenario assumes that the low-level SF present in some ETGs is fueled
by newly accreted external gas, i.e., that the original molecular gas
is no longer available. This mechanism is especially interesting in
the light of recent work that attempts to establish the importance of
the so-called ``cold accretion'' in supplying the gas to galaxies in
general \citep{keres,dave11}. Note that we use the term
``intergalactic'' without implying the precise provenance of the gas
(which before it enters the galaxy could be circum-galactic, i.e., it
may accumulate beyond the stellar disk before it becomes available for
SF), but rather to indicate a relatively smooth, possibly prolonged
process that stands in contrast to a bursty satellite
accretion\footnote{While both the minor merger accretion and IGM
  accretion represent {\it accretion}, to avoid confusion we will use
  this term in connection with the latter mechanism alone.}. If, as
recent studies indicate, the IGM accretion is the dominant mode by
which galaxies acquire gas, then its regulation determines if and when
a galaxy leaves the blue cloud and becomes optically red in the first
place. This scenario was mostly considered in the context of normal,
actively star forming galaxies (e.g., \citealt{keres2}). It has
recently received more attention in connection with SF in ETGs
(\citealt{kauff07,donovan,cortesehughes,thilker10,oosterloo}, SR2010).

We distinguish two flavors of this scenario, differing in SF history:
intermittent IGM accretion and quasi-static IGM accretion. Both are
illustrated in Figure \ref{fig:scheme}. In the first type the IGM
accretion has resumed after some period of true quiescence. The period
of quiescence may be the result of the same process that led to the
quenching in the first place. That would be the case if the preventive
feedback was temporary or is intermittent. SF proceeds when the
feedback is not active. In the second type the quenching shuts down most
of the SF and removes most of the original gas, but there is no
mechanism in place to prevent new, low-level IGM accretion. Note that
in such a case the galaxy never reaches true passivity, so it can remain
in the green valley for a very long period of time, which is why we
call it quasi-static.

Both minor mergers and IGM accretion have been proposed as mechanisms
leading to XUVs, relating them to inside-out disk building. The
phenomenon was originally described in star-forming spirals
\citep{thilker07} but has recently been extended to ETGs
\citep{lemonias,moffett}, and so becomes relevant in the context of
this study. Furthermore, both minor mergers and accretion might be
associated with external reservoirs of \HI\, such as those found
around some ETGs having ongoing SF (e.g.,
\citealt{donovan,cortesehughes,thilker10}). Thus, the evidence for one
mechanism versus another must rely on the additional information
beyond the presence of neutral gas, such as the UV and optical
morphologies explored here, or the inferred SF histories (Paper II).

In relation to the above processes the term ``rejuvenation'' is
sometimes encountered in the literature
\citep{hawarden,shapiro,thilker10}. If rejuvenation is taken to mean
any recent SF that can lower the mean age of the stellar population
all of the processes except fading would fall into this category. On
the other hand, if rejuvenation is taken to mean the resumption of SF
after it has been completely absent (as in Paper II) only the minor
merging and the resumed/intermittent IGM accretion would qualify under
this term.

So far, we have discussed SF mechanisms without considering the two
types of ETGs (Es and S0s). Fading assumes that SF happens in a disk,
and therefore it is only applicable to S0s or later types. Minor mergers
could in principle operate on either ellipticals or S0s, but the
resulting SF morphologies could be quite different in the two
cases. In the case of spheroids, a merger may lead to SF that
preferentially happens in the centers \citep{peirani} or forms
morphologically very distinct polar ring galaxies \citep{polar}, while
for the disk galaxies significant merger-induced SF may happen in an
existing disk as well \citep{mihos_hernquist,tjcox}. A difference in
the way in which SF is induced in Es and S0s may also exist for IGM
accretion. In S0s gradual accretion could lead to SF in the existing
stellar disk, perhaps in its outskirts. On the other hand, the IGM
accretion onto ellipticals may result in a detached rotating ring, as
in the case of Hoag's galaxy \citep{schweizer,finkelman}, without any
gas cooling onto the spheroid itself.

Finally, note that the fading SF scenario implicitly assumes that S0s
form from spirals via some non-destructive mechanism (e.g.,
strangulation), while the other three scenarios do not imply any
specific mechanism for the original transformation of a progenitor
into an S0 galaxy.

\subsection{SF mechanisms in the \hst\ sample}

With the above review in hand, we now evaluate the SF mechanisms
against the morphological evidence for each UV class described in \S\
\ref{sec:uv}. We start by considering ETGs with extended SF
(ESF-ETGs).

\subsubsection{Regular wide rings}

Galaxies with {\it regular wide rings} represent the most numerous
single class in our sample, and half of all ESF-ETGs. They have fairly
undisturbed morphologies both in the UV and in the optical suggesting
no gas-rich mergers with mass ratios smaller than 10:1 in the last 1
to 2 Gyr \citep{lotz}.  These rings are very different from the
collisional rings and could instead be secular rings where a
non-axisymmetric feature such as a bar leads to the formation of rings
\citep{buta_combes}. Regular morphology argues strongly against
mergers or interactions, which are actually believed to be able to
destroy the outer rings \citep{elmegreen}. A possible exception is
SR18, which shows what looks like an incomplete UV ring coincident
with a compact optical source, and has some extended features along the
major axis of the optical disk. The presence of rings is compatible
with any of the remaining mechanisms (fading, recycling, IGM
accretion). The lack of features in rings and the lack of optical
spiral structure, as is basically the case here (7 S0s and 1 Sa),
argues against the fading of the original SF which would presumably
preserve more structure in either the optical or the UV. Absence of
large-scale structure would then favor either recycling or IGM
accretion.

As recycling should be present in {\it all} S0s, especially in inner
regions, and bars that would facilitate the transport of gas from
inner regions to the co-rotation radius are also common, one would
expect most barred S0s to exhibit a {\it star-forming ring}. We test
this hypothesis by extracting 89 prominently barred ($L_{\rm
bar}\geq0.5$) S0 ($-3\leq T \leq-1$) galaxies from the EFIGI catalog
of 4458 nearby galaxies with detailed morphological information
\citep{efigi}. From these we randomly select 20 that are observed by 
\galex\ and check their UV images. We find that
only two exhibit any signs of extended FUV emission: NGC4643 has a
narrow ring at the co-rotation radius, and NGC5052 has a somewhat
small offset ring. From this quick analysis we conclude that while
recycled gas may be present in the majority of ETGs it alone will
rarely lead to SF even when a driving mechanism such as a strong bar
is present. Since all S0s must be accumulating internal gas in their
disks, evidently there is some sort of ubiquitous ``cleaning''
mechanism that is rendering this gas inactive.  To explain the
ESF-ETGs this way would require that the mechanism be failing in most
S0s.  But why would that happen?  There is no difference in the
properties of these objects that we can see that points to a
difference in inherent cleaning efficiency.  We therefore assume that
cleaning operates in all galaxies and that the presence of excess gas
in the ESF-ETGs must be due to extra gas from other sources, not to a
pile-up from internal sources.  Moreover, this argument applies to all
types of ESF-ETGs, not just those with wide rings, and thus we do not
consider possibility (2) (SFR from recycled gas) any further in the
rest of the analysis.

What evidence is there that the wide UV rings in the \hst\ sample are
the result of IGM accretion? A possible indicator is that in half of
these cases a bar is not seen. This makes it unlikely that some other
global dynamical process can completely remove the {\it internal} gas
from the central regions. In contrast, {\it external} accretion may
not need a bar to settle in a ring, since its natively high angular
momentum would cause it to remain in the outskirts. Alternatively, the
external gas is prevented from reaching the central regions even with
a very weak non-axisymmetric potential, such as a dissolved bar that
would not be visible in our optical images. A feature that may suggest
such scenario is that the UV rings in this class are wide and usually
extend from the inner regions (presumably a co-rotation radius) out to
the very edges, or in some cases, beyond the optical disk. Further
evidence for a mechanism that does not involve a bar is the fact that
in well studied nearby resonance rings in late-type spirals, where
secular processes shuffle the existing disk gas, the co-rotation ring
is narrow and is oftentimes accompanied by an additional, outer
Lindblad resonance ring, which we do not see. In NGC2974 (type S0/a),
the only galaxy from the nearby SAURON sample that has UV rings (also
embedded within the stellar disk), \citet{jeong07} posit the existence
of a bar that leads to ring structures through resonances, although
the bar is not visible in this galaxy. Remarkably, NGC2974 is
surrounded by a ring-shaped reservoir of \ion{H}{1} coincident with
the UV rings \citep{weijmans}, an arrangement that fits the picture
presented here in which a conspicuous bar may not be needed if the
source of gas is external.

If IGM accretion is the process that leads to SF in these galaxies, it
is probably occurring at a rate or time scale that may lead to some
enhancement of the existing disk. Figure \ref{fig:optsize_mass} shows
that most members of this subclass lie above the expected size for
their mass, with the most extreme case by far being SR02, with its
wide, flocculent ring. Such enhancement in size and the overall
regularity of UV structures may point towards a prolonged accretion,
as in the quasi-static scenario described above (see also Paper II).

In summary, IGM accretion appears as the most viable mechanism for
fueling SF in ESF-ETGs with wide rings. We exclude SF due to recycled
gas because it is not seen universally among barred S0s. The
morphologies are too regular to imply gas-rich mergers and too
unstructured to suggest fading spirals.

\subsubsection{Disks with small central holes}

Turning now to the three ESF-ETGs classified as {\it disks with small
  central holes} the same arguments hold against the minor merger
scenario: non-disturbed UV and optical morphologies. Of the remaining
two mechanisms, here the fading of the original SF is perhaps more
likely than IGM accretion. Indeed, these three galaxies with UV
structures reminiscent of tightly wound arms and relatively small
central clearings are the best candidates for the fading scenario in
our sample. Two, perhaps even all three, have optical signatures of
spiral arms. In SR14 a bar is present and it may be responsible for
the central clearing. In the other two cases the clearing is very
small, so an equally small bar cannot be fully ruled out from our
imaging. \galex\ Atlas of Nearby Galaxies \citep{gildepaz07} contains
one galaxy, NGC2841, a flocculent Sb with a classical bulge, that has
a central hole in the UV that is very similar to galaxies in this
class, especially SR14. \citet{youngscoville} have suggested that the
clearing in NGC2841 could be the result of the gas exhaustion in the
region of the bulge. \citet{fisher} finds that SF is often times
suppressed in spirals with classical bulges (see also
\citealt{lacknergunn}) but speculates that it is due to an AGN
feedback.  Classical bulges are believed to be the result of mergers
and share many properties with elliptical galaxies including the
presence of a supermassive black hole \citep{droryfisher}. Classical
bulges could then be associated with holes in the UV without a need
for a bar. Interestingly, NGC2841, like galaxies in our sample, has
low SFR for its mass compared to other spirals, with a curious absence
of HII regions in its disk \citep{crockett}. \citet{droryfisher}
suggest that all galaxies with classical bulge have had their current
disks {\it re-formed} after the merger that formed the bulge and
presumably destroyed an earlier disk. Some evidence that subsequent
disk accretion is taking place may lie in the fact that NGC2841
exhibits an XUV disk \citep{thilker07}. In that case, galaxies with a
classical bulge may represent extreme cases of rejuvenation that
proceeds at a relatively slow pace since the overall optical color of
galaxies with classical bulges is red
\citep{droryfisher,lacknergunn}. Nevertheless, we will consider
objects in our sample having small central holes to exhibit SF from
the original, disk-building phase that is currently fading onto the
red sequence. Such a conclusion is also supported by the fact that the
optical disk sizes of these three galaxies do not exceed what is
expected for their mass, consistent with this being the original SF
episode.

\subsubsection{Narrow rings}

The next category is ESF-ETGs with narrow rings, which are somewhat
less regular than the rings in the previous group. Because of this
varying level of regularity the SF in this group may have
heterogeneous causes. SR05, with its offset outer ring, offset optical
disk and a faint companion, is a possible merger candidate. SR28 may show
evidence of UV arms, so it could be another example of the fading
category. Its normal optical size supports this. SR29 also has a
normal optical size, but interestingly, we again fail to detect a bar
that maintains its ring.

\subsubsection{Irregular rings in extreme disks and Giant LSBs} \label{sssec:glsb}

The two galaxies that we classify as {\it irregular rings in extreme
  disks} (SR01 and SR07) deserve special attention because they have
whopping UV disks ($D\sim 70$ kpc) and their optical sizes by far
exceed what is typical for their mass, which makes them potentially
very rare objects. SR01 and SR07 are very similar to nearby {\it giant
  low surface brightness galaxies} (LSBs), such as Malin 1, Malin 2
and UGC6614 \footnote{Giant LSBs are different from more common {\it
    dwarf} LSBs. The latter are diffuse, isolated dwarf galaxies,
  while the former are massive, have bulges, and feature more regular
  appearances, and it is only their disks that have low surface
  brightness.} . SDSS color composite images of Malin 2 and UGC6614
show dominant bulges. Malin 2 in addition has arcs similar to those
seen in the $R$ image of SR01. Some arcs are optically red, while some
are blue, indicating more intense SF. UGC6614 shows an inner blue
ring, which also marks the inner boundary of its extended disk. In the
UV, \galex\ images of both galaxies show spectacular structures,
especially in the far UV. Both resemble SR01, except that Malin 2 has
a less well-defined inner ring, while the inner ring of UGC6614 is
somewhat more regular than that of SR01.

The formation of giant LSBs has presented a puzzle ever since their
discovery. Originally, giant LSBs were considered to represent spirals
with forming disks \citep{malin1}. Recent work breaks free of trying
to fit all giant LSBs into an `unusual spiral' category. Studies of
Malin 1 noted the presence of a smaller, relatively normal {\it
  early-type} disk \citep{barth} with a normal inner rotation curve
\citep{lelli}. Previously, \citet{quillen} remarked on the extremely
red disk colors of UGC6614 and Malin 2, suggestive of ETG
populations. However, this connection with ETGs may not even hold for
all {\it giant} LSBs. Of the 10 giant LSBs studied in the UV by
\citet{wyder09} we notice that only Malin 1, Malin 2 and UGC6614 have
central NUV$-r$ colors consistent with old populations. Furthermore,
SDSS spectra of these three galaxies reveal high D4000 index values
typical of ETGs and not found among normal spirals (1.91, 1.94 and
1.77)\footnote{From DR7 data processed by the MPA/JHU group.}. While
strong emission lines are present in all three, the line ratios place
them on the AGN branch of the BPT diagram, so no central SF is
indicated. These data and our observations present an alternative
picture in which some giant LSBs are relatively normal S0 galaxies
imbedded in extreme disks, possibly similar to the XUV disks of some
spirals. The lack of optical blue light in SR01 and SR07 with respect
to Malin 2 and UGC6614 may, however, indicate a more mature phase in
the evolution of such ``S0+LSBs''. \citet{donovan} in their study of
the nearby S0 galaxy ESO 381--47 with a star-forming ring note that it
may represent an ETG on its way to becoming a giant
LSB. Interestingly, the far-UV morphology of ESO 381--47 observed by
\galex\ resembles that of SR07, with a main irregular ring surrounded
by more diffuse and filamentary emission outside of it.

The connection between giant LSBs and S0s, or between LSBs and XUVs,
does not by itself resolve the question of the source of star-forming
gas. \citet{noguchi} presents numerical simulations which show that a
normal high surface brightness disk can double its scale length due to
secular processes involving bar formation in an unstable disk. An
alternative model involving galaxy collisions (head-on mergers) was
introduced by \citet{mapelli}. They propose that large disks are the
aftermath of collisions that early on (100-200 Myr after the
collision) appear as classical collisional rings (e.g., the Cartwheel
galaxy) but later (0.5--1 Gyr after the collision) take on the form of
extended disks. In both of these scenarios it is assumed that the gas
is already present in the original, smaller disk. The ``ruffled'' UV
appearance of SR01 and SR07 is strongly suggestive of interactions and
less indicative of secular processes. However, the overall UV
morphology of these two galaxies, as well as other ``S0+LSBs'', does
not appear quite like the model predictions of \citet{mapelli}. Most
notably, there is no evidence of the radial structures (spokes) that
are seen in the simulations. On the other hand, the overall arc-like
structure and the inner irregular rings are features seen in more
conventional simulations of minor mergers involving a disk galaxy and
a companion on a parabolic orbit \citep{tjcox}. These simulations
generally show that minor mergers can lead to SF on galaxy-wide scales
on the first passage, and not just in the centers as they undergo a
final plunge. Perhaps under some circumstances such mergers can even
lead to the very extended SF that are observed here.

\subsubsection{Tidal arms and pseudoarms}

The two galaxies in the UV arm class differ in appearance
and likely have had different formation mechanisms. The arms in SR11 are part
of a pseudoring, which is known from theoretical considerations to arise from a
bar-induced resonance at the outer Lindblad resonance radius
\citep{schwarz_pasa,athanassoula}. The source of gas in these
simulations is assumed to be internal. The presence of a full set of
resonance features, and not just one wide ring as in the {\it regular
wide ring} class, makes it more likely that the source of gas in SR11
is internal as well. There is one caveat to this. What is different
between SR11 and most other nearby galaxies with pseudorings is that
the latter are usually found in gas-rich late-type spirals
\citep{buta_combes}. SR11, on the other hand, is dominated by old
stellar populations, like the rest of our sample.

Of all the galaxies in our sample SR17 most clearly shows UV and
optical signatures indicative of an interaction, with asymmetric arms
that appear tidal in origin. The companion with which it is likely
interacting is the galaxy to the upper right in the optical image
(Fig.\ \ref{fig:panel6}). It is an S0 galaxy with no extended
SF. Therefore, while the gas in SR17 is probably stirred to SF by the
interaction, it does not seem to originate from the companion, but
could instead be internal.

To summarize, both galaxies in this category may best fit the
fading scenario.

\subsubsection{Summary of SF mechanisms in ETGs with extended SF}

Overall, the mechanisms that drive SF in ETGs with extended UV
structures appear to be diverse. ETGs with regular wide rings probably
form stars from the gas accreted relatively smoothly from the IGM,
possibly in a prolonged quasi-static way. In ETGs with central holes
the gas may be from the original blue-cloud SF. ETGs in the narrow
rings category may be acquiring their gas through various mechanisms
including minor mergers. The merger scenario is perhaps more likely
than accretion also for ETGs having irregular rings and large
disks. Finally, the gas in the two ETGs with UV arms could again come
from the original, internal gas supply that is not yet fully
quenched. A tentative tally based on the analysis of each UV
morphology group suggests that IGM accretion is the leading cause,
being the likely mechanism in 55\% of the ESF-ETG sample. As to the
form of IGM accretion, we cannot exclude the quasi-static in which
some accretion has been present ever since the major phase of SF
ended, i.e., these galaxies may never have been entirely passive. In
some 25\% of the ESF-ETGs the UV and optical morphologies and optical
sizes are consistent with the fading of the original SF. In the
remaining 20\%, non-axisymmetric UV structures could result from the
gas supplied by minor mergers. The relative scarcity of merger-induced
SF is at odds with the findings of \citet{kaviraj09} who use numerical
simulations to explain the entire population of star-forming ETGs as
resulting from minor mergers. We emphasize that our results apply to
extended SF with little or no central SF, while minor merging would
preferentially lead to nuclear SF. Therefore the fraction of ETGs with
SF that is due to mergers must be larger than the 20\% that we
find. In any case it is impossible for all SF to be explained in this
way. We suggest that the result of \citet{kaviraj09} is critically
sensitive to the timescale over which a minor merger leads to enhanced
SFR. If the actual timescales are shorter by a factor of few than what
their specific simulations predict ($\sim 2$ Gyr), the merger
mechanism would drop in significance from the dominant one to being a
contributing mechanism.

Our results indicate that an external origin of gas is more common in
our sample of ETGs with extended SF than an internal origin. However,
we have to take into account that our selection misses 1/3 of the ESF-ETG
population with stronger central emission. If we assume that all of
them are fading disks, the shares of ESF-ETGs with internal and
external source of gas would become equal. 

The fraction of ESF-ETGs that undergo fading of the original SF as
opposed to subsequent ``rejuvenation'' could also be determined by
measuring their neutral H gas content. Detection of \HI\ reservoirs
around 2/3 of nearby field ETGs makes the external origin of gas a
plausible scenario \citep{oosterloo}. Further information is obtained
by studying \HI\ fractions. Fading galaxies would more likely be \HI\
deficient (low specific \HI\ mass, i.e., \HI\ fraction). In contrast,
rejuvenated ETGs (due to either the IGM accretion or merging) would
have higher \HI\ fractions, more similar to those of the blue cloud
galaxies. A result pointing in the same direction as ours, that the
green valley is being comprised of two distinct populations, was
obtained by \citet{cortesehughes} who showed that the nearby galaxies
with normal \HI\ content for their mass, yet in the green valley or on
the UV-optical red sequence, frequently (6 out of 12 cases) have UV
rings. They also show that part of this dichotomy is driven by
environment--\HI-deficient, and thus presumably fading, galaxies favor
clusters and groups. As to the form of accretion (mergers vs.\
smooth), \citet{wang_kauff} using the GASS survey \citep{catinella}
find no correlation between the \HI\ fraction and the level of optical
asymmetry, thus favoring smooth IGM accretion over satellite
accretion.

\subsubsection{Small-scale star forming ETGs}

For six ETGs with small-scale star forming regions, the situation is
more straightforward. There the SF proceeds in smaller patches, thus
excluding both fading and IGM accretion. If, as our estimate confirms,
most of these patches are physically associated with the main optical
source, then we are probably witnessing minor gas-rich mergers in
various stages of coalescence. SR27 may be an early phase, with
its companion being quite undisturbed\footnote{Unless it is a chance
  superposition, see \S\ \ref{ssec:sssf}.}. The companion looks more
distorted in the UV in the case of SR30, and even more so in SR19. In
SR13 the coalescence is almost complete. Finally, SR15 and 25 show
central UV emission. Given their low mass, SR15 and SR25 could
represent the local analogs of the high-redshift ``blue spheroidals'' of
\citet{im_faber}.

Because the optical images of galaxies in this class do not show much
evidence of interaction, but simply look like superpositions of the
main galaxy and the companion, these must be very minor mergers (e.g.,
with mass ratios greater than 10:1,
\citealt{bournaud07}). Recall that our sample is selected against
significant central SF, so a more complete exploration of the
mechanisms of small-scale SF would be better addressed using different
selections.

\subsection{Is extended SF in ETGs an S0 phenomenon?}
\label{ssec:s0phen}

None of the 19 galaxies with extended SF in our sample is an
elliptical galaxy, and the analysis in \S\ \ref{sec:s0_vs_e} shows
that the frequency of ETGs with extended UV excess follows trends that
are consistent with it being a phenomenon that occurs only in
S0s. Extended SF that are confidently known to be ellipticals appears
to be rare even among the well studied nearby galaxies. Curiously, we
find only two such cases in the literature. One is Hoag's galaxy (type
E0) with its detached star-forming ring for which \citet{finkelman}
recently argued for a formation through cold accretion. Another is an
HI-rich NGC5173 (type E1) with clumpy SF regions $\sim 4$--10 kpc from
the center, possibly resulting from a merger \citep{vader}. NGC5173 is
one of only two ellipticals with confident CO detection among the
volume complete sample of 56 ellipticals within 42 Mpc
\citep{young11}. There is no evidence for SF in the \galex\ FUV or NUV
images of the other elliptical with CO detection (NGC2768,
\citealt{jeong09}). We tentatively show in \S\ \ref{sec:s0_vs_e} that
even the central SF, expected from mergers, is also rarer in Es than
in S0s. These results are corroborated by the wealth of information on
the stellar populations in local ETGs that show that ellipticals do
not have young populations (e.g., \citealt{kuntschner,combes,temi}).

To understand why SF, and particularly {\it extended} SF, may be an
exclusive S0 phenomenon in the context of external source of
star-forming gas, it is important to understand why ETGs (particularly
S0s) stop (or greatly reduce) forming stars in the first place. This
question is the focus of many recent studies that try to explain the
galaxy color bimodality and mechanisms behind it. An emerging
consensus is that some feedback mechanism is required to quench the SF
and bring the galaxy onto the red sequence, and that possibly {\it
  another} mechanism is required to {\it keep} the galaxy there, i.e.,
to maintain its quiescence by preventing the infall and/or cooling of
fresh gas. \citet{croton1} suggest that a weak, radio-type AGN can
serve this latter purpose. The need for such non-environmental, {\it
  maintenance} feedback is exacerbated by the existence of isolated
quiescent galaxies \citep{croton2}. One possibility then why
ellipticals maintain their quiescence while many S0s do not could be
that the former have more effective preventive feedback
mechanisms\footnote{Alternatively, in the fading scenario, some S0s
  have SF simply because they have not yet attained
  quiescence. Ellipticals, being typically older, would have acquired
  quiescence earlier in the cosmic history.}. Considering that radio
AGN are primarily found in massive ellipticals this may not be
surprising. Another, possibly related reason why ellipticals may be
better at preventing SF,+ is that many of them, especially the so
called ``slow'' rotators, have hot X-ray halos (whatever the heating
mechanism is), which may prevent gas from cooling and forming stars
\citep{nipoti,sarzi07,kormendy}.

A classification of ETGs into fast and slow rotators
\citep{emsellem07,kormendy} was recently suggested as more fundamental
than the S0 vs.\ E classification \citep{emsellem}. Considering that
both the radio AGN and the X-ray gas are prevalent in slow rotators it
would be natural to expect the presence or absence of SF to follow
this kinematical division rather than the S0 vs.\ E division. The
kinematic split appears to be corroborated by some recent results. For
example, \citet{shapiro} find that central SF (based on 8 $\mu$m PAH
emission) is found only in fast rotators among the nearby SAURON
galaxies. Similar absence of SF signatures (but in CO emission) among
the slow rotators is seen for the larger ATLAS$^{\rm 3D}$ sample
\citep{young11}. Note, however, that basically all (94\%) of S0s are
fast rotators, as well as most (66\%) Es \citep{cappellari}, yet the
CO detection rate is consistent with zero for {\it all} ellipticals in
ATLAS$^{\rm 3D}$, whether fast or slow.  \citet{shapiro} seemingly
find two SAURON galaxies with PAH emission classified as ellipticals,
but NGC2974 is certainly not an E4 (RC3 classification), but rather an
S0/a \citep{naim95,buta10}, while NGC5845 is classified as an
``uncertain E'' because it contains a small stellar disk
\citep{ebneter,kormendy05}. Altogether, studies of nearby ETGs offer
only sporadic evidence for star formation in either slow or
fast-rotating ellipticals, suggesting that SF follows the traditional
E vs.\ S0 divide.

In our study we see ample evidence for E vs.\ S0 dichotomy when it comes
to {\it extended} SF. The situation is less clear for small-scale SF,
which may be present in some (perhaps less massive) ellipticals. In
any case we believe that it is premature to abandon the classical
distinction between Es and S0s in favor of the ``new'', kinematical
one. In other words, calling fast rotating Es ``misclassified
lenticulars'' \citep{cappellari} may not be fully justified since
ellipticals, whether fast or slow, do not show evidence of extended
SF, while many S0s of similar mass do.

The reason why some {\it field} S0s experience SF on the optical red
sequence and others do not may be related to the absence, or the
periodic nature of the maintenance feedback mechanisms (e.g., the AGN
activity). In the first case the SF was quenched during the
transformation of a spiral into an S0 which involves the buildup of a
bulge (e.g., ``morphological quenching'', \citealt{martig}), but no
mechanism was in place to make the quiescence permanent. The second
possibility (periodic preventive feedback) may be the result of
intrinsically intermittent AGN activity in either Es or S0s.
\citet{gabor} showed that periodic AGN feedback provides better
agreement between cosmological simulations and observations. While the
periodic AGN activity would lead to periodic prevention of SFs in S0s,
the ``accumulation'' of AGN energy in X-ray gas in ellipticals could
render the SF prevention permanent \citep{kormendy}.

One could argue that any elliptical that has experienced prolonged
extended SF subsequent to its morphological transformation would
re-grow a disk (e.g., \citealt{governato,delucia}) and consequently be
classified as an S0 (or later type) by definition. Whether such a
scenario actually happens is by no means obvious because we do not
seem to observe cases where the process of disk reformation is
starting in or around ellipticals. Also, such a scenario would be
conceivable only for fast-rotating ellipticals, as there are almost no
slow-rotating S0s. Second, since there are no S0s as massive as some
ellipticals, this E to S0 transformation would still be restricted
only to less massive ellipticals (i.e., on the less massive side of
the 'E--E' dichotomy \citealt{kormendy}).

The above discussion on the prevention of SF in Es as opposed to some
S0s makes sense only if we assume that a galaxy is not completely
detached from the supply of external gas. In the case of fading
scenario such detachment may be what is causing the galaxy to be
fading in the first place, so considerations of preventive feedback
are not needed.

%%%
\section{Conclusions} \label{sec:conclusions}
%%%

This study explores the morphology of $z\sim0.1$ optical red sequence
galaxies (primarily early-type) with UV-detected star formation. ETGs
were selected to have strong UV excess yet weak central ionized
emission. Such selection encompasses the large majority of ETGs with
extended, galaxy-scale SF. Here are the main findings:

\begin{enumerate}

\item There are two modes of SF in our sample. The dominant one is
  extended on scales similar to or larger than the optical extent of a
  galaxy (19 of 25, or 76\%).  The secondary mode is one in which SF
  is concentrated in regions smaller than the optical extent.  Such
  small-scale SF may represent cases in which the ETG assimilates a
  gas-rich dwarf. The remaining conclusions refer only to 19 ETGs with
  extended SF (ESF-ETGs).

\item None of the ESF-ETGs is optically classified as a true
  elliptical galaxy. All show the presence of stellar disks, usually
  of S0 type.  Based on the analysis of the general SDSS/\galex\
  sample (for which Hubble types are not available) the incidence of
  extended SF is highest among the ETGs with $10.6<\log M_*<11.1$ and
  declines at both higher and lower masses; it also declines in ETGs
  with higher optical concentrations. These trends independently
  suggest that the extended SF is primarily a phenomenon of central
  (non-satellite) S0 galaxies. Tentatively, extended SF is estimated
  to be present in no more than 2\% of massive ellipticals (with
  fraction closer to zero not being excluded), as opposed to $\sim
  20$\% for S0s. The latter is similar to the incidence rate of XUV
  disks in spirals, hinting at similar fueling mechanism(s).

\item In all but one case the SF in ESF-ETGs takes place in UV rings
  with diameters of tens of kpc. The star-forming rings are not
  conspicuous in the optical (except one) and less than a half of
  hosts have an optical stellar bar. The latter indicates that the UV
  rings are not necessarily maintained by bar resonances. The
  dominance of UV rings is partially due to our selection
  criteria. However, even if we allow all ETGs with extended SF not
  covered by our selection to have non-ring morphology (e.g.,
  in-filled disks), the UV ring incidence would still dominate
  ($\sim 2/3$). The morphology of UV rings varies and may indicate
  different sources of star forming gas or modes of SF.

\item The morphological analysis suggests that the recent or ongoing
  IGM accretion is the likeliest dominant mechanism for the source of
  gas in ESF-ETGs having wide UV rings, while the fading of the
  original SF (the latest stage of quenching) may be responsible for
  SF in disks with central UV holes. The latter also show signatures of
  optically fading spiral structure. Galaxy interactions may fuel SF
  in several other cases, including the two extreme disks (diameters
  $\sim 70$ kpc), but in general the disturbances are not visible in
  the optical but only in the UV, suggesting at most very minor
  mergers. The ESF-ETGs with extreme disks resemble the giant LSB
  galaxies Malin 2 and UGC6614 and together with them may represent a
  distinct class of giant LSBs, which we call S0+LSBs.

\item Analysis of disk sizes shows that in roughly half of the
  ESF-ETGs the optical disks are $\sim 50\%$ larger than in the
  underlying population of quiescent ETGs of the same mass, suggesting
  that significant disk build-up occurred in ESF-ETGs after the galaxy
  concluded the original epoch of SF.  Such disk enhancement is
  consistent with long-lasting and relatively smooth accretion from
  the IGM and not the more recent merger-supplied gas. ESF-ETGs,
  whether their disks have been enhanced in the optical or not, on
  average have a larger ratio of UV to optical size than the
  comparable green valley galaxies, suggesting that the source of gas
  is more likely to be external (assuming that most green valley
  galaxies are fading onto the red sequence due quenching that cuts
  off the infalling gas).

\item Altogether the IGM accretion may be responsible for SF in 55\%
  of the sample, followed by fading of the original SF in 25\% and
  minor mergers in the remaining 20\%. External origin of gas (from
  accretion and minor mergers) thus together accounts for 3/4 of the
  sample. The IGM accretion in these galaxies may have been present
  ever since the major SF ended, i.e., these galaxies may never have
  been truly passive but are instead suspended in the green valley
  (quasi-static IGM accretion). This is consistent with the fact that
  the plane of new SF coincides with the old disk.

\end{enumerate}

The results at which we arrive here point to a range of mechanisms
driving SF in ETGs. Keeping in mind that our sample does not include
filled-in disks, we find that the extended SF in most (but not all)
cases looks like it is the result of a gradual, non-merger
process. This is in general agreement with the results presented in
Paper II, where it is shown that UV and optical colors do not favor
bursty SF for most of our sample. Our conclusions primarily hold for
extended SF. Small-scale central or circumnuclear SF, which our selection
disfavors but could be as frequent as the extended SF, might instead
preferentially be the result of minor mergers.

The widespread occurrence of galaxy-scale SF in ETGs reported in this
paper may appear to contrast the results from the SAURON sample, in
which the extended SF is reported for only two out of 48 galaxies
(NGC2974 and NGC2685). The primary reason behind this is in SAURON
sample selection--which is 1/2 S0s and 1/2 ellipticals by design, with
a disproportionate fraction being cluster galaxies. When taken into
account that the extended SF is common only in {\it field} S0s, the
small numbers of extended SF in the SAURON sample are not
surprising. The incidence of galaxy-wide SF will certainly be higher
in the ATLAS$^{\rm 3D}$ sample, which includes all ETGs within 42 Mpc,
but even then it will be the UV, rather than the emission line maps,
that will have the requisite surface brightness sensitivity to detect
low levels of SF.

The absence of true ellipticals from our sample of ESF-ETGs is
intriguing and underlines the importance of distinguishing between
ellipticals and S0s, especially when considering SF. Results from
other studies seem to support this: ellipticals of intermediate and
high mass, whether rotating fast or slow, do not show strong evidence
of either extended or central current SF. The traditional view of all
non-dwarf ellipticals as being ``red and dead'' is therefore basically
upheld even in field populations. In contrast, many (perhaps even the
majority, if the process is episodic) field S0s maintain or reacquire
some level of SF activity.

\acknowledgments S.~S.\ wishes to thank Ronald Buta, T.~J.\ Cox,
Alister Graham, Janice C.\ Lee and Antonietta Marino for useful
discussions and comments, and Aaron Barth for help regarding his
excellent ATV.pro image display tool.  S.~S.\ also acknowledges
observing assistance provided by WIYN/KPNO staff and Steven
Janowiecki. Special thanks to Karim Salim for producing Figure
\ref{fig:scheme}. J.~J.~F. acknowledges financial support from a UCSC
Regent's Fellowship and HST grant GO-11175. Based on observations made
with the NASA/ESA Hubble Space Telescope under program GO-11158.

%%%
\appendix
\section{Sample selection and the population of ETGs with extended SF}
%%%

As discussed in \S\ \ref{sec:sample} the original selection criteria
to obtain the \hst\ sample were geared towards ETGs with no evidence
for SF based on spectroscopic criteria. In contrast, the sample
exhibits widespread SF. This extended SF is the focus of the present
work. However, since the original selection criteria were not
optimized to select a full population of ETGs with extended SF, it may
have inadvertently excluded those that also have significant SF in the
central ($\sim 5$ kpc) regions. Therefore, it is important to
establish what fraction of ETGs with extended SF are {\it not}
centrally quiescent and therefore not probed by our sample. Such
galaxies may have significantly different morphology than the galaxies
we selected which could affect the tally of different star-forming
processes that we explore.

\begin{figure}
\epsscale{1.2} \plotone{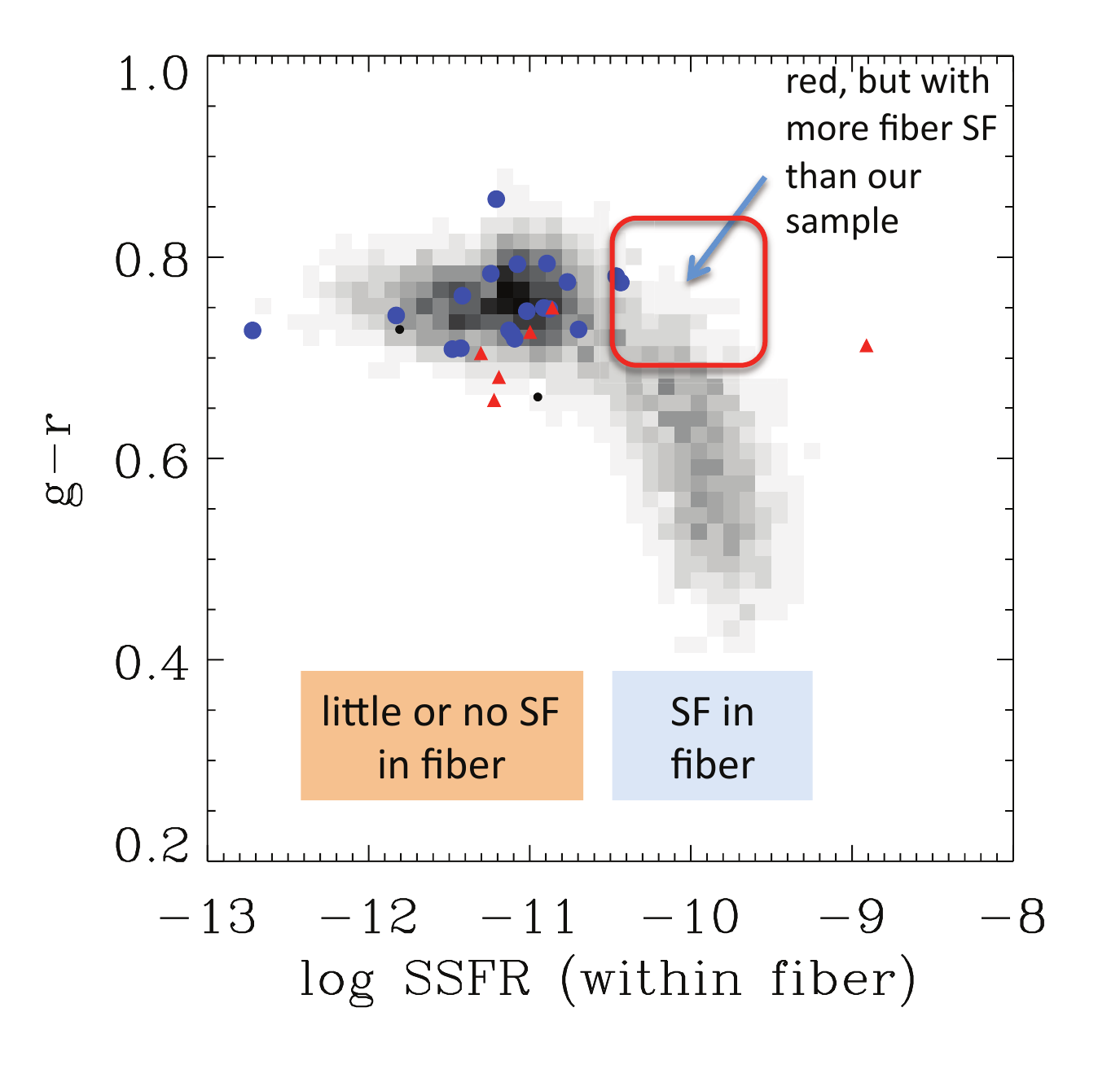}
\caption{Relation between the global optical color ($g-r$,
  K-corrected) and the star formation activity within the SDSS fiber
  that samples the inner $\sim 5$ kpc region of a galaxy. Fiber
  specific SFRs are obtained from SDSS DR4 MPA/JHU
  catalogs. Underlying population (grayscale) has the redshift and
  mass range of the \hst\ sample (symbols) and omits edge-on
  (reddened) galaxies. Galaxies with low SF in fiber also tend to be
  globally red, but some red sequence galaxies can have higher fiber
  specific SFR than probed by our sample (red rectangle).}
\label{fig:gr_fibssfr}
\end{figure}

To probe red sequence ESF-ETGs in a completely unbiased way we would
need to replace the spectroscopic quiescence criterion (number 5 in
\S\ \ref{sec:sample}) with a criterion that ensures integrated
(global) red optical colors and a criterion that selects only extended
SF. Optical red color did not feature as the original selection
criterion and yet the sample galaxies were all red. The reason behind
this can be seen in Figure \ref{fig:gr_fibssfr} where we plot the
global optical color $g-r$ vs.\ the specific SFR within the SDSS fiber
(3$"$). The underlying population (shown as grayscale) are galaxies
with the mass and redshift range of the \hst\ sample ($10.2<\log
M_*<11.2$, $0.08<z<0.12$), with the edge-on ($b/a>0.5$) presumably
dusty disks removed. We see that galaxies with little or no SF within
the fiber ($\log \sfr/M_*\lesssim -10.5$) are invariably red ($g-r>0.7$),
which explains why our \hst\ sample (symbols) and especially the ETGs
with extended SF (blue dots) are on the red sequence). Optically blue
galaxies also tend to have higher specific SFRs within the fiber, but
there do exist some galaxies which are on the red sequence, but
whose fiber SF is higher that the low levels allowed by our original
spectroscopic criterion.

\begin{figure}
\epsscale{1.2} \plotone{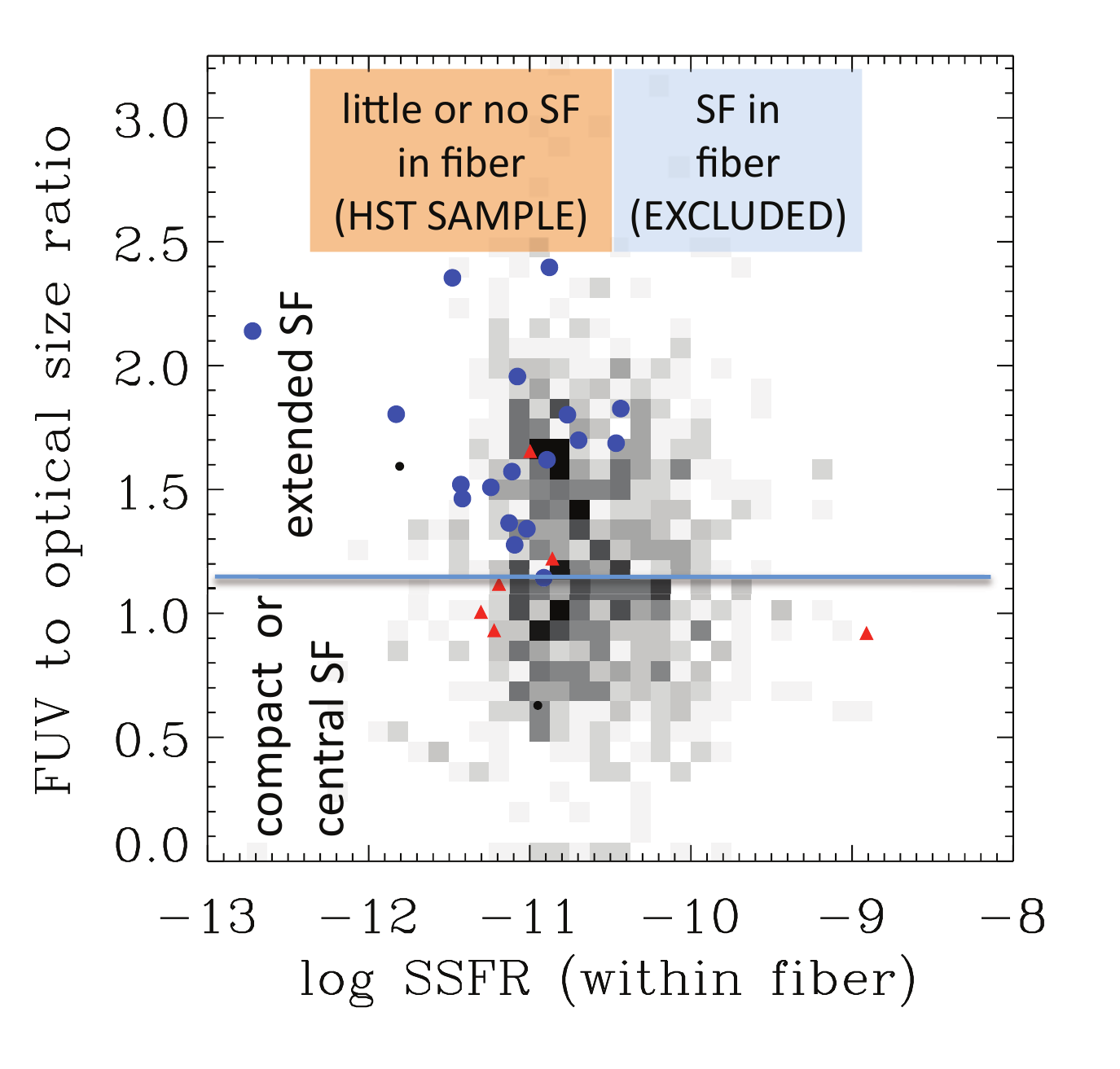}
\caption{The level of the extendedness of SF (the FUV to optical size
  ratio) against the fiber specific SFR. ETGs with extended SF tend to
  be above size ratio of 1.15 (horizontal line). The \hst\ sample
  probes extended SF in ETGs with lower fiber specific SFR, which
  represents 2/3 of the population.}
\label{fig:sizerat_fibssfr}
\end{figure}

To see what fraction of red sequence ETGs have high fiber SF {\it and}
are extended in the UV we show UV-to-optical size ratio against the
specific SFR in the fiber, but now only of red ($g-r>0.7$) ETGs
($C>2.5$), and again without highly inclined systems (Figure
\ref{fig:sizerat_fibssfr}). The information on UV size comes from
\galex\ and is therefore not very precise, but as discussed in \S\
\ref{sec:uv_opt} it allows selection of ETGs with extended SF. The
\hst\ sample of ESF-ETGs (blue dots) has size ratios $>1.15$ and fiber
specific SFRs of $\log \sfr/M_*< -10.5$. There are 296 underlying ETGs
with characteristics like that and therefore, likely similar
morphologies as the \hst\ sample. On the other hand, there are 157
ETGs which are also extended (size ratio $>1.15$) but with higher
fiber specific SFRs. This represents $35\%$ of all ESF-ETGs. If all of
these ESF-ETGs had very different morphologies from centrally
quiescent ESF-ETGs, then certain morphologies (e.g., infilled disks)
would be underrepresented in our sample in favor of others (e.g.,
rings), which may affect the relative importance of the various SF
mechanisms that we discuss. 

To summarize, at least 2/3 of ETGs with extended SF likely have
similar UV morphologies to our ESF-ETGs, while only up to 1/3 may be
different. Thus our conclusions are relevant for the majority of the
ETG population with ESF.

\clearpage

\LongTables

%\begin{landscape}
\begin{deluxetable}{l l r r r r r r r l l l r} 
%\rotate
 \tablecaption{Properties of the \hst\ sample of ETGs with strong UV excess \label{table}}
%\scriptsize
\tiny
\tablewidth{0pt}
\tablenum{1}
 \tablehead{
   \colhead{Name}      &
   \colhead{IAU Designation}      &
   \colhead{RA(J2000)} &
   \colhead{Decl.(J2000)} &
   \colhead{Redshift} &
   \colhead{$\log M_*$} &
   \colhead{log SFR} &
   \colhead{$t_{\rm exp}$ ($R$)} &
   \colhead{$t_{\rm exp}$ (NB)} &
   \colhead{Hubble type} &
   \colhead{UV class} &
   \colhead{UV subclass} &
   \colhead{UV size} \\
   \colhead{} &
   \colhead{} &
   \colhead{(deg)} &
   \colhead{(deg)} &
   \colhead{} &
   \colhead{} &
   \colhead{} &
   \colhead{(sec)}  &
   \colhead{(sec)} &
   \colhead{} &
   \colhead{} &
   \colhead{} &
   \colhead{(kpc)} \\
   }
\startdata
SR01  &SDSS J204524.60$-$044634.2&311.352510& $-$4.776174&
0.119&10.99$\pm$  0.09&  0.10$\pm$  0.50&      3000&      2400$^*$&S0/a
&Extended SF&Irreg.\ ring w/ extreme disk&    75\\
SR02  &SDSS J231557.10$-$000848.8&348.987945& $-$0.146896&  0.081&10.19$\pm$  0.06& $-$0.58$\pm$  0.04&      2400&      2400$^\dagger$&SBa      &Extended SF&Regular wide ring     &    35\\
SR03  &SDSS J162021.48+491609.0&245.089529& 49.269172&  0.057&10.55$\pm$  0.04& $-$0.78$\pm$  0.08&          &          &SB0      &Extended SF&Regular wide ring     &    30\\
SR04  &SDSS J084612.82+522101.4&131.553432& 52.350410&  0.114&10.65$\pm$  0.06& $-$0.48$\pm$  0.14&          &      1200&S0       &Extended SF&Narrow ring(s)        &    31\\
SR05  &SDSS J080804.34+501256.0&122.018109& 50.215561&  0.101&10.58$\pm$  0.07& $-$0.09$\pm$  0.19&       600&       900&SB0      &Extended SF&Narrow ring(s)        &    26\\
SR06  &SDSS J104716.72$-$004737.9&161.819701& $-$0.793885&  0.095&10.78$\pm$  0.07& $-$0.27$\pm$  0.18&       600&      1200&SB0      &Extended SF&Regular wide ring     &    37\\
SR07  &SDSS J001156.72$-$104934.7&  2.986350&$-$10.826323&
0.107&10.73$\pm$  0.07& $-$0.43$\pm$  0.21&      1800&          &S0
&Extended SF&Irreg.\ ring w/ extreme disk&    59\\
SR08  &SDSS J021946.93$-$090521.1& 34.945579& $-$9.089211&  0.111&10.75$\pm$  0.07& $-$0.48$\pm$  0.17&      1200&      2400&S0       &Extended SF&Regular wide ring     &    27\\
SR09  &SDSS J230619.97$-$094447.0&346.583222& $-$9.746410&  0.092&10.38$\pm$  0.08& $-$0.60$\pm$  0.31&      1800&          &SB0      &Extended SF&Regular wide ring     &    25\\
SR10  &SDSS J024041.36$-$084947.4& 40.172352& $-$8.829841&  0.111&11.22$\pm$  0.06&  0.09$\pm$  0.17&       600&          &S0       &Extended SF&Disk w/ hole          &    42\\
SR11  &SDSS J140216.66+025902.7&210.569453&  2.984093&  0.077&10.66$\pm$  0.09&  0.05$\pm$  0.21&       600&      2200$^\dagger$&SBab     &Extended SF&Arms (pseudoring)     &    34\\
SR12  &SDSS J014214.66$-$091103.6& 25.561091& $-$9.184346&  0.114&11.12$\pm$  0.07&  0.36$\pm$  0.18&      1800&      3600$^*$&Sa       &Extended SF&Disk w/ hole          &    46\\
SR13  &SDSS J160830.21+534107.9&242.125866& 53.685539&  0.107&10.76$\pm$  0.06& $-$0.10$\pm$  0.19&      1800&          &E/S0     &Small-scale SF&Off-center            &     9\\
SR14  &SDSS J161646.04+010255.4&244.191861&  1.048746&  0.085&11.00$\pm$  0.08& $-$0.07$\pm$  0.17&      1800&      2400$^*$&SB0/a    &Extended SF&Disk w/ hole          &    34\\
SR15  &SDSS J101807.52$-$004857.7&154.531362& $-$0.816040&  0.033& 9.49$\pm$  0.09& $-$1.58$\pm$  0.37&          &          &E        &Small-scale SF&Central               &     5\\
SR16  &SDSS J041246.35$-$055129.4& 63.193153& $-$5.858179&  0.100&10.82$\pm$  0.08& $-$0.54$\pm$  0.46&          &          &S0 (pec) &Unresolved    &                      &\\
SR17  &SDSS J225131.65$-$102218.9&342.881890&$-$10.371922&  0.118&11.00$\pm$  0.07& $-$0.10$\pm$  0.19&      2400&      2400$^\dagger$&Sab      &Extended SF&Arms (tidal)          &    38\\
SR18  &SDSS J161041.32+512600.4&242.672188& 51.433444&  0.116&10.74$\pm$  0.07& $-$0.04$\pm$  0.56&      1200&          &S0       &Extended SF&Regular wide ring     &    34\\
SR19  &SDSS J002950.16$-$090833.5&  7.459033& $-$9.142660&  0.118&10.85$\pm$  0.07& $-$0.47$\pm$  0.20&          &          &E/S0     &Small-scale SF&Off-center            &    11\\
SR20  &SDSS J084424.13+520122.0&131.100575& 52.022785&  0.114&10.92$\pm$  0.07& $-$0.04$\pm$  0.22&          &          &S0       &Extended SF&Regular wide ring     &    32\\
SR21  &SDSS J133743.93+030646.8&204.433061&  3.113001&  0.112&10.57$\pm$  0.07& $-$0.03$\pm$  0.22&          &          &E        &Unresolved    &                      &\\
SR22  &SDSS J041105.03$-$053412.8& 62.770969& $-$5.570247&  0.066&10.56$\pm$  0.07& $-$2.46$\pm$  0.65&          &          &E        &No UV excess&                      &\\
SR23  &SDSS J085208.61+552847.9&133.035874& 55.480000&  0.114&11.10$\pm$  0.07& $-$0.41$\pm$  0.19&          &          &S0       &Extended SF&Regular wide ring     &    34\\
SR25  &SDSS J084949.45+570713.7&132.456064& 57.120499&  0.042& 9.65$\pm$  0.08& $-$2.08$\pm$  0.69&          &          &E        &Small-scale SF&Central               &     4\\
SR26  &SDSS J030415.99$-$084057.1& 46.066662& $-$8.682551&  0.117&10.87$\pm$  0.07& $-$0.44$\pm$  0.27&          &          &E/S0     &No UV excess&                      &    35\\
SR27  &SDSS J140540.19+034710.9&211.417487&  3.786375&  0.092&10.36$\pm$  0.08& $-$1.50$\pm$  0.63&       600&      2400&E/S0     &Small-scale SF&Off-center            &    10\\
SR28  &SDSS J235959.36$-$111131.5&359.997370&$-$11.192087&  0.104&10.72$\pm$  0.07& $-$0.37$\pm$  0.26&      1800&          &SB0      &Extended SF&Narrow ring(s)        &    25\\
SR29  &SDSS J161521.62+491841.5&243.840122& 49.311544&  0.059&10.38$\pm$  0.09& $-$0.38$\pm$  0.59&      2400&      2400&S0       &Extended SF&Narrow ring(s)        &    12\\
SR30  &SDSS J143607.80+041507.4&219.032527&  4.252061&  0.114&10.86$\pm$  0.07& $-$0.69$\pm$  0.27&       860&      1950&S0       &Small-scale SF&Off-center            &    15\\

\enddata
\tablenotetext{}{Coordinates and redshifts are taken from SDSS
  DR4. Stellar masses are in $M_{\odot}$ and SFRs in $M_{\odot}{\rm
    yr}^{-1}$, derived in \citet{s07}, and given for Chabrier
  IMF. Exposure times of WIYN 3.5 m $R$ band and \ha\ narrow-band
  imaging is given as $t_{\rm exp}$ ($R$) and $t_{\rm exp}$
  (NB). Asterisk next to the narrow band exposure time indicates that
  the \ha\ emission was confidently detected, while $\dagger$
  indicates tentative detection.  Optical and UV classification is
  explained in the text. UV sizes correspond to a visually estimated
  maximum span of UV emission in \hst\ ACS/SBC images. SR24 has not
  been successfully observed by the \hst\ and is not listed here.}
\normalsize
\end{deluxetable}
\clearpage
%\end{landscape}

\end{document}